\documentclass[11pt, a4paper]{article}
\usepackage[11pt]{extsizes}
\usepackage[utf8]{inputenc}
\usepackage{float}
\usepackage{graphicx}
\usepackage{amsthm}

\usepackage{latexsym}
\usepackage{amsmath}
\usepackage{amssymb}
\usepackage{amsfonts}
\usepackage{multirow}
\usepackage{amsthm}

\usepackage{enumerate}
\usepackage{graphics}
\usepackage{fullpage}
\usepackage{color}
\usepackage{comment}

\usepackage{bussproofs}

\usepackage{booktabs, tabularx}

\usepackage{xspace}
\usepackage{xcolor,colortbl}

\usepackage{mathrsfs}

\usepackage{amsfonts}

\usepackage{tikz}
\usetikzlibrary{trees,arrows,automata,shapes,decorations}

\newtheorem{thm}{Theorem}[section]
\renewcommand{\thethm}{\arabic{section}.\arabic{thm}}

\renewcommand{\thethm}{\arabic{section}.\arabic{thm}}
\newtheorem{defi}[thm]{Definition}
\newtheorem{lem}[thm]{Lemma}
\newtheorem{cor}[thm]{Corollary}
\newtheorem{rem}[thm]{Remark}
\newtheorem{exam}[thm]{Example}
\newtheorem{prop}[thm]{Proposition}
\newtheorem{fact}[thm]{Fact}

\newcommand{\ot}{\leftarrow}

\newcommand{\RN}[1]{%
  \textup{\uppercase\expandafter{\romannumeral#1}}%
}

\newcommand{\bM}{{\mathbf M}}
\newcommand{\mb}{\mathbf}

\newcommand{\myComment}[1]{}

\newcommand{\dual}{\mathbb{D}}
\newcommand{\edual}{\mathbb{E}}

\usepackage{fancyhdr}
\fancypagestyle{firstpage}{
    \fancyhead[L]{}
    \fancyhead[R]{\vspace{-2cm} \footnotesize{Published in {\it Journal of Philosophical Logic} (Springer)\\
\mbox{doi:10.1007/s10992-020-09588-z}\\
\textcopyright The Authors, 2021}}
}

\fancypagestyle{stylereset}{
\fancyhf{}
\fancyhead[L]{}
\fancyhead[R]{}
}

\begin{document}

\title{A Simple Logic of Functional Dependence}
%\subtitle{ }


%\author{XXX}
\date{}

\author{Alexandru Baltag\footnote{A. Baltag, Institute for Logic, Language and Computation (ILLC), University of Amsterdam, The Netherlands.} \, and
        Johan van Benthem\footnote{J. van Benthem, Institute for Logic, Language and Computation (ILLC), University of Amsterdam, The Netherlands \& Stanford University, Department of Philosophy \& Tsinghua University, Department of Philosophy.}}

\maketitle

\thispagestyle{firstpage}
\bigskip
\bigskip

\begin{abstract}

This paper presents a simple decidable logic of functional dependence LFD, based on an extension of classical propositional logic with dependence atoms plus dependence quantifiers treated as modalities,  within the setting of generalized assignment semantics for first order logic. The expressive strength, complete proof calculus and meta-properties of LFD are explored. Various language extensions are presented as well, up to  undecidable modal-style logics for independence and dynamic logics of changing dependence models. Finally, more concrete settings for dependence are discussed: continuous dependence in topological models, linear dependence in vector spaces, and temporal dependence in dynamical systems and games.
\smallskip\par\noindent
{\bf keywords}: Functional dependence \and Generalized assignment semantics \and Modal logic \and Epistemic logic \and Logics of dependence.
\end{abstract}

\bigskip
\bigskip

\section{\large{\textbf{Introduction: Toward a logic  of local dependence}}}

Dependence is a ubiquitous notion, pervading areas from probability to reasoning with quantifiers, and from informational correlation in databases to causal connections or interactive social behavior. How the Moon moves depends on how the Earth moves, and vice versa. What you will do in our current Chess game depends on how I play. And dependence, or independence, matters. Whether variables are dependent or not is crucial to probabilistic calculation. And as for qualitative reasoning, dependence is at the heart of quantifier combinations in logic.

\medskip

Now ubiquity does not mean unity: there need not be one coherent notion behind all talk of dependence in science or daily life.\footnote{There may not even be one uniform conception of dependence in logic running from, say, dependent choices in quantifier combinations as mentioned above to independence of sets of axioms. In particular, the latter widespread sense, studied for instance in \cite{deJChagrova}, \cite{Humber}, is not what is at issue in this paper.} Still, over the last century, various proposals have been made for a basic logic of reasoning about dependence and independence, witness publications such as \cite{Armstrong}, \cite{Lambal}, \cite{ABN},  \cite{HintSand}, \cite{Narens}, \cite{Vaana}. While some of these logics are weak calculi of pure dependence statements, others are very strong and second-order. And most of them are  non-classical: the propositional connectives break classical laws such as Tertium Non Datur, while the semantics differs radically from that of First Order Logic (FOL); being either a game semantics, or some higher-order version of first-order semantics, evaluating formulas on sets of assignments.

\bigskip

In this paper we explore one more perspective,  minimalistic in its simplicity: a  Logic of Functional Dependence between variables (LFD), based on classical logic. LFD is obtained by adding local dependence atoms to a known generalization of the usual semantics of First Order Logic, namely, the logic CRS of generalized assignment models, \cite{ABN}. CRS treats quantifiers as modalities over accessibility relations between assignments, and LFD adopts this view also for further modalities for reasoning about dependence.

\medskip

This new calculus, which offers a simple base level for analyzing reasoning about functional dependence, is decidable and yet reasonably expressive. Unlike the other approaches mentioned, it focuses on a `local' sense of dependence, which may be more fundamental than the usual global version. This locality brings LFD close to modal logic, but still allows it to cover many senses of dependence, both `ontic' in terms of linked behavior in the world, and `epistemic' in terms of information: learning something about one thing implies learning about another. Taking this minimal modal perspective, one can then take a fresh look at the surplus of many richer notions of dependence and correlation, and design further  logics for reasoning about these.

\pagestyle{stylereset}

\subsection{\textbf{Global and local dependence in a complete database}}

As an example, consider relational databases with tuples of values assigned to attributes, cf. \cite{AbitHV}. %\cite{tCKol}.
%The following  highly simplified database will serve our purpose here.
%introduces some key features of the models and notions to be used in this article.

\begin{exam}\label{Restaurant}  \emph{Here is a simple information structure.
The table below is meant as a \emph{complete database}, i.e., a full description of the restaurant situation in a small town, sorted by four variables (so-called `attributes'):
%\footnote{Variables are usually called ``attributes" in Database Theory, assignments are called ``tuples".}
Restaurant name, Food type, Price range, and Location.}

%\vspace{-3mm}

\begin{center}
\begin{tabular}{|l l l l|}
  \hline
  Restaurant & Food & Price & Location\\
  \hline
   Roma & Italian & Moderate & Center\\
   Hasta La Pasta & Italian & Cheap & Center\\
   Mama Makan & Indonesian & Moderate & South\\
   Bunga Mawar & Indonesian & Cheap & West\\
   Wilde Zwider & Dutch & Expensive & East\\
   Greetje & Dutch & Expensive & West\\
   \hline
\end{tabular}
\end{center}

%\vspace{-3mm}
\medskip

\noindent \emph{A variable $v$ \emph{depends} on a variable $u$ (or on a \emph{set} $U$ of variables) if, whenever the value of $u$ is the same in two rows (or the values of all variables in $U$ are the same in the two rows), so is the value of $v$. This is in line with the intuitive sense of dependence: the value of $u$ \emph{determines} the value of $v$. From the table, we see that each of the attributes $Food$, $Price$ and $Location$ depends on $Restaurant$ (each restaurant offers a unique type of food, has a unique price range and a unique location). But neither $Restaurant$ nor $Price$ nor $Location$ depend on $Food$ (e.g., there are two Indonesian restaurants, with different price ranges and locations),
and $Food$ does not depend on $Price$ either (there are cheap Italian and Indonesian places). These facts imply others. E.g.,  $Restaurant$ does not depend on $Price$:  if it did, then by transitivity and the fact that $Food$ depends on $Restaurant$, $Food$
would depend on $Price$, quod non. In fact, $Restaurant$ does not even depend on $Price$ and $Food$ taken together (i.e. on the set $\{Price, Food\}$): both Wilde Zwider and Greetje are expensive Dutch restaurants. But $Restaurant$ does depend on $Price$ and $Location$ taken together:
%indeed,
for every possible price range and location, there is at most one restaurant offering that food type in that location.}

\medskip

\emph{What was listed so far are \emph{global} dependencies. However, underlying these are many \emph{local} dependencies in the table: given a current row, a variable $v$ locally depends on $u$ if, every row matching the current value of $u$, also matches the current value of $v$. For example, in the fifth row of the table, $Price$ depends on $Food$
(since all Dutch restaurants are expensive),
$Food$ locally depends on $Price$
(all expensive restaurants are Dutch), and  $Restaurant$ locally depends on $Location$ (the only restaurant located in the East is Wilde Zwider). Again, some of these things follow from others. E.g., still at the fifth row, the fact that $Food$ locally depends on $Location$  follows by transitivity from the local dependence of \emph{Restaurant} on \emph{Location} plus the global (and hence also local) dependence of $Food$   on $Restaurant$.}\footnote{Databases may also satisfy \emph{non-functional} dependencies. E.g., the above table satisfies the rule `if not Dutch, then  not Expensive'. Such constraints, too, can be expressed in the dependence logic of this paper.}
\end{exam}

A set of variable assignments as above, with a designated current assignment, connects in an obvious way with the semantics of first-order logic. At the same time, the distinction local vs. global is crucial to the essentially \emph{modal} approach taken in this paper.

\subsection{\textbf{Dependence in  first-order semantics: CRS logic}}

%As announced, our aim is to stick as close as possible to the classical Tarskian semantics of first order logic.
%So we start our investigation with the following natural question: \emph{why} can't first order logic capture directly the phenomenon of functional dependency? In fact, we know that it can express dependence in an \emph{indirect} manner, via complex functional terms.\footnote{Moreover, we will see later that, when the number of variables is finite, there is an alternative indirect way to capture it in a purely relational version of FOL, by assigning a privileged role to a special relation that binds all the variables.} But why not capture directly statements of the form ``$x$ is functionally determined by $y$ and $z$"?

What does dependence have to do with first-order logic? In classical FOL, distinct variables have no intrinsic meaning and are fully inter-changeable. No correlations between their values are allowed, beyond the fact that they all range over the given domain. Accordingly, first-order models are `full': all possible variable assignments are available, and the FOL quantifiers can arbitrary reset the values of any variables, while keeping the values of all the other variables fixed. This effectively amounts to a strong \emph{independence} assumption,  reflected for instance in the commutation law $\exists x \, \exists y \, \varphi \leftrightarrow \exists y\, \exists x \,\varphi$. Often seen as a  triviality, this validity is in fact a very strong symmetry principle, which is entailed by the above strong independence assumption.

\medskip

To allow for dependencies between variables, a simple solution is to just drop the `fullness' assumption, while keeping the good features of FOL such as its perspicuous syntax and compositional truth definition. This approach, known as \emph{generalized assignment semantics}, was developed in the 1990s. A `generalized assignment model' $(M, A)$ consists of a first order model $M$ and a family $A$ of `admissible' variable assignments $s:V\to O$ (with $V$ the  variables and $O$ the objects in $M$), circumscribing the global states that can occur when evaluating the first-order quantifiers. These models widen the scope of applicability of FOL to settings with significant correlations, or even functional dependencies, between variables. As stated in \cite{ABN}, p. 46, generalized assignment semantics {\it ``models the natural phenomenon of \emph{dependencies}
between variables: which occurs when changes in value for one variable
$x$ may induce, or be correlated with, changes in value for another
variable $y$. (...) Dependence cannot be modeled
in standard Tarskian semantics, which modifies values for variables
completely arbitrarily."} In this perspective, classical FOL describes the special case of `full models' in which all possible assignments are available, i.e. $A=O^V$.
%\footnote{Sets of assignments are called `teams' in later approaches to dependence logic: cf. Hodges 1998, V{\"a}{\"a}n{\"a}nen 2001.}

%Classical FOL semantics thus corresponds to restricting the semantics only to full models. But in every non-full model, there will be some correlations between variables, and this will be reflected in the satisfiability of formulas contradicting some of the standard $FOL$ principles. Hence, the name ``dependency models", used occasionally [REF] to refer to generalized-assignment models.\footnote{The connection with the modern terminology in Dependence Logic becomes obvious if we notice that a dependency model is a just pair of a FOL model $M$ together with a `team' $A$, in the sense of Hodges [REF].}

\medskip

What is the meaning of quantifiers in these generalized models? The original generalized assignment semantics, known as the logic CRS\footnote{The technical name CRS stands for `cylindric relativized set algebra', referring to algebraic origins in \cite{Nem85}.}, simply restricts the usual Tarskian definition to the family $A$ of admissible assignments. Unlike in FOL, polyadic quantifiers such as $\forall x y\, \varphi$ can no longer be reduced to iterated monadic ones $\forall x \forall y \, \varphi$. Hence, CRS takes polyadic quantifiers $\forall X \, \varphi$ as a primitive notion, for every finite \emph{set} of variables $X\subseteq V$ (while defining monadic quantifiers $\forall x\, \varphi$ as just an abbreviation for $\forall \{x\} \varphi$): for any assignment $s\in A$, we put
$$s\models \forall X \varphi \,\, \, \mbox{ iff } \,\,\,
 t\models \varphi \mbox{ for every $t\in A$ satisfying $s(y)=t(y)$ for all $y\in V-X$}.$$

As we have seen, dependencies between variables are present in non-full models. In fact, the language can spot these dependencies in an \emph{implicit}  way: via the \emph{failure} of some classical FOL validities in the weaker logic CRS. For instance, if a dependence model invalidates the above law $\exists x\exists y \phi \to \exists y \exists x \phi$, then there exist some non-trivial correlations between variables.

\medskip

A key goal of generalized assignment semantics was analyzing the causes of the undecidability of  validity for FOL. The intent was to decouple the desideratum of a compositional semantics for the first-order language from additional mathematical assumptions (about existence of all possible functional assignments) that increase complexity. Indeed, while CRS semantics is clearly compositional, the set of validities is decidable, forming roughly a core calculus of monotonicity and persistence reasoning inside full predicate logic.\footnote{Further  axioms such as the above commutation law $\exists x \, \exists y \, \varphi \leftrightarrow \exists y\, \exists x \,\varphi$ then impose a confluent Church-Rosser structure on the set of assignments, leading to undecidability arguments via encoding tiling problems, \cite{Marx}.} Thus, CRS makes a distinction between general simple inferences inside FOL and more complex reasoning relying on special mathematical existence assumptions.\footnote{For much more information on CRS and related modal logics, cf. \cite{Venema}, \cite{ELD}, \cite{MarxVenema}.}

\smallskip

This lower complexity may be understood as a result of `modalization', \cite{ELD}. The above analysis also works on abstract state models for the first-order language without underlying objects, where first-order logic becomes a modal logic. This modal perspective will be significant in what follows, as it explains how a logic of dependence can be decidable.

\smallskip

Still, from a dependence perspective, the CRS quantifiers have some peculiar features. Notably, the  \emph{Locality} property of FOL fails: the truth value of a CRS-formula $\varphi$ need not depend only on the values of its free variables, it may well depend on values of variables that do not even occur in $\varphi$. This `dependence on irrelevant variables' is an artifact of the specific way in which CRS generalizes FOL semantics by letting only the values of $X$ vary, \emph{keeping the values of all other variables fixed}, including the ones not occurring at all in the given formula.

\medskip

This problem was noticed early on in the CRS literature, leading to an alternative proposal for generalizing FOL quantifiers.\footnote{See e.g. Marx \cite{Marx}, who attributes the proposal to Venema.} Since these alternative operators do satisfy Locality, we will call them \emph{local quantifiers},  denoted here by $\forall_X \varphi$:
$$s\models \forall_X \varphi \,\, \, \mbox{ iff } \,\,\,
 t\models \varphi \mbox{ for every $t\in A$ with $s(y)=t(y)$ for all $y\in Free(\forall_X \varphi)=Free(\varphi)-X$},$$
where $Free(\varphi)$ is the set of free variables in $\varphi$. This fixes only the values of the \emph{actually occurring} free variables that do \emph{not} belong to $X$, allowing all the others to vary.

\smallskip

Note that in full models (with $A=D^V$), both $\forall_X\varphi$ and $\forall X \varphi$ collapse to classical FOL quantifiers; so they are both entitled to play the role of generalized FOL quantifiers. Even so, both versions of CRS still have a major drawback: there is no explicit way to say that a variable $x$ functionally depends on other variables. Moreover, no new validities are added that capture interesting laws of dependence. For this, further steps are needed, to be previewed now.

\begin{rem} \emph{The language of CRS also supports modalities for \emph{substitutions}.
A formula $[y/x]\varphi$ ($\varphi$ with all free occurrences of $x$ replaced by $y$, where no substituted $y$ becomes bound) is true at an assignment $s$ if there is an available assignment $t$ in the model equal to $s$ except that
$t(x) = s(y)$ with   $\varphi$ true at $t$. There is also a natural extension for simultaneous substitutions $[{\bf y}/{\bf x}]\varphi$, which do not reduce to iterated single ones. The usual recursive definition of syntactic substitution in FOL now expresses various substantial properties of the (in general, partial) semantic substitution function on assignments and its interactions with CRS quantifiers, cf. \cite{ELD}. For the proof theory of this modal view of substitution,
cf. \cite{MarxVenema}.}
\end{rem}

%the generalized assignment semantics has the advantage that it lowers the complexity: when considered on dependency models FOL (in either of the above two versions) becomes a \emph{decidable} logic (see [REF: Andreka, van Benthem and Nemeti]). We can understand this move as a ``modalization" of FOL: by restricting quantifier range to the ``accessible" worlds (a.k.a. assignments), our quantifiers become Kripke-style modalities, and as a result the complexity goes down.

\subsection{\textbf{Explicit logic of local dependence}}

As we saw, CRS is an  `implicit' logic of dependence. In this paper, we add the explicit syntactic {\it atomic dependence formulas} $D_X y$ of \cite{Vaana}, now read locally as: \emph{$X$ locally determines (the value of) $y$}, or \emph{$y$ locally depends on $X$}.
%(or \emph{$y$ locally depends on $X$})
%\footnote{This natural next step seems to have never been taken in the literature on generalized assignment semantics.}
These atomic formulas are interpreted at assignments $s\in A$ using the local dependence relation $D_X^sy$, saying that
all admissible assignments that keep the values of $X$ fixed to the \emph{current} ones also fix the value of $y$:
$$s\models D_X y \,\, \, \mbox{ iff } \,\,\, s(y)=t(y)
\mbox{ holds for every $t\in A$ satisfying $s(x)=t(x)$ for all $x\in X$}.$$
Next, we reconsider the quantifiers. From a dependence perspective, it is natural to introduce \emph{dependence modalities} or \emph{dual quantifiers} $\dual_X \varphi$, which 'fix' the values of $X$ to the current ones.  More precisely, like the dependence atoms, these talk about all the assignments that keep $X$ equal to its current value(s), saying that they also fix the truth value of $\varphi$ to `true':
$$s\models \dual_X \varphi \,\, \, \mbox{ iff } \,\,\, t\models \varphi
\mbox{ holds for every $t\in A$ satisfying $s(x)=t(x)$ for all $x\in X$}.\footnote{The analogy between $D_X y$ and $\dual_X \varphi$ is made more precise in \cite{Baltag2016}:  introducing natural Boolean variables $?\varphi$,  with value $1$ at any assignment $s\in A$ satisfying $\varphi$ (and $0$ at all others), validates the equivalence $\dual_X \varphi \leftrightarrow ( \varphi\wedge D_X ?\varphi)$. This can be turned into a definition, thus unifying the two types of LFD dependence statement.}$$
We read $\dual_X \varphi$ as \emph{$X$ locally determines the truth of $\varphi$}. Recall that in standard FOL, `free' variables are the ones whose current values are kept fixed (while the values of `bound' variables are ignored as irrelevant). This fixing the values of $X$ explains why we  sometimes  call dependence modalities $\dual_X \varphi$    `dual quantifiers': they `free' the variables in $X$ (rather than binding them), while \emph{binding all the other variables} (in $V-X$, regardless of whether they occur in the formula).

\medskip

Like the local universal quantifiers $\forall_X \varphi$, dependence modalities do satisfy Locality.
But they appear to be more fundamental: indeed, $\forall_X\varphi$ is simply definable via the equivalence $\forall_X\varphi \leftrightarrow\dual_{Free(\varphi)-X}\varphi$, whereas the converse is not as straightforward.\footnote{One can indeed go the other way around, but via a more complicated formula. Let $\top_X$ be an abbreviation for any  tautology whose free variables are exactly the ones in $X$. Then $\dual_X\varphi$ is equivalent to $\forall_{Free(\varphi)-X}(\varphi\wedge \top_X)$. As for non-local CRS quantifiers, they are equally expressive to the dependence modalities when $V$ is \emph{finite}, via the equivalences
$\forall X \varphi\leftrightarrow\dual_{V-X} \varphi$
and $\dual_X \varphi \leftrightarrow \forall (V-X) \varphi$. When $V$ is infinite, the two notions seem to be independent of each other (at least with our syntax, allowing for quantifiers only over \emph{finite} sets of variables).} Note also here that, like both FOL and CRS quantifiers $\forall X$ (but in contrast to local quantifiers $\forall_X$), dependence modalities validate the standard Distribution axiom $\dual_X(\varphi\to \psi)\to (\dual_X\varphi\to\dual_X\psi)$.\footnote{See the footnote to Example \ref{valid} for a counterexample to Distribution for $\forall_X$. The deeper reason for this difference is that, as we will see, the FOL and CRS quantifiers, as well as the dependence modalities, are in fact normal \emph{relational modalities}, quantifying over assignments that are accessible via some accessibility relation ($=_{V-X}$ or $=_X$), while the local quantifiers are not modalities of this kind.} Dependence modalities can also  quantify over \emph{all} assignments in $A$: taking $X$ to be the empty set yields the \emph{universal modality} $\rotatebox[origin=c]{180}{$\forall$}\varphi := \dual_\emptyset \varphi$, saying that \emph{all} admissible assignments satisfy $\varphi$. As a consequence,  \emph{global dependence} of $y$ on $X$ can be expressed as $\rotatebox[origin=c]{180}{$\forall$} D_Xy$.

%\noindent One might call these operators $\dual_X$ \emph{dual quantifiers}, or ``fixing" quantifiers. In a sense, they ``free" variables $X$ (rather than binding them), but they also bind (quantify) all the other variables.\footnote{This dual formulation has in fact been proposed in the CRS tradition. Marx 199{\bf x} studies quantifiers interpreted as follows: $s\models \forall_X \varphi \,\, \, \mbox{ iff } \,\,\,
 %t\models \varphi \mbox{ for every $t\in A$ satisfying $s(y)=t(y)$ for all $y\in Free(\varphi)-X$}.$}

\medskip

The resulting logic of functional dependence LFD is more expressive than may meet the eye, as will become clear in what follows. Also, while capturing the main properties of functional dependence,  it retains all classical Boolean operators with their standard laws; thus demonstrating that dependence is not an intrinsically non-classical phenomenon. Neither is basic reasoning about dependence necessarily complex, LFD is simple and well-behaved,
with transparent axiomatizations and good meta-properties: decidability, forms of the finite model property, compactness, strong interpolation, and a form of cut elimination. Of course, this does not come for free. As always in logic, system design involves a balance between expressive power and other nice system properties. The more expressive the language, the more complex the validities -- or stated conversely, the more well-behaved the logic, the less expressive the language.  On the minimal basis language of LFD, however, one can  analyze just which additional features in modeling dependence (and independence) force greater complexity for a logical system. Moreover, the \emph{modal} flavor of LFD brings interesting connections with epistemic logics \cite{FHMV,vEGWang,Baltag2016}, interrogative and inquisitive logics \cite{BentMin,CGRoe,DepQ}, and situation-theoretic logics of informational correlations, \cite{BentMart}. Finally, as we shall show, LFD offers a platform for studying concrete notions of dependence in many fields in a way that imports only a  minimum of logical complexity.

\subsection{\textbf{Structure of this article}}

Section 2 defines our models, giving a structural characterization of dependence. Section 3 introduces the logic LFD, together with a translation into FOL, a discussion of the differences between LFD quantifiers and the classical ones, and an equivalent modal relational semantics. The tandem of first-order and modal views will recur throughout the paper. Section 4 proves the decidability of LFD using object-free `type models', while Appendix A has proofs of decidability and completeness using standard modal techniques. Section 5 presents a Hilbert-style axiomatization and a sequent calculus admitting a form of cut elimination, as well as interpolation and Beth definability results (with proofs in Appendix B). Section 6 explores extensions of LFD, including function terms, identity, independence,  informational correlation, and dynamic modalities over changing dependence models. Section 7 looks at richer settings for dependence: including vector spaces, topological models, and dynamical systems. Section 8 draws comparisons with  other approaches, including some discussion of their expressive surplus over LFD and questions raised by this. Conclusions and further prospects are found in Section 9.

\section{\large{\textbf{State spaces, dependence graphs, functions}}}

%The models for dependence to be used in this article were proposed in van Benthem 1995, following an earlier algebraic literature  (cf. Nemeti 1985, Venema 199x). They arose in a semantic analysis of first-order logic with models having a designated set of `available assignments' (functions from variables to objects) freed from the usual independence assumptions for assigning values to variables, leading to a decidable set of validities. In particular, the existence of `gaps' in the full function space of all assignments over a model can make variables $x, y$ dependent: changing the value for $x$ in the given set of assignments may only be possible by also changing the value of $y$.

The starting point of this paper are the basic properties of semantic dependence relations, which
will be determined here. Also a natural duality will emerge with explicit functional definitions for
dependence, as well as appealing connections with consequence relations.

\vspace{-2mm}

\subsection{\textbf{Dependence models}}

Throughout this paper, we assume given a set of \emph{variables} $V$ and a relational vocabulary $(Pred, ar)$,  where $Pred$ is a set of \emph{predicate symbols} and $ar: Pred\to N$ is an \emph{arity} map, associating to each predicate $P\in Pred$ a natural number $ar(P)$.

\begin{defi}[\textbf{Dependence models, agreement, local dependences}] A \emph{dependence model} $\bM$ is a pair $\bM=(M, A)$ of a (relational) FOL model $M=(O,I)$ with a domain $O$  of \emph{objects} and interpretation map $I$ (sending each predicate symbol $P\in Pred$ of arity $n$ into a set $I(P)\subseteq O^n$ of $n$-tuples of objects), together with a set $A\subseteq O^V$ of \emph{admissible assignments} of objects to variables.

A dependence model is  \emph{full} if all possible assignments are admissible, i.e., if $A=O^V$.
For assignments $s\in A$ and sets $X\subseteq V$, we put $s\hspace{-0.1cm}\upharpoonright\hspace{-0.1cm} X$ for the \emph{restriction} of $s$ to domain $X$.\end{defi}

\begin{defi}[\textbf{Agreement, local dependence, atoms}]\label{local dep} In dependence models, we define three basic relations: (a) for each set $X\subseteq V$ of variables, an \emph{agreement relation} $s=_X t$ on assignments $s,t\in A$, (b) for each  $s\in A$, a \emph{local dependence} relation $D_X^s y$ between sets $X\subseteq V$ and variables $y\in V$, and (c) for each $n$-ary predicate $P$ and each assignment $s\in A$, an $n$-ary relation $P^s\subseteq V^n$ on variables (where we use the notation $=_y$ for $=_{\{y\}}$):
\vspace{-1mm}
$$s=_X t \mbox{} \,\,\, \mbox{ iff } \,\,\, s\hspace{-0.1cm}\upharpoonright\hspace{-0.1cm} X=t\hspace{-0.1cm}\upharpoonright\hspace{-0.1cm} X,$$
$$D_X^s y \mbox{} \,\,\, \mbox{ iff }  \,\,\, s=_X t \mbox{ implies } s=_y t \mbox{ for all $t\in A$},$$
$$P^s x_1 \ldots x_n  \,\,\, \mbox{ iff }  \,\,\, I(P)(s(x_1), \ldots, s(x_n)) \mbox{ holds},$$

\vspace{-3mm}

If $s=_X t$, $s$ and $t$ are said to \emph{agree on $X$}, and
if $D^s_X y$
we say that \emph{$y$ locally depends on $X$ at $s$}. For any $Y\subseteq V$, we write $D^s_X Y$ if $D^s_X y$ holds for \emph{all} $y\in Y$. Finally, we skip the set brackets for singletons, writing $D^s_xY$ for $D^s_{\{x\}}Y$, and $D^s_xy$ for $D^s_{\{x\}}\{y\}$.
\end{defi}

\begin{defi}[\textbf{Global dependence}]\label{global dep}
The \emph{global dependence} relation $D^{\bM} \subseteq {\mathcal P}(V)\times V$ quantifies over all assignments in $A$: \emph{$y$ depends on $X$ in $\bM$}, written $D^{\bM}_X y$, if $D_X^s y$ holds locally at all  assignments $s\in A$. As for local dependence, this notation is extended to sets $Y\subseteq V$, by writing $D^{\bM}_X Y$ if $D^\bM_Xy$ holds for  \emph{all} $y\in Y$; and again, set brackets are skipped for singletons. When the context is clear,  superscripts $\bM$ for current models will be dropped.  %e.g., writing $D^{\bM}_xy$ for $D^{\bM}_{\{x\}}y$.
\end{defi}
%\vspace{-2mm}

Note that our global dependence statement $D^{\bM}_X y$ matches the semantic clause for the so-called \emph{dependence atom} $=(X;y)$ introduced in V{\"a}{\"a}n{\"a}nen's Dependence Logic \cite{Vaana}, when interpreted on the `team' $A$ of all admissible assignments.\footnote{More generally, structures resembling our dependence models occur in areas such as epistemic logic, \cite{Baltag2016}, temporal logic,  \cite{PaRam}, and situation theory, \cite{BarSel}, \cite{LDII}.}

\subsection{\textbf{Dependence graphs}}
The basic structural properties of dependence relations are as follows.

%%%%% SONJA CHANGED definition into defi in the code below when opening and closing the definition
\begin{defi}\label{struct prop} Let $R \subseteq {\mathcal P} (V)\times V$ be a relation between sets of variables and variables. Using the same conventions as for the dependence relation $D$ above (writing $R_Xy$ instead of $(X,y)\in R$, $R_XY$ as an abbreviation for $\bigwedge_{y\in Y} R_Xy$, and skipping set brackets for singletons), we say that:

\vspace{-1mm}

\begin{itemize}

\vspace{-1mm}
\item\quad $R$ satisfies \emph{Reflexivity} if $R_xx$ holds for all $x \in V$
\vspace{-1mm}
\item\quad $R$ satisfies \emph{Transitivity} if  $R_XY$ and $R_YZ$ imply $R_XZ$
\vspace{-1mm}
\item\quad $R$ satisfies \emph{Monotonicity }if $R_Xy$ and $X\subseteq Z$ imply $R_Zy$
\vspace{-1mm}
\item\quad $R$ satisfies the \emph{Projection} property if $R_Xx$ holds for all $x\in X$
\vspace{-1mm}
\item\quad $R$ satisfies the \emph{Inclusion} property if $R_XY$ holds for all $Y\subseteq X$
\vspace{-1mm}
\item\quad A variable $y\in V$ is an \emph{$R$-constant} iff $R_\emptyset y$ holds.
\end{itemize}
\vspace{-2mm}
\end{defi}

\vspace{-2mm}

The following is easy to see:

\begin{fact}\label{projection}
If $R\subseteq {\mathcal P} (V)\times V$  satisfies Transitivity, then the following are equivalent:
\begin{enumerate}
\item\quad $R$ satisfies Reflexivity and Monotonicity;
\item\quad $R$ satisfies the Projection property;
\item\quad $R$ satisfies the Inclusion property.
\end{enumerate}
\end{fact}

It is well-known that the combination of Reflexivity, Transitivity and Monotonicity provides a characterization of classical logical consequence, cf. \cite{Scott71}. The following two results show that the same three properties characterize the relation of (local and global) dependence:\footnote{However, in our formal axiomatizations in Section 5, we will use the equivalent combination of Projection and Transitivity, cf. Fact \ref{projection}.}

\begin{fact} For every dependence model $\bM=(M, A)$ and assignment $s\in A$, both global dependence $D_X^{\bM} y$ and local dependence $D_X^s y$ satisfy Reflexivity, Transitivity and Monotonicity. For both relations $R\in \{D^{\bM}, D^s\}$, the $R$-constants are  exactly the variables $y\in V$ whose value is the same for every assignment in $A$.
\end{fact}

Fact 2.5 follows immediately from Definitions \ref{global dep} and \ref{local dep}. The converse takes more work:

%\footnote{Monotonicity can be dispensed with given the remaining properties, but we are not aiming for  minimality here.}

%Turning things way around, here is a representation result for the preceding properties.

\begin{prop}\label{Armstrong completeness}
\vspace{-1mm}
\begin{enumerate}

\item\quad For every relation $R \subseteq {\mathcal P} (V)\times V$ satisfying Reflexivity, Transitivity and Monotonicity, there is a dependence model $\bM$ whose \emph{global} dependence relation $D^\bM$ coincides with $R$.  Moreover, if $V$ is finite, then $\bM$ can be taken to be finite as well, of size bounded by $2^{|V|}$.

\item\quad For every relation $R \subseteq {\mathcal P} (V)\times V$ satisfying Reflexivity, Transitivity and Monotonicity, there is a dependence model $\bM_R$ where $R$ coincides with  \emph{all the local} dependence relations $D^s$ (at \emph{all} assignments $s\in A$), and hence it also coincides with the global dependence $D^\bM$. Moreover, if $V$ is finite, then $\bM_R$ can be taken to be finite as well, of size bounded by $2^{2^{|V|}}$.

\item\quad Let ${\mathcal R}\subseteq {\mathcal P} (V)\times V$ be a family of relations satisfying Reflexivity, Transitivity and Monotonicity, s.t. all relations ``agree on constants" (i.e.,  $R_\emptyset y$ iff $R'_\emptyset y$, for all $y\in V$ and $R,R'\in {\mathcal R}$). Then  ${\mathcal R}$ coincides with the family $\{D^s: s\in A\}$ of \emph{all local dependence relations} of some dependence model $\bM_{\mathcal R}$.
%Moreover, the global dependence relation $D^\bM$ on $\bM:=\bM_{\mathcal R}$ is the intersection $\bigcap {\mathcal R}$.
Moreover, if $V$ is finite then $\bM_{\mathcal R}$ can be taken to be finite.
\end{enumerate}
\end{prop}

\begin{proof}\quad \,\,
For a start, we need some preliminary notations and results. Let $R \subseteq {\mathcal P} (V)\times V$ be a relation satisfying Reflexivity, Transitivity and Monotonicity. A subset $X\subseteq V$ is \emph{$R$-closed} if we have $y\in X$ for all $y\in V$ satisfying $R_Xy$. Let $\Gamma$ be the family of all $R$-closed subsets of $V$. Note that $\Gamma$ is closed under arbitrary intersections\footnote{Let $\{X^i:i\in I\}\subseteq \Gamma$ be a family of $R$-closed sets, with $X:=\bigcap_{i\in I} X^i$. To show that $X$ is $R$-closed, let $R_Xy$ for some $y\in V$. By Monotonicity, $R_{X^i}y$ for all $i\in I$, and so by $R$-closure, $y\in X^i$ for all $i\in I$, i.e. $y\in X$.}. [Note: for greater readability in what follows, we have put the simple proof of this and some later auxiliary statements in footnotes.] Also, it is immediate that the family $\Gamma$ contains the set $V$ of all variables. We put $\tilde{X}:=\{y\in V: R_Xy\}$ for \emph{the $R$-closure of $X$}, which is  the least $R$-closed set s.t. $X\subseteq \tilde{X}$.\footnote{$X\subseteq \tilde{X}$ follows from the fact that $R_Xx$  holds for all $x\in X$, by Reflexivity and Monotonicity. To see that $\tilde{X}$ is $R$-closed, let $z\in V$ be s.t. $R_{\tilde{X}}z$. This, together with the fact
that $R_X \tilde{X}$ (by the definition of $\tilde{X}$) yields
$R_{X}z$ (by Transitivity), i.e., $z\in\tilde{X}$. Finally, if $Y$ is any  $R$-closed set with $X\subseteq Y$, we show that $\tilde{X}\subseteq Y$. Let $y\in \tilde{X}$, i.e., $R_Xy$. Then $R_{Y}y$ (by Monotonicity and $X\subseteq Y$), and therefore $y\in Y$ (by the $R$-closure of $Y$).}

\smallskip

\emph{Proof of Part 1.} Let $R \subseteq {\mathcal P} (V)\times V$ satisfy Reflexivity, Transitivity and Monotonicity. Consider the model $\bM=(O, I, A)$ with (a) $O=V\cup{\mathcal P}(V)$,  (b) the interpretation map $I$ makes all predicates \emph{false}, and (c) the family $A=\{s_X: X\in \Gamma\}$ consists of assignments $s_X$, one for each $R$-closed set $X$, with

\vspace{2mm}

$s_X(x)=x$ if $x\in X$, and $s_X(x)=X$ if $x\not\in X$.

\vspace{2mm}

\noindent Note that $|A|=|\Gamma|\leq |{\mathcal{P}}(V)|=2^{|V|}$.
The model $\bM$ validates the following two claims, for all $Y,Z\in \Gamma$ and $U\subseteq V$:

\vspace{1mm}

(a) $s_Y=_U s_Z$ \,iff \,either $Y=Z$ or $U\subseteq Y\cap Z$.

\vspace{1mm}

(b) $D^{\bM}_X y$ holds\, iff \, $R_X y$ holds.

\vspace{2mm}

\emph{Claim (a)}: This follows from the definition of the assignments $s_X$ via the following sequence of equivalences:  $s_Y=_U s_Z$  \, iff \, $s_Y(x)=s_Z(x)$ for all $x\in U$ \, iff \,  either $Y=Z$ or $s_Y(x)=s_Z(x)=x$
for all $x\in U$\,  iff \, either $Y=Z$ or $x\in Y\cap Z$ for all $x\in U$.

\vspace{1mm}

\emph{Claim (b)}: From left to right, let  $D_X^\bM y$, and consider the assignments  $s_{\tilde{X}}$ and $s_V$, with $V$ the set of all variables. Note that $s_V(y) = y$ for all variables $y$. By (a), we have $s_V=_X s_{\tilde{X}}$ (since $X\subseteq \tilde{X}=V\cap\tilde{X}$). Therefore, since $D_X^\bM y$, $s_V=_y s_{\tilde{X}}$, and this means by the definition of the two assignments that  $s_V(y)= s_{\tilde{X}}(y) = y$. In particular, then,  $y \in \tilde{X}$, i.e., $R_Xy$.

\vspace{1mm}

%but that we do \emph{not} have $R_X y$. Then $y\not\in \tilde{X}$ (by the definition of the $R$-closure $\tilde{X}$). Take the assignments $s_V$ (where $V$ is the set of all variables) and $s_{\tilde{X}}$. By (a), we have $s_V=_X s_{\tilde{X}}$ (since $X\subseteq \tilde{X}=V\cap\tilde{X}$)
%but $s_V\not=_y s_{\tilde{X}}$ (since $y\not\in \tilde{X}=V\cap\tilde{X}$), contradicting the assumption that $D_X^\bM y$ holds.

From right to left, assume that  $R_X y$. To show that $D_X^\bM y$ holds, let $s_Z, s_U\in A$ (with $Z,U\in \Gamma$) be any two assignments with $s_Z=_X s_U$. By (a), $s_Z=_X s_U$ implies that either $Z=U$ or $X\subseteq Z\cap U$. In the first case, $Z=U$ immediately gives  $s_Z=_y s_U$, as desired. In the second case,  $R_X y$ and $X\subseteq Z\cap U$ imply  $R_Z y$, $R_U y$ by Monotonicity, which means by the $R$-closure of $Z, U$ that  $y\in Z$, $y \in U$, By  definition then $s_Z(y) = s_U(y) = y$.\footnote{Closer inspection of this argument shows that the local dependence relation at the special assignment $s_V$, with $V$ the set of all variables, actually equals the given relation $R$.}

\vspace{1mm}
%hence $y\in Y\cap Z$ (since $Y\cap Z$ is $R$-closed, being the intersection of two $R$-closed sets $Y,Z\in \Gamma$), and thus by (a) we have $s_Y=_y s_Z$, as desired.

\medskip

\emph{Proof of Part 2.} Let $R \subseteq {\mathcal P} (V)\times V$ satisfy Reflexivity, Transitivity and Monotonicity. For each $x\in V$, we define a binary relation $\sim_x$ on \emph{families} of $R$-closed sets
$\mathcal{A},\mathcal{B}\in\mathcal{P}(\Gamma)$,  by putting:
$$\mathcal{A}\sim_x \mathcal{B} \,\, \mbox{ iff } x\in \bigcap (\mathcal{A}\bigtriangleup \mathcal{B}),$$
where $\mathcal{A}\bigtriangleup \mathcal{B}:= (\mathcal{A}-\mathcal{B})\cup (\mathcal{B}-\mathcal{A})$ is the symmetric difference of the two families.

It is easy to check that each $\sim_x$ is an equivalence relation.\footnote{Reflexivity follows since $\bigcap(\mathcal{A}\bigtriangleup \mathcal{A})=\bigcap\emptyset=V$. Symmetry follows from the commutativity of symmetric difference.  Transitivity follows from the fact that  $\mathcal{A}\bigtriangleup \mathcal{C}\subseteq (\mathcal{A}\bigtriangleup \mathcal{B})\cup (\mathcal{B}\bigtriangleup \mathcal{C})$, which implies that $\bigcap (\mathcal{A}\bigtriangleup \mathcal{C})\supseteq \bigcap(\mathcal{A}\bigtriangleup \mathcal{B})\cap
\bigcap (\mathcal{B}\bigtriangleup \mathcal{C})$. If $\mathcal{A}\sim_x \mathcal{B}\sim_x \mathcal{C}$, then $x\in \bigcap (\mathcal{A}\bigtriangleup \mathcal{B}) \cap \bigcap (\mathcal{B}\bigtriangleup \mathcal{C})$, hence $x\in \bigcap (\mathcal{A}\bigtriangleup \mathcal{C})$, i.e. $\mathcal{A}\sim_x \mathcal{C}$.}
For any family $\mathcal{A}\subseteq \Gamma$ and variable $x\in V$, we  denote by $[\mathcal{A}]_x$ the equivalence class of $\mathcal{A}$ modulo $\sim_x$.

We construct now a model $\bM_R=(O_R, I_R, A_R)$ with: $O_R=\{[\mathcal{A}]_x: \mathcal{A}\subseteq\Gamma, x\in V\}$, i.e., all the equivalence classes modulo all the relations $\sim_x$, the interpretation $I_R$ makes all predicates \emph{false}; and $A_R=\{s^{\mathcal A}: \mathcal{A}\subseteq \Gamma\}$ consists of assignments $s^{\mathcal A}$ with $s^{\mathcal A}(x):=[\mathcal{A}]_x$ for all $x\in V$. Note that, if $V$ is finite, then $\bM_R$ is finite as well, and in fact $|A_R|\leq |\mathcal{P}(\Gamma)|\leq |\mathcal{P}(\mathcal{P}(V))|=2^{2^{|V|}}$.

This model validates the following claims, for all $s^{\mathcal A},s^{\mathcal B}\in A_R$ and $U\subseteq V$:
\vspace{2mm}

(a) $s^{\mathcal A}=_U s^{\mathcal B}$ \,iff\, $U\subseteq \bigcap (\mathcal{A}\bigtriangleup \mathcal{B})$.

\smallskip

(b)   $D_X^{s^{\mathcal A}} y$ holds in $\bM_R$ \, iff \, $R_X y$.

\vspace{2mm}

\emph{Claim (a)}: This follows directly from the definition of the assignments $s^{\mathcal A}$, via the following sequence of equivalences:  $s^{\mathcal A}=_U s^{\mathcal B}$  \, iff\, $[\mathcal{A}]_x=[\mathcal{B}]_x$ for all $x\in U$ \, iff\, $\mathcal{A}\sim_x \mathcal{B}$ for all $x\in U$\, iff \, $x\in \bigcap (\mathcal{A}\bigtriangleup \mathcal{B})$ for all $x\in U$.

\emph{Claim (b)}: From left to right, suppose that $D_X^{s^{\mathcal A}} y$ holds in $\bM_R$. Take the family $\mathcal{B}:=\mathcal{A}\cup \{\tilde{X}\}$. Case (i): $\tilde{X} \not\in \mathcal{A}$. Then we have
$\mathcal{A}\bigtriangleup \mathcal{B}=\{\tilde{X}\}$, hence $\bigcap (\mathcal{A}\bigtriangleup \mathcal{B})=\tilde{X}$. Thus, by (a) we have $s^{\mathcal A}=_X s^{\mathcal B}$ (since $X\subseteq \tilde{X}$). It follows by the truth of $D^s_Xy$ at $s^{\mathcal A}$ that  $s^{\mathcal A}=_y s^{\mathcal B}$. But this means by the  already proved equivalence (a)    that $y\in\tilde{X}$, i.e., $R_X y$. Case (ii): $\tilde{X} \in {\mathcal A}$. Repeat the preceding argument, but now w.r.t. the families ${\mathcal A}$ and ${\mathcal A} - \{\tilde{X}\}.$

%but that we do \emph{not} have $R_X y$. Then $y\not\in \tilde{X}$ (by the definition of the $R$-closure $\tilde{X}$). Take the family $\mathcal{B}:=\mathcal{A}\bigtriangleup \{\tilde{X}\}$. Then we have
%$\mathcal{A}\bigtriangleup \mathcal{B}=\{\tilde{X}\}$, hence $\bigcap (\mathcal{A}\bigtriangleup \mathcal{B})=\tilde{X}$. Thus, by (a) we have $s^{\mathcal A}=_X s^{\mathcal B}$ (since $X\subseteq \tilde{X}$) but $s^{\mathcal A}\not=_y s^{\mathcal B}$ (since $y\not\in \tilde{X}$), contradicting the assumption that $s=s^{\mathcal A}$ satisfies $D_X^s y$.

From right to left, assume that $R_Xy$, and consider any assignment $s^{\mathcal A}\in A_R$. Let  $s^{\mathcal B}\in A_R$ be any admissible assignment s.t. $s^{\mathcal A}=_X s^{\mathcal B}$. By claim (a),  $X\subseteq \bigcap (\mathcal{A}\bigtriangleup \mathcal{B})$. Putting this together with $R_Xy$, we obtain by Monotonicity that $R_{\bigcap (\mathcal{A}\bigtriangleup \mathcal{B})}y$. But $\bigcap (\mathcal{A}\bigtriangleup \mathcal{B})$ is $R$-closed (being the intersection of a family of $R$-closed sets), and therefore, $y\in \bigcap (\mathcal{A}\bigtriangleup \mathcal{B})$. Applying claim (a) again, we conclude that $s^{\mathcal A}=_y s^{\mathcal B}$. Thus  $s^{\mathcal A}$ satisfies $D_X^s y$, as desired.

\smallskip
The desired conclusion follows immediately from the second claim.\footnote{There may be a way of proving Part 2 using some general product construction on the simpler models produced by in the proof for Part 1, but we have not yet been able to find one.}

\medskip

\emph{Proof of Part 3}. Let ${\mathcal R}$ be a family of binary relations on $V$ satisfying Reflexivity, Transitivity and Monotonicity, and agreeing on constants.
For each $R\in {\mathcal R}$, put $C=\{y\in R: R_\emptyset y\}\subseteq V$ for the common set of $R$-constants. Construct all the models $\bM_R$ as in Part 2, for every $R\in {\mathcal R}$. Then each $R$ is both the local and the global dependence in the corresponding $\bM_R$.

The only remaining step for our main proof involves the following general disjoint union construction on dependence models. Define a new model $\bM_{\mathcal R}=(O_{\mathcal R},I_{\mathcal R},A_{\mathcal R})$, where (a)
$O_{\mathcal R}:=C+\sum_{R\in {\mathcal R}} O_R =C\cup \bigcup_{R\in {\mathcal R}} \{R\}\times O_R$
is the \emph{disjoint union} of the common set of constants and all  sets of objects of the models $\bM_R$, (b) the interpretation $I_{\mathcal R}$ makes all predicates \emph{false}, and (c)
$A_{\mathcal R}:=\{s_R: R\in {\mathcal R}, s\in A_R\}$ consists of new assignments $s_R$, each associated to an old assignment $s\in A_R$ with $R\in {\mathcal R}$, with

\vspace{2mm}

$s_R(x):= (R,s(x))\in \{R\}\times O_R$ for $x\in V-C$, and $s_R(x):=x$ for $x\in C$.

\vspace{2mm}

\noindent Note that in this model, $s_R=_X s'_{R'}$ for all $X\subseteq C$ and all $s_R, s'_{R'}\in  A_{\mathcal R}$. Also, for $X\not\subseteq C$,
$s_R=_X s'_{R'}$ holds in $\bM_{\mathcal R}$ iff $R=R'$ and $s=_X s'$ holds in $\bM_R$. Using these facts, it is easy to see that the local dependence statement $D^w_X y$ holds in $\bM_{\mathcal R}$ at a state $w=s_R\in A_R$  iff it holds at $s$ in the corresponding component $\bM_R$, and so the global dependence statement $D^{\bM}_X y$ holds in $\bM_{\mathcal R}$ iff it holds in all components.
It follows that ${\mathcal R}$ coincides with the family of all local dependence relations within $\bM_{\mathcal R}$, and that $\bigcap {\mathcal R}$ is the global dependence relation on $\bM_{\mathcal R}$.\footnote{This representation argument can be turned into a proof of completeness and finite model property for a simple logic of dependence atoms plus the universal modality over available assignments, a precursor to the completeness proof for the richer  language of LFD in Section \ref{Axiomatize}.}\end{proof}

The preceding representation method uses a large number of objects in general. What happens when one restricts the available objects that can be assigned to variables?

\begin{exam}\label{Numerical}  \emph{Consider a dependence model given by the table below:}

%\vspace{-2mm}

\begin{center}
\begin{tabular}{|l c r|}
  \hline
  $x$ & $y$ & $z$ \\
  \hline
   0 & 1 & 0\\
   1 & 1 & 0\\
   2 & 0 & 0 \\
   \hline
\end{tabular}
\end{center}
\end{exam}

%\vspace{-1mm}

\noindent This table uses three values to represent a \emph{strict} linear dependence order of three variables $x, y, z$: we have global dependencies $D_x y$ and $D_y z$, but not the other way around. But as is easy to see, this cannot be done with only two objects.\footnote{One can fill in the table for $y$ and $z$ to get the right failures of dependence, but then, by functionality, a third value must be assigned somewhere to $x$.} To state the underlying observation positively, the following
%principle
is \emph{valid on two-valued
models}:

 \smallskip

 \indent \emph{if $D_xy$ and $D_yz$, then $D_yx$ or $D_zy$}.

 \smallskip

 \noindent More generally, the following can be shown:

 \smallskip

 \emph{Arbitrarily high finite numbers of values are needed to represent arbitrary finite linear orders.}

 \medskip

 \noindent What are minimal sets of objects for representing given dependence graphs? How can one axiomatize the structural dependence properties for each fixed finite set of objects?

\begin{rem}[Dependence and consequence] \emph{As already mentioned, the three stated structural properties (Reflexivity, Transitivity and Monotonicity) are known to be characteristic for the relation of classical {\it logical consequence}, \cite{Scott71}.
%, and unsurprisingly others related it to \emph{inquisitive implication}, \cite{G&S97}.
But the preceding observations  show one essential difference. To represent a three-element linear sequence of variables ordered by dependence, three objects were needed in Example \ref{Numerical}. But to represent an analogous sequence of strict consequences, only two truth values are needed, e.g.:}
\end{rem}

\vspace{-4mm}

\begin{center}
\begin{tabular}{|l c r|}
  \hline
  $p$ & $q$ & $r$ \\
  \hline
   0 & 1 & 1\\
   0 & 0 & 1\\
   \hline
\end{tabular}
\end{center}

\vspace{-1mm}

\noindent In fact, any finite acyclic graph can be represented in terms of classical logical consequence.

All this suggests a move to non-classical consequence relations without a simple truth value semantics. In fact, the format $D_X y$, with multiple `premises' in $X$ and a single `conclusion' $y$, resembles Gentzen-style sequents for intuitionistic logic, and  dependence has been related to intuitionistic implication, \cite{Abr}. Also, given the analogies between dependence and implication between questions to be discussed in Section \ref{EpisInq}, dependence has been related to notions of implication in interrogative and  inquisitive logics, \cite{Baltag2016}, \cite{DepQ}.

The analogy between dependence and consequence can also be extended in other ways. For instance, adopting a classical sequent format, one can study dependencies $D_XY$ read disjunctively in the set $Y$. Or, softening the strict universal quantification over assignments in our semantic notion, one   obtains new non-monotonic varieties of dependence where the dependence only holds `under normal circumstances', by analogy with non-monotonic logics, \cite{Brewka}.

\subsection{\textbf{Explicit function definitions}}\label{Functions} Our semantic definition makes dependence $D_Xy$ a form of implicit definability, as fixing  the values of the dependent variable $y$ by fixing the values of the variables in $X$.   But there is also a broad alternative intuition of dependence, viz. as $y$  being {\it definable} in terms of $X$ using some repertoire of available operations.\footnote{Explicit functional dependence relies crucially on the available operations. For instance, in the set $\{1, 2\}$, 2 depends on 1 if addition is present, but not if the only operation is multiplication. Another typical example are dependent vectors that are special \emph{linear combinations} of other vectors.}
The two views are connected. In mathematics,  implicit semantic definability justifies the explicit introduction of a corresponding function. %For instance, after proving the existence of a unique function $f$ that is equal to its own derivative and satisfies $f(1)=0$, one can introduce the natural exponential function.

This discussion suggests the following general line.

\begin{defi} Given a dependence model $\bM$, a set $X\subseteq V$ of variables and a variable $y\in V$,  let $F^y_X$ be the partial function from $X$-indexed tuples in $O^X$ to objects in $O$, satisfying for all tuples $\mb{u}\in O^X$: $F^y_X(\mb{u})= o$ iff $o$ is the unique object in $O$ s.t. $s(y)=o$ holds for some $s\in A$ with $s\hspace{-0.1cm}\upharpoonright\hspace{-0.1cm} X=\mb{u}$; if no such object exists, $F^y_X(\mb{u})$ is undefined. In other words: $F^y_X(\mb{u})= o$ holds iff we have both (1) $\mb{u}=s\hspace{-0.1cm}\upharpoonright\hspace{-0.1cm} X$ for some assignment $s\in A$, and (2) for all assignments $t, t'\in A$, $t\hspace{-0.1cm}\upharpoonright\hspace{-0.1cm} X=t'\hspace{-0.1cm}\upharpoonright\hspace{-0.1cm} X$ implies $t(y)=t'(y)=o$.  We denote by $dom(F^y_X)$ the domain of this function. The expansion of $\bM$ with all these partial functions $F^y_X$ is called the  \emph{induced function model} $F(\bM)$.
\end{defi}
The partial functions introduced in this Skolemization-like manner make explicit the functions that underlie local and global dependencies:

\begin{fact}
Induced function models satisfy the following two equivalences:

\vspace{1mm}

{} \indent $D^s_Xy$ \, iff \, $s\hspace{-0.1cm}\upharpoonright\hspace{-0.1cm} X\in dom(F^y_X)$ \, iff \, $s(y) = F^y_X(s\hspace{-0.1cm}\upharpoonright\hspace{-0.1cm} X)$

\vspace{1mm}

{} \indent $D^\bM_Xy$ \, iff \, $dom (F^y_X)=\{s\hspace{-0.1cm}\upharpoonright\hspace{-0.1cm} X: s\in A\}$  \, iff \, $s(y) = F^y_X(s\hspace{-0.1cm}\upharpoonright\hspace{-0.1cm} X)$ for all $s\in A$.\end{fact}
Explicit definability is a
natural companion to our semantic view as implicit determination.\footnote{An operational view  may also underlie dependence notions in logic. Dependence in quantifier combinations $\forall x \exists y$ means that a value for $y$ can be produced given one for $x$, something that can be made concrete by a Skolem function. And dependence in a set of formulas like $\{p, p \lor q\}$ may mean that some proof  `produces' $p \lor q$ from $p$.}
Developing an abstract purely operational  approach matching our semantic view of dependence may be worthwhile, and some concrete instances of how such an approach might work can be found in Section \ref{Vectors} on the notion of linear dependence in vector spaces.

\section{\large{\textbf{The logic of functional dependence}}}
We now introduce the language of our logic LFD of functional dependence.
\subsection{\textbf{Syntax and semantics of LFD}}
\vspace{-2mm}

%\subsection{Syntax, semantics and validities}
\begin{defi}
Given a vocabulary $(V, Pred, ar)$, the \emph{language LFD} is recursively given by:

\vspace{-3mm}

$$
\begin{array}{c cc cc ccc cc cc cc cc}
\varphi :: =  & P\mb{x}   &|& \neg \varphi &|& \varphi\wedge\varphi         &|&  \dual_X \varphi &|& D_{X} y
\end{array}
$$
where $y\in V$ is any variable, $P$ is any predicate symbol, $\mathbf{x}=(x_1\ldots x_n)$ is a finite string of variables of length $n=ar(P)$ and $X\subseteq V$ is a finite set of variables.
\end{defi}

%\subsection{Semantics}

\begin{defi} (\emph{Semantics}) \emph{Truth of a formula} $\varphi$ in a dependence model $\bM=(M,A)$ at an assignment $s \in A$ (written $\bM, s \models \varphi$, with the index $\bM$ dropped when the model is understood) is defined by the following clauses:

\vspace{-2mm}

\[
\begin{array}{llll}

\vspace{1mm}

s\models P\mb{x} \  \ & \mbox{ iff } \ \ & I^{\bM}(P) \mbox{ holds for the tuple s(\bf{x})}\\

\vspace{1mm}

s\models \neg\varphi \  \ & \mbox{ iff } \ \ & s\not\models\varphi\\

\vspace{1mm}

s\models \varphi\wedge\psi \  \ & \mbox{ iff } \ \ & s\models\varphi \mbox{ and } s\models\psi\\

\vspace{1mm}

s \models \dual_X\varphi \  \ & \mbox{ iff } \ \ & t \models \varphi \mbox{ holds for all $t\in A$ with $s =_X t$}\\
s \models D_Xy \  \ & \mbox{ iff } \ \ & D^s_Xy  \mbox{ holds in $\bM$.}
\end{array}\]
\end{defi}

\noindent As announced in the Introduction, the \emph{dependence modality} $\dual_X \varphi$ is read as ``$X$ locally determines $\varphi$":  \emph{the current values of $X$ determine the truth of $\varphi$}. Similarly, $D_X y$ is read as ``$X$ locally determines $y$": it says that \emph{the current values of $X$ determine the value of $y$}.\footnote{As already mentioned in the Introduction, the obvious analogy between $\dual_X \varphi$ and $D_X y$ can be made precise by introducing Boolean variables $?\varphi$ as in \cite{Baltag2016}, that record the truth-values of formulas $\varphi$, and then \emph{defining} the dependence quantifiers as $\dual_X \varphi := \varphi\wedge D_X ?\varphi$. Here, we choose to take these quantifiers as primitive, as they have an independent logical motivation and in our view they play an equally important role as the dependence atoms in the study of (partial) dependencies and correlations between variables.}

One important notion in LFD is that of \emph{free variables}. Here we have to be careful. As in FOL, we want the free variables to be those whose current values \emph{determine the truth value of a formula}; indeed, binding a variable is a way of ``forgetting" its value as irrelevant, while the specific value currently assigned to a free variable \emph{is relevant} for the meaning of the formula.
But the definition is subtly different from FOL, since the dual quantifiers $\dual_{X}\varphi$ explicitly list the variables that are left \emph{free} (rather than listing the bound ones, as do the usual quantifiers).

\begin{defi}
\emph{Free($\varphi$)} is defined by the following recursion:

\medskip

(a) \emph{Free}($Px_1\ldots x_n$) = $\{x_1, \ldots, x_n\}$,

\vspace{1mm}

{} \indent (b) \emph{Free}($\neg \varphi$) = \emph{Free}($\varphi$), (c) \emph{Free}($\varphi \land \psi$) = \emph{Free}($\varphi$) $\cup$ \emph{Free}($\psi$),

\vspace{1mm}

{} \indent (d) \emph{Free}($\dual_{X}\varphi$) = $X$, (e) \emph{Free}($D_{X}y$) = $X$.
\end{defi}

Now we can check that indeed the values of the free variables occurring in a formula uniquely determine its truth value. In other words, like FOL (and unlike CRS), LFD is `local':
%\footnote{Interestingly, the Locality Lemma is itself a statement of functional dependence: this time, of the Boolean truth value of the formula $\varphi$ on some set of object variables occurring in $\varphi$.}

\begin{fact}\label{locality} (\emph{Locality.}) If \emph{Free}$(\varphi) \subseteq X$, and $s =_X t$, then $s \models \varphi$   \emph{iff}  $t \models \varphi$.
\end{fact}

\begin{proof}\quad \,\,
%It is enough to show that $s=_{Free(\varphi)} t$ and $s\models \varphi$ imply $t\models \varphi$ (since the converse follows by symmetry).
The proof is by induction on $\varphi$. The atomic and Boolean cases are entirely straightforward. \emph{Dependence modalities}: Assume that $s=_X t$ and $s\models \dual_{X}\varphi$. To show that $t\models \dual_{X}\varphi$, consider any $w\in A$ with  $t=_X w$. Here $s=_X t$ and $t=_X w$ imply $s=_X w$. This together with $\dual_{X}\varphi$ then yields $w\models \varphi$ as desired, by the semantics of $\dual_X$. \emph{Dependence atoms}: Assume that $s=_X t$ and $s\models D_X y$. Then $s=_y t$ by the semantics of $D_X$. To show that $t\models D_X y$, consider any $w\in A$ with $t=_X w$. Here
$s=_X t$ and $t=_X w$ imply $s=_X w$, which together with $s\models D_X y$ gives that $s=_y w$. From this and $s=_y t$, it follows that $t=_y w$ as desired.\end{proof}

Here is a useful immediate consequence in the presence of additional local dependencies.

\begin{cor} If two assignments $s, t$ agree on all formulas with free variables in $X$, then they also agree on all formulas with free variables in the extended set $\{y \mid s \models D_X y\}$.\footnote{The set $\{y \mid s \models D_X y\}$ includes $X$, as a consequence of Reflexivity and Monotonicity.}
\end{cor}

\par\noindent\textbf{Abbreviations}. Boolean connectives $\top, \bot, \varphi\vee \psi, \varphi\to \psi, \varphi\leftrightarrow \psi$ are defined as usual. We use $D_x y$ for $D_X y$ when $X=\{x\}$ is a singleton, and same for $\dual_x \varphi$. Other abbreviations are:

\smallskip

$(a) \quad \rotatebox[origin=c]{180}{$\forall$}\varphi \,\, :=\,\, \dual_{\emptyset} \varphi   \,\,\, \mbox{ (universal modality)}  $ \quad
$(b) \quad \edual_X \varphi \,\, :=\,\, \neg \dual_X  \neg \varphi $ \quad
$(c) \quad \rotatebox[origin=c]{180}{$\exists$}\varphi \,\, :=\,\, \neg \rotatebox[origin=c]{180}{$\forall$}\neg\varphi$

\smallskip

$(d) \quad C y \,\, :=\,\, D_{\emptyset} y \,\,\, \mbox{ (constant value)}  $ \quad
$(e) \quad =(X;y) \,\, :=\,\,  \rotatebox[origin=c]{180}{$\forall$} D_X y \,\,\, \mbox{ (global dependence)} $

\smallskip

$(f) \quad   D_X Y  \,\, :=\,\, \bigwedge_{y\in Y} D_X y  \,\,\, \mbox{ (multi-variable dependence) } $ \quad

\smallskip

$(h) \quad   \forall_X \varphi  \,\, :=\,\, \dual_{Free(\varphi)-X} \varphi  \,\,\, \mbox{ (universal quantifier) }  $ \quad

\smallskip

  $(i) \quad \exists_X \varphi  \,\, :=\,\, \neg \forall_X\neg \varphi  \,\,\, \mbox{ (existential quantifier)}  $

\medskip
%\medskip

\noindent These defined connectives behave as expected. E.g., syntactically, we have that:
$Free(\forall_X \varphi)=Free(\exists_X \varphi)=Free(\varphi)-X$.
Semantically, e.g.,  $\rotatebox[origin=c]{180}{$\forall$} \varphi$ means that \emph{all} assignments satisfy $\varphi$, etc.:
\[
\begin{array}{llll}

\vspace{1mm}

s\models \rotatebox[origin=c]{180}{$\forall$} \varphi  \  \ &  \mbox{ iff } \ \ &
 t\models \varphi \mbox{ for all $t\in A$}\\

\vspace{0.7mm}

s\models \edual_X \varphi \  \ & \mbox{ iff } \ \ &  t\models \varphi \mbox{ for some $t\in A$ with $s=_X t$}\\

\vspace{0.7mm}

s\models C y \ \ & \mbox{ iff } \ \ & t=_y s \mbox{ for all $t\in A$}\\

\vspace{0.7mm}

s\models =(X; y)  \ \ & \mbox{ iff } \ \ & D^\bM_X y \mbox{ holds }\\

\vspace{0.7mm}

s\models \forall_X \varphi \ \ & \mbox{ iff } \ \ &  t\models \varphi \mbox{ for all $t\in A$ with $s=_{Free(\varphi)-X} t$}.
\end{array}\]
Note that our defined formula $\forall_X\varphi$ matches the semantics of the `local' version of the universal quantifier (in the sense of satisfying the Locality principle from Fact \ref{locality}), as given in the Introduction. Recall that these amount to the standard FOL quantifiers on full models, and are their closest analogue on arbitrary dependence models. Thus, LFD  \emph{contains the first-order quantifiers},  generalized from their standard models to the larger realm of dependence models. For further discussion of the meaning of LFD quantifiers, cf. Section \ref{quantification}.

\begin{rem}[Informational interpretation]\label{Inform}
\emph{The set $A$ of admissible assignments in a dependence model is a
`complete database', as in Example \ref{Restaurant}, and can be interpreted as an \emph{information structure}, encoding the `knowledge base' of an (anonymous) agent: a full list of all the tuples that are consistent with the agent's background information. The underlying assumption is that only one tuple (the `current assignment') represents the actual state of the world, but that tuple is typically unknown: the agent can only narrow down the possibilities to the set $A$. The universal modality $\rotatebox[origin=c]{180}{$\forall$} \varphi := \dual_\emptyset \varphi$ then captures the agent's \emph{information}:
$\rotatebox[origin=c]{180}{$\forall$} \varphi$ means that \emph{$\varphi$ is `known'}. The constant-value formula $Cx:= D_\emptyset x$ says that the value of $x$ is `known'. Dependence quantifiers capture a form of \emph{conditional knowledge}:
$\dual_X \varphi$ means that the agent can know that $\varphi$ if she is given the current values of $X$. Analogously, dependence atoms $D_X y$ express a form of \emph{conditional knowledge of a value}: the agent can know the value of $y$ if given the values of $X$. Finally, global dependence $\rotatebox[origin=c]{180}{$\forall$} D_X y$ captures \emph{known correlations}: the agent knows how to determine the value of $y$ from the values of $X$. }
\end{rem}

\begin{rem}[Further notions of dependence]
\emph{We can also formalize a common alternative intuition of dependence, \cite{ELD}, as `changing $x$ involves changing $y$': this is just $D_yx$. Moreover,  weaker notions of dependence can be defined, such as `restricting the value of $x$ to property $P$ restricts the value of $y$ to have property $Q$' (cf. Remark 1.2). This is expressed by $\rotatebox[origin=c]{180}{$\forall$}(Px \rightarrow Qy)$. Yet another definable notion of dependence is that the current values of $X$ restrict the value of $y$ to have property $Q$, captured by the formula $\dual_X Qy$.}
\end{rem}

%Semantic consequence and validity in LFD are defined as usual.

\begin{exam}\label{valid} \emph{Here are some illustrations of valid and invalid consequences:}
%\footnote{Semantic consequence and validity in LFD are defined as usual.}
\vspace{-1mm}

\begin{enumerate}
\item\quad (a) $\varphi \rightarrow \dual_X \varphi$ is valid if $Free(\varphi) \subseteq X$. (b) $Px \rightarrow \dual_{y} Px$ is \emph{not} valid.
\item\quad (a) $\dual_{X\cap Y}\varphi \rightarrow \dual_X \dual_Y\varphi$ is valid. (b) $\dual_X \dual_Y\varphi \rightarrow \dual_{X\cap Y}\varphi$ is \emph{not} valid.
\item\quad \, Let $X \cap Y = \emptyset$. (a) $(\edual_X\varphi \land \edual_Y\psi) \rightarrow \edual_{X\cup Y}(\varphi \land \psi)$ is
\emph{not} valid.  However, (b) if $Free(\varphi) \subseteq X$ and $Free(\psi) \subseteq Y$, the preceding implication is valid, and in fact we have the stronger validity
   $(\edual_X\varphi \land \edual_Y\psi) \rightarrow (\varphi \land \psi)$.\footnote{To check this: assume that $s\models \edual_X\varphi \land \edual_Y\psi$. Then there are $s'=_X s$ and $s'' =_Y s$ with $s'\models\varphi$ and $s''\models \psi$. Using these facts and the assumptions that $Free(\varphi) \subseteq X$ and $Free(\psi) \subseteq Y$, we obtain $s\models\varphi$ and $s\models\psi$ by Locality, Fact \ref{locality}. It follows that $s\models \varphi\land \psi$.}
\item\quad \, The Distribution axiom is sound for dual quantifiers: $\dual_X(\varphi\to\psi)\to (\dual_X \varphi\to \dual_X\psi)$ is valid. However, (b) Distribution is not sound for local quantifiers: $\forall_X(\varphi\to\psi)\to (\forall_X \varphi\to \forall_X\psi)$ is \emph{not} valid.\footnote{A counterexample, for $V=\{x,y,z\}$, is given by taking $O=\{0,1\}$, $I(P)=\{(1,1)\}$ and $A=\{(1,1,1), (1,1,0), (0,0,1)\}$, where a triplet of values denotes the corresponding assignment on $(x,y,z)$. The assignment $(1,1,1)$ satisfies $\forall_x (Pxy\to Pxz)$ and $\forall_x Pxy$, but fails to satisfy $\forall_x Pxz$.}
\item\quad (a) $\dual_X\varphi\to \varphi$ and $\forall_X\varphi\to\varphi$ are valid. However, (b) the classical elimination rule for the universal quantifier is \emph{not} sound: $\forall_x \varphi \to [y/x]\varphi$ is \emph{not} valid.\footnote{Here, $[y/x]\varphi$ is the result of substituting $y$ for $x$ in the formula  $\varphi$.}
\end{enumerate}

\end{exam}

\noindent The last non-validity is explained by the fact that in LFD variables are no longer arbitrary placeholders, but have an individual meaning, denoting specific quantities (as commonly done in the empirical sciences, where e.g. $t$ stands for time, etc). This means that, unlike in FOL, bound alphabetic variants may have \emph{different} truth values: $\forall_x Px$ can be true in a model while $\forall_y Py$ is false. On the other hand, LFD still allows for a formulation of the intuition behind bound variants that the choice of variables is arbitrary: though no longer holding inside one given model, the invariance under renaming still holds \emph{across models}.

\begin{fact}\label{Renaming} (\emph{Renaming Lemma}.) Consider any dependence model $\bM=(M,A)$ and  \emph{LFD}-formula $\varphi$. Let $\sigma$ be a permutation of all variables, and let $\sigma(\varphi)$ be the result of replacing in $\varphi$ every occurrence of any variable $x\in V$ by $\sigma(x)$. Moreover, for every $s\in A$, let $s^\sigma$ be the assignment given by putting $s^\sigma(x) := s(\sigma^{-1}(x))$ for all variables $x$, and let $\bM^{\sigma}=(M, A^\sigma)$ be the dependence model obtained by taking $A^\sigma=\{s^\sigma:s\in A\}$). Then the following equivalence holds:
$$\bM,s \models \varphi \,\, \mbox{ iff } \,\, \bM^{\sigma}, s^\sigma \models \sigma(\varphi).$$
As a consequence, \emph{validity is invariant under variable renaming}: $\varphi$ is valid iff $\sigma(\varphi)$ is valid.\end{fact}

\begin{proof}\quad \,\,  The proof of the first claim is by induction on $\varphi$.

\smallskip

\emph{Atomic formulas}. Note that, for all $s\in A, {\bf x}\in V$ we have $s(\mb{x})=s(\sigma^{-1}(\sigma(\mb{x}))=s^\sigma (\sigma(\mb{x}))$, and also
$\sigma(P\mb{x})=P\sigma(\mb{x})$. Using these and the truth definition, we obtain the  equivalences: $\bM,s\models P\mb{x}$ iff
$s(\mb{x})\in I(P)$ iff $s^\sigma (\sigma(\mb{x}))\in I(P)$ iff $\bM, s^\sigma\models P\sigma(\mb{x})$ iff $\bM, s^\sigma\models \sigma(P\mb{x})$.

\smallskip

\emph{Boolean combinations}. This inductive step follows immediately by the truth clauses for Boolean operations, the induction hypothesis and the fact that permutations satisfy $\sigma(\neg\varphi)=\neg \sigma(\varphi)$ and $\sigma(\varphi\wedge\psi)=\sigma(\varphi)\wedge \sigma(\psi)$.

\smallskip

\emph{Dependence modalities}. First note the following equivalence: $s=_X t$ iff $s^\sigma=_{\sigma(X)} t^\sigma$. Using this, the truth clause for the universal dependence modality and the induction hypothesis, we obtain the following sequence of equivalences: $\bM,s\models \dual_X \varphi$ iff $\forall t\in A\, (s= _X t \, \Rightarrow \, \bM,t\models \varphi)$ iff $\forall t\in A\, (s^\sigma =_{\sigma(X)} t^\sigma \, \Rightarrow \, \bM^\sigma,t^\sigma\models\sigma(\varphi))$ iff $\forall w\in A^\sigma\, (s^\sigma=_{\sigma(X)} w \, \Rightarrow \, \bM^\sigma,w\models\sigma(\varphi))$ iff $\bM^\sigma,s^\sigma\models \dual_{\sigma(X)} \sigma(\varphi)$ iff
$\bM^\sigma,s^\sigma\models \sigma(\dual_X \varphi)$.

\smallskip

\emph{Dependence atoms}. Using the same observation as in the previous case, together  with the fact that $\sigma(D_X y)=D_{\sigma(X)}\sigma(y)$, as well as truth clause for dependence atoms and the induction hypothesis, we obtain the sequence of equivalences:  $\bM,s\models D_Xy$ iff $\forall t\in A\, (s=_X t \, \Rightarrow \, s=_y t)$ iff $\forall t^\sigma\in A^\sigma\, (s^\sigma=_{\sigma(X)} t^\sigma \, \Rightarrow \, s^\sigma=_{\sigma(y)} t^\sigma)$ iff $\forall w\in A^\sigma\, (s^\sigma=_{\sigma(X)} w \, \Rightarrow \, s^\sigma=_{\sigma(y)} w)$ iff $\bM^\sigma, s^\sigma \models D_{\sigma(X)}\sigma(y)$ iff  $\bM^\sigma, s^\sigma \models \sigma (D_Xy)$.

\smallskip

Finally, the second claim follows immediately from the first, by quantifying over both admissible assignments and dependence models.
\end{proof}

\subsection{\textbf{Discussion: Quantification over objects in LFD}}\label{quantification}

Having entered the world of LFD with its special behavior of variables, one might ask whether  the above quantifier companions $\forall _X  \varphi$ of the dependence modalities are `true' quantifiers. This question calls for some distinctions. First, as we saw earlier, both the CRS-style quantifiers $\forall X \varphi$ and their local versions $\forall_X\varphi$  are simply generalizations of the FOL quantifiers to a broader class of models, and the dependence modalitities of LFD are their close duals. However, one might require a true quantifier to be a semantic operator, acting on \emph{objects}, so that the precise variable used in its syntax does not matter.

Now in one sense, this is true in LFD: variable names do not matter when we look \emph{across models}. What can be said in one model using $x$ can be said in another model using another variable $y$, by the Renaming Principle \ref{Renaming}, which underpins, for instance, the use of alphabetic variants for proofs in our axiomatic systems of Sections 5.1, 6.2. But \emph{locally} within one model, the existence of non-trivial dependencies gives rise to \emph{asymmetries} of behavior between variables: as already observed, variables in a fixed dependence model acquire `individuality'. As a result, in a given model,
quantifiers in LFD quantify over admissible assignments, \emph{not over tuples of objects} like the first-order quantifiers. Thus, existential quantifiers in LFD do not seem at first sight to be obviously related to the usual Skolem functions in the semantics of FOL.

But more can be said. In fact,  a string of LFD quantifiers \emph{does induce a semantic operator over tuples of objects}, albeit one that, in contrast to its classical counterpart: (a) quantifies over a restricted range of tuples (the ones that are in the range of admissible joint values for the given variables),  and (b) imposes further constraints on the corresponding Skolem function, requiring it to behave well with respect to the admissible assignments. These additional features are both natural and informative in generalized assignment semantics. Indeed, quantifier combinations in LFD play a twofold role, giving information  both about \emph{objects} and about \emph{variable ranges and dependencies}.

\smallskip

More precisely, let us compare the meaning of some quantifiers and quantifier combinations in LFD with their classical meanings in FOL. To do this, we need  some notation. Given a tuple of variables $\mb{x}=(x_1, \ldots, x_n)\in V^*$, let $X:=\{x_1, \ldots, x_n\}$ be the set of its variables.
Also, for any given  dependence model $\bM=(O,I,A)$, we denote by
$$O^{\mb{x}}:=\{s(\mb{x}): s\in A\}=\{(s(x_1), \ldots, s(x_n)): s\in A\}$$
the \emph{range of admissible $\mb{x}$-values}, as a subset of $O^n$. As a special case, we have
$$O^{(x)}=\{s(x):s\in A\}\, \subseteq O.$$

\medskip

For a start, with the given notational convention, consider the LFD formula
$$\forall_X \, P\mb{x}$$
This holds in a full model (with $A=O^V$) iff we have
$$ O^{n}\subseteq I(P)$$
So on \emph{full models}, $\forall_X$
captures universal quantification, exactly as $\forall X$ does in FOL.
But over an \emph{arbitrary} dependence model $\bM=(O,I,A)$, the same formula holds iff we have
$$O^{\mb{x}} \subseteq I(P)$$
This is clearly universal quantification, but only over the restricted range $O^{\mb{x}}$ of admissible simultaneous $\mb{x}$-values. Though weaker  than the  FOL formula $\forall {\bf x} P {\bf x}$, this is precisely the  natural meaning in a dependence model, where each variable $x$ or tuple of variables $\mb{x}$ has its own range of (tuples of) values. The universal LFD quantifier simply quantifies over  that range.

\medskip

Our second,  perhaps more telling, example is a quantifier  combination expressing a functional dependence.\footnote{The analysis here is also reminiscent of the discussion in Section \ref{Functions}, but we forego details.} It is easy to see that the LFD formula
$$\forall_x \exists_y \, Pxy$$
holds in a full model iff there is a function witnessing this fact:
$$\exists F:O\to O \, \forall o\in O: \, (o, F(o))\in I(P)$$
This matches the usual Skolem-type meaning of the FOL formula $\forall x \exists y Pxy$.
But spelling out the LFD semantics for the defined dependence quantifiers, over an arbitrary dependence model $\bM=(O,I,A)$, the same formula $\forall_x \exists_y \, Pxy$ holds iff we have
$$\exists F:O^{(x)} \to O \, \forall o\in O^{(x)}\, (o,F(o))\in I(P)\cap O^{(x,y)} \,\,\, \footnote{In more detail, starting from any assignment, the first quantifier $\forall x$ ranges over all admissible assignments $s$ and any value for $x$ there, the second  $\exists y$ then ranges over all admissible assignments $t$ with $s =_x t$.}$$
This statement is \emph{neither weaker, nor stronger} than the one expressed by the FOL formula. On the one hand, the domain of $F$ is restricted to the admissible $x$-values, which is a weakening. But on this restricted range, we have a stronger statement: not only do all resulting pairs $(o,F(o))$ satisfy $P$, but they are all realized by admissible simultaneous assignments of values to $(x,y)$. Once again, this is a natural statement: in the earlier terms, the combination $\forall_x \exists_y$ gives information about  objects \emph{and}  on how these objects can be accessed by variables.

This twofold nature is shared by all quantifier combinations in LFD, making them meaningful in a broader realm than the classical quantifier combinations to which they reduce on full models. Alternatively, they can also be viewed as just being restricted versions of the classical quantifier combinations, but with the added value that they are now forced to also give information beyond their traditional comfort zone (about constraints and correlations on variable ranges).

\medskip
Summing up the presentation so far, LFD has both dependence modalities and  quantifiers in one setting. But one can also view the system at a higher  level. Modalities can be seen as quantifiers, as is well-known in modal logic, \cite{BRV}, and conversely, a system like CRS shows how first-order quantifiers can be seen as modalities. Thus, two perspectives are possible on LFD: it is both a first-order logic and a modal logic.
This interplay will continue throughout this paper, as it allows for borrowing notions and techniques from both sides. In the remainder of
this section, the two intertwined perspectives are taken a bit further, starting with a connection of LFD to standard FOL in terms of translation between languages and semantics.

\subsection{\textbf{First-order translation}}

As is the case for modal logic, the preceding language can be translated faithfully into a first-order language. But before doing so, it is important to be clear in which sense this is meant. LFD can be seen as a weak, decidable first-order logic over generalized models. But  sometimes, a language interpreted over generalized models can be translated into a fragment of that same language interpreted over the original standard models.\footnote{Cf. the analysis of two-way connections between CRS and the Guarded Fragment of FOL in \cite{ELD}, \cite{Bent2005}.}

By the Locality of LFD, it is enough to consider a finite set $V=\{v_1, \ldots, v_n\}$ of $n$ variables, with a given enumeration. First take fresh copies of these variables $V'= \{v_1', \ldots, v_n'\}$. Also introduce a new $n$-ary predicate $A$ where intuitively, $A (v_1, ..., v_n)$ encodes the fact that the tuple of values assigned to $v_1, \ldots, v_n$ belongs to the admissible assignments $A$ of the current dependence model. Now consider FOL with variables in $V\cup V'$ and predicates in $Pred\cup\{A\}$.

For each dependence model $\bM$, there is an \emph{associated FOL model} $T(\bM)$ for this extended language, having the same domain and the same interpretation of the old predicate symbols, and with the new predicate $A$ interpreted as above. Conversely, \emph{every FOL model for the extended language is the translation $T(\bM)$ of some dependence model}.

\begin{defi}
The \emph{first-order translation} $tr(\varphi)$ from \emph{LFD}-formulas $\varphi$ to first-order formulas in the above finite vocabulary  is defined as follows.

\medskip

(a) $tr(P{\bf x}) = P{\bf x}$, (b) $tr(\neg\varphi) = \neg tr(\varphi)$,
(c) $tr(\varphi \land \psi) = tr(\varphi) \land tr(\psi)$,

\medskip
(d) $tr(\dual_X\varphi) = \forall {\bf z} (A{\bf v}  \rightarrow tr(\varphi))$, where ${\bf v}$ is the enumeration of all the variables in $V$ \\ \indent and ${\bf z}$ is the enumeration of all the variables in $V-X$,

\medskip

(e) $tr(D_Xy) = \forall {\bf z} \forall {\bf z'} ((A {\bf v} \land A {\bf v}[{\bf z'}/{\bf z}]) \rightarrow y = y')$, where ${\bf v}$ and ${\bf z}$ are as in part (d), \\ \indent ${\bf z'}$ and $y'$ are the corresponding fresh $V'$-copies of ${\bf z}$ and respectively $y$, and \\ \indent $A {\bf v}[{\bf z'}/{\bf z}]$   is the result of replacing the variables ${\bf z}$ by ${\bf z'}$ in the formula $A {\bf v}$.

\end{defi}

\emph{Free} and \emph{bound occurrences} of a variable in a FOL formula are defined as usual. Variables are allowed to occur both free and bound in different parts of the same formula, so that freely occurring variables can be reused in quantification.
The \emph{free variables of a FOL formula} are also defined as usual: as those variables that occur free at least once in the formula.

It is easy to see from the above translation that, for every formula $\varphi$ of LFD over $V$, \emph{the set of free variables
of its FOL translation $tr(\varphi)$ is exactly $Free(\varphi)$}.

\begin{fact}\label{translation} For all \emph{LFD} models $\bM$ and LFD formulas $\varphi$, we have:
$$\bM, s \models \varphi \,\, \mbox{ iff  } \,\, T(\bM), s \models tr(\varphi),$$
where $tr(\varphi)$ is the above FOL translation of $\varphi$, and $T(\bM)$ is the FOL model associated to $\bM$.
\end{fact}

The proof is a simple induction following the idea of the stated translation.

\begin{cor}\label{RE} The validities of \emph{LFD} are recursively enumerable.\footnote{
By Fact \ref{translation} (and the above observation that every FOL model for the extended language is the translation of a dependence model), a LFD formula is satisfiable iff its first-order translation is. The statement then follows from the completeness theorem for FOL and the effectiveness of the above translation. Since the above translation is easily extended to all  other dependence logics  considered in this article, the  corollary holds for all of these.}
\end{cor}

Further benefits of the above translation include immediate transfer of the fundamental Compactness and L{\"o}wenheim-Skolem properties of FOL to LFD.

%\footnote{A few relevant notions and arguments will be found in the sections to come.}

\subsection{\textbf{Modalization of LFD: standard relational semantics}}\label{StandardRelationalSemantics}

Next, we elaborate the \emph{modal} perspective on LFD. An equivalent semantics is obtained by abstracting away the assignments from their concrete set-theoretical interpretation as functions and treating them as
abstract possible worlds. This eliminates all references to values assigned to variables, and replaces identity of values $s=_x t$ by abstract equivalence relations $\sim_x$.
%\footnote{Thus, modal-style relational models are `object-free', something which has occurred as a desideratum in some abstract approaches to first-order semantics.}

\begin{defi} A \emph{standard relational model} is a triple ${\mathcal M}=(W,\sim,\|\bullet\|)$, consisting of: (a) a set $W$ of  worlds or `states'; (b) a map $\sim: V\to \mathcal{P}(W\times W)$ associating to each variable $x\in V$ an equivalence relation $\sim_{x}$ on $W$; and (c) a \emph{valuation} $\|\bullet\|$ associating to each formula of the form $P\mb{x}$ a set of worlds $\|P\mb{x}\|\subseteq W$.
It is useful to introduce \emph{auxiliary relations} $\sim_X$ on $W$, for \emph{sets} of variables $X\subseteq V$, defined by taking intersections $\sim_X:=\bigcap_{x\in X}\sim_x$. With this notation, the valuation is required to satisfy the following \emph{additional condition}:

\smallskip

\quad \emph{if $w\sim_{X} v$ and $w\in \|Px_1\ldots x_n\|$ for some $x_1, \ldots, x_n\in X$, then $v\in \|Px_1\ldots x_n\|$}.
\end{defi}

We interpret dual quantifiers $\dual_X\varphi$ as universal modalities for the relation $\sim_X$, while dependence atoms $D_Xy$ capture a local inclusion (every $\sim_X$-successor is also a $y$-successor):

\begin{defi} In a standard relational model ${\mathcal M}=(W,\sim,\|\bullet\|)$, the notion of truth ${\mathcal M}, s\models\varphi$ (with the index ${\mathcal M}$ dropped when the model is fixed) is given by the valuation for atomic formulas $P\mb{x}$, by the usual recursive clauses for the Boolean operators, and by:

\smallskip

$w\models \dual_{X}\varphi \quad \mbox{ iff }  \quad \forall v\in W \, \left(  \, w\sim_{X} v \mbox{ implies }  v\models \varphi\, \right)$
\smallskip

$w\models D_{X} y \quad \mbox{ iff } \quad \forall v\in W \, \left(  \, w\sim_{X} v \mbox{ implies } w\sim_{y} v\, \right)$
\end{defi}

%The LFD logic of standard relational models is the same as the logic of dependence models. Thus the basic dependence language does not talk primarily about values of variables, its essential relations are `agreement on values' of $X$, `dependence of $y$ on $X$', and the like.
%The first step is to look at the special case of standard models.
\noindent The two kinds of models introduced so far are closely related: we can show that \emph{the standard relational semantics is equivalent to the dependence-model semantics}.

One direction is given by the following observation:

%\vspace{-1mm}
\begin{fact}\label{Standard1}
Every dependence model $\bM=(O,I,A)$ induces a \emph{standard} relational model  $rel(\bM)=(W,\sim, \|\bullet\|)$, whose possible worlds are the admissible assignments $A$ (so $W:=A$), the accessibility relations $s\sim_x t$ are given by pointwise equality of $x$-values $s=_x t$ (as already defined in dependence models, for both individual variables $x$ and sets of variables $X$), and the valuation is given by $\|P\mb{x}\|=\{s\in W: s(\mb{x})\in I(P)\}$.  Moreover, \emph{the dependence-model semantics agrees with the relational semantics on the induced model}: for all $s\in A=W$ and formulas $\varphi$ of \emph{LFD},

\vspace{-1mm}

$$\bM,s\models \varphi \mbox{ iff } {rel(\bM)},s\models \varphi.$$

\vspace{-1mm}
\end{fact}

This construction is so tight, that its adequacy should be clear without further proof.

A slightly less routine construction yields the opposite direction:

\vspace{-1mm}

 \begin{defi}
Every standard relational model ${\mathcal M}=(W,  \sim, \|\bullet\|)$ induces a dependence model $dep({\mathcal M})=(O^\sim, I^\sim, A^\sim)$, obtained by taking:

\begin{description}
\item[(a)]\quad $O^\sim=\{(x,[w]_x): w\in W, x\in V\}$, where objects are pairs $(x,[w]_x)$ of a variable and an equivalence class $[w]_x=\{v\in W: w\sim_x v\}$

\item[(b)]\quad $A^\sim=\{w^\sim: w\in W\}$, with the admissible assignments  $w^\sim(x)=(x,[w]_x)$ for all $x\in V$

\item[(c)]\quad  the interpretation $I^\sim$ maps each $n$-ary predicate $P$ to the set
$$I^\sim (P):=\{(x_1,[w]_{x_1}), \ldots, (x_n,[w]_{x_n})): w\in W, x_1, \ldots, x_n\in V \mbox{ with } {\mathcal M},w\models Px_1\ldots x_n\}.$$
\end{description}
\end{defi}

Note that $w^\sim (x)=v^\sim(y)$ implies that $x=y$ and $w\sim_x v$. Using this, one  easily checks that $I^\sim (P)$ is well-defined on objects, i.e., independent of the choice of representatives for the  equivalence classes. Moreover, the construction preserves  truth of LFD formulas:

\begin{fact}\label{Standard2} Given a standard relational model ${\mathcal M}=(W,  \sim, \|\bullet\|)$, \emph{the relational semantics on ${\mathcal M}$ agrees with the dependence-model semantics on $dep({\mathcal M})$}: that is, for all worlds $w\in W$
and all formulas $\varphi$ of \emph{LFD}, we have
$${\mathcal M},w\models \varphi \,\, \mbox{ iff }\,\,  {dep(\mathcal M)},w\models \varphi.$$
%Hence, in particular, ${\mathcal M}$ and $dep({\mathcal M})$ are modally equivalent.
\end{fact}
\begin{proof}\quad \,\,
The proof is by induction on $\varphi$. The atomic case holds by the definition of $I^\sim$ in $dep({\mathcal M})$. Boolean cases are routine. The inductive cases for $\dual_X \varphi$ and $D_X y$ follow easily from the semantic definitions, together with the following simple  fact: $w \sim_x v$ in ${\mathcal M}$ \, iff\, $w^\sim =_x v^\sim$ in $dep({\mathcal M})$.
\end{proof}

\smallskip

The two constructions can also be intertwined, with outcomes such as the following.

\begin{fact}\label{Homomorphism}
For every standard relational model ${\mathcal M}=(W, \sim, \|\bullet\|)$, the function $w\mapsto w^\sim$ is a surjective homomorphism from ${\mathcal M}$ to $rel(dep({\mathcal M}))$.
\end{fact}

%To make observations like this into a full categorical duality, we would need a converse, but a notion of homomorphism for dependence models is still to be explored.

%Standard relational models suggest connections of the system LFD with other modal logics, in particular, with epistemic logic:

\begin{rem} \emph{An obvious next desideratum is a natural notion of modal bisimulation for LFD, capturing its precise range within the first-order language over standard models. One lead here might be the connection with generalized assignment semantics for FOL. A natural analogue to modal bisimulation for FOL is \emph{potential isomorphism}, using partial assignments from finite sets of variables to objects.
%\cite{BentBon}.
The crucial back-and-forth clauses of a potential isomorphism $F$ are easily adapted to generalized assignment semantics.
%\footnote{More in detail, the key condition is that, if $F$ connects $s$ to $t$, and a variant $s' \sim_X s$ is available for one of the models compared, then there must be some variant $t' \sim_X t$ for the other model with $t'$ connected to $s'$ by $F$.}
}\end{rem}

\noindent\emph{{\bf Open problem} \, Find a bisimulation invariance theorem characterizing LFD.}\footnote{In response to a preprint version of this paper, Koudijs \cite{Koudijs} defined notions of \emph{dependence bisimulation} for our dependence models as well as their modal relational versions, and proved a Characterization Theorem for LFD as a fragment of FOL that is invariant under dependence bisimulations. A similar characterization of LFD was found independently in P\'{u}tzst\"{u}ck \cite{Aachen1}.}

\subsection{\textbf{Other interpretations: information, knowledge, questions }}\label{EpisInq}
The relational semantics, and its equivalence with the dependence-models semantics, shows that
the actual values of variables do not play an essential role in LFD: what is important are the \emph{relations} of `agreement on values' of $X$, and `dependence of $y$ on $X$'. This suggests other, non-variable-based interpretations of our logic.  Three such interpretations will be outlined here (epistemic, interrogative, and mixed), all \emph{information-based}, like the informational interpretation in Remark \ref{Inform}. The informational perspective is ubiquitous: one often talks informally about even ontic dependence in the real world as knowing the value for one variable implying knowing that of the other, or as answers to some questions implying answers to other questions.
%Indeed, a variable carrying information about another can be interpreted as knowledge of a fact by an agent. In inquisitive logics, dependence means that answers to a question yield answers to another question.

%\emph{
A straightforward \emph{epistemic} reading of LFD re-interprets the variables $x\in V$ as \emph{agents}, while the equivalence relations $\sim_x$ represent the agents' \emph{uncertainty relations}. Then the modal statement
$\dual_x \varphi$ captures agent $x$'s \emph{individual knowledge}, while  $\dual_X \varphi$ expresses \emph{distributed knowledge} among the group of agents $X$, \cite{FHMV}. Dependence atoms $D_xy$ express \emph{knowledge subsumption}:
`agent $x$ knows at least as much as agent $y$', \cite{DitmarschEtAl},
while atoms $D_XY$ stand for the analogue notion of group subsumption.\footnote{Even so, some natural epistemic notions lack an obvious match in LFD. What is a dependence counterpart to \emph{common knowledge} $C_G\varphi$, or other epistemic fixed-point notions?}
%}
%\cite{vBvEK}?}
%The valid analogies noted here can be turned into formal translations between full-fledged epistemic and dependence languages, but details are not relevant in this article.

%\emph{
Next, since an equivalence relation is also a partition as used in the traditional semantics of \emph{questions}, \cite{G&S97}, dependence models also have an \emph{interrogative interpretation}. Variables $x$ represent \emph{basic questions}, and sets of variables $X$ are \emph{joint questions}  asking for the answers to all the given questions). The dependence modality $\dual_x\varphi$ is the `interrogative modality' $Q \varphi$  of \cite{BentMin},
%saying that $\varphi$ is entailed by the (true) answer to question $x$ (at the current state);
while $\dual_X\varphi$ extends this to joint questions. Dependence atoms $D_Xy$ are local versions of `inquisitive implication' between questions, see \cite{CGRoe,DepQ} for modern versions.

Finally, in \emph{mixed readings},   some  variables stand for agents, others    denote objects, while yet others  represent questions. Such mixtures greatly enhance the range of LFD. For instance, the logic for mixed readings in \cite{Baltag2016}   captures a group's distributed knowledge of the value of a variable, as well as individual or group knowledge of a dependence between variables.

\section{\large{\textbf{Decidability via type models}}}\label{Decidability}

In this section, we show that LFD is decidable, using \emph{type models}. These `models' are just syntactic constructs, with no explicit objects, resembling the `quasi-models' used in \cite{ABN}, \cite{ELD} to investigate the Guarded Fragment. Sparse models like this have an independent interest, and they yield a bare-bones proof of decidability. However, the price of this directness is a certain amount of ad-hoc syntactic construction. In Appendix A, we use general semantic methods from modal logic to give a more elegant (though less direct) proof of decidability for LFD.

\subsection{\textbf{Syntactic type models}}\label{TypeModels}
 Consider any finite set $F$ of LFD formulas, and let $V_F$ be the finite set of all variables occurring in $F$. Add to $F$ all formulas $D_X Y$ for all sets of variables $X, Y\subseteq V_F$.  Close the resulting set under subformulas, as well as one round of negations, where explicit negations themselves are left as they are. Call the resulting finite set $\Phi=\Phi_F$. This set will be fixed henceforth, and models and arguments about them will only involve these formulas.

 \begin{defi} A subset $\Sigma\subseteq \Phi$ is a \emph{Hintikka set for $\Phi$} (also occasionally called a  syntactic `type') if it satisfies the following conditions,
 where all formulas mentioned run over $\Phi$ only:
\medskip

\indent (a)\, $\neg\psi \in \Sigma$ iff $\psi \not\in \Sigma$, \quad (b) $(\varphi \land \psi) \in \Sigma$ iff $\varphi \in \Sigma$ and $\psi \in \Sigma$

\smallskip

\indent (c) if $\dual_X\psi\in \Sigma$, then $\psi \in \Sigma$, \quad (d) $D_X x\in \Sigma$ for all $x\in X$

\smallskip
(e) if $D_X Y, D_Y Z\in \Sigma$, then $D_X Z\in \Sigma$.
%\indent (g) if $\edual_X\psi \in \Sigma$, then there is some $Y\supseteq X$ s.t. $Y$ is dependence-closed wrt $\Sigma$ and $\edual_Y\psi \in \Sigma$.\footnote{One can show that in fact $Y:= D_X^\Sigma$ has this property, but we will not need this.}
\end{defi}

Note that there are only finitely many Hintikka sets for a given finite set $\Phi$. Moreover, the property of being a Hintikka set for a set $F$ of bounded size $\leq N$ is clearly \emph{decidable}.

% \begin{defi} Given a Hintikka set $\Sigma\subseteq \Phi$, a set of variables $X\subseteq V_F$ is \emph{dependence-closed with respect to $\Sigma$} if for every $y\in V_F$, $D_X y\in \Sigma$ implies $y\in X$.
%\end{defi}

\begin{defi}\label{DepClosure}
For every Hintikka set $\Sigma\subseteq \Phi$ and every set of variables $X\subseteq V_F$, the \emph{dependence closure of $X$ wrt $\Sigma$} is the set of variables $D_X^\Sigma := \{y\in V_F: D_X y\in \Sigma\}$. \end{defi}
The terminology `closure' is justified by the following  observations. First, clause (d) on  Hintikka sets implies that the dependence closure $D_X^\Sigma$ contains $X$; second, clauses (b), (e) together imply that $D_X^\Sigma$ is closed under adding variables $z$ with $D_Yz \in \Sigma$ for any $Y\subseteq D_X^\Sigma$; third, $D_X^\Sigma$ is the smallest set (in the sense of set inclusion) of variables satisfying the first two properties.  If $Z$ is any other set satisfying the two properties, then $D_X^\Sigma\subseteq Z$. For, let $z\in D_X^\Sigma$, so $D_Xz\in \Sigma$. We have $X\subseteq Z$ by the first property, and so $z\in Z$ by the second property.

%includes $X$ itself and is dependence-closed with respect to $\Sigma$.
%\footnote{In fact, $Y=D_X^\Sigma$ is the smallest set with these two properties, but  this fact is not needed in what follows.}
%\end{fact}

%\begin{proof}\quad \,\,
%$X\subseteq D_X^\Sigma=Y$ follows from condition (d) on Hintikka sets. To show that $Y$ is dependence-closed with respect to $\Sigma$, note first that, by the definition of $Y=D_X^\Sigma$, $D_X y\in \Sigma$ holds for all $y\in Y$, and hence $D_X Y\in \Sigma$, by condition (b) on Hintikka sets and the closure of $\Phi$ under all $D_XY$. Let now $z\in V_F$ with the property that $D_Y z\in \Sigma$. This, together with $D_X Y\in \Sigma$ and condition (e) on Hintikka sets, gives $D_X z\in \Sigma$, i.e., $z\in D_X^\Sigma=Y$, as desired.
%\end{proof}

%The next definition uses the notion of free variables introduced in Definition 3.3.

\begin{defi}\label{Xequivalence}
For Hintikka sets $\Sigma, \Delta\subseteq \Phi$ and $X\subseteq V_F$,
$$\Sigma\sim_X \Delta \,\, \,\, \mbox{ iff }\,\,\,\, \mbox{ $\Sigma$ and $\Delta$ have the same formulas
$\psi\in \Phi$ with $Free(\psi)\subseteq D_X^\Sigma$}.$$
\end{defi}

\begin{fact}\label{propequiv}
The following hold for all Hintikka sets $\Sigma,\Delta$, and sets of variables $X,Y\subseteq V$:
\begin{enumerate}
    \item\quad $\Sigma\sim_X \Delta$ implies $D_X^\Sigma=D_X^\Delta$,
    \item\quad $\sim_X$ is an equivalence relation,
    \item\quad $\Sigma\sim_X \Delta$ and $D_XY\in\Sigma$ imply $\Sigma\sim_Y \Delta$.
\end{enumerate}
\end{fact}

\begin{proof}\quad \,\,  For the first item: if  $\Sigma\sim_X \Delta$, then the sets $\Sigma, \Delta$  contain the same dependence atoms $D_Xy$ -- since the latter have only free variables $X$, and $X \subseteq D_X^\Sigma$. It follows that $D_X^\Sigma = D_X^\Delta$.

For the second item:  $\sim_X$ is evidently reflexive, by its definition. Symmetry and transitivity also follow immediately from the definition of $\sim_X$ together with the first item (the invariance of $D_XY$ under $\sim_X$).

The third item follows from the fact that $D_XY\in \Sigma$ implies that $D_Y^\Sigma\subseteq D_X^\Sigma$. Indeed, if $z\in D_Y^\Sigma$ and $D_XY\in \Sigma$, then we have $D_Yz, D_XY\in \Sigma$, thus $D_Xz\in \Sigma$ by property (e) of Hintikka sets, and hence $z\in D_X^\Sigma$.
\end{proof}

%Each $\sim_X$ is obviously an equivalence relation, and moreover we have:

%\begin{fact}\label{DependenceSet} If $\Sigma\sim_X \Delta$ then:
%$X$ is dependence-closed wrt $\Sigma$ iff it is dependence-closed wrt $\Delta$.
%\end{fact}

%\begin{proof}\quad \,\,  From the preceding definition of the relations  $\sim_X$ and the fact that $Free(D_X y)=X$ for every $y\in V_F$, it follows that $D_X y$ belongs to $\Sigma$ iff it belongs to $\Delta$. Using this fact plus the definition of dependence-closure, the desired equivalence is obvious. \end{proof}

Next we define a syntactic notion capturing key aspects of the families of Hintikka sets that can occur together in one dependence model. Here Clause (f) reflects the witnessing for existential dependence modalities in the model, and Clause (g) the fact that constants (i.e., variables $x$ for which $D_\emptyset x$ holds) behave uniformly in the model.

 \begin{defi}
A \emph{type model for $\Phi$} is a family $\mathfrak{M}$ of Hintikka sets for $\Phi$ obeying the following two conditions. The first is an additional `witness condition' for existential modalities:

\medskip

\indent (f)\, if $\edual_X\psi \in \Sigma\in \mathfrak{M}$, then there exists a set
%$Y\supseteq X$ and
$\Delta\in \mathfrak{M}$, such that (i) $\psi\in \Delta$, (ii) $\Sigma \sim_X\Delta$.

\medskip
\noindent The second condition expresses uniformity for constants:

\medskip

\indent (g)  $\Sigma\sim_\emptyset\Delta$ (as given in Definition \ref{Xequivalence}) holds for all $\Sigma, \Delta\in \mathfrak{M}$.\footnote{I.e., $\sim_\emptyset$  is the universal relation on $\mathfrak{M}$. This means that all Hintikka sets in $\mathfrak{M}$ contain the same atomic statements $D_\emptyset x$ (if any) and the same formulas whose free variables are all in the set of these constant $x$.}

%\smallskip
%\indent \quad \quad and (iii) the set $Y$ is dependence-closed with respect to $\Sigma$.
\end{defi}
\smallskip
Once again, for a given finite set $F$, there are only finitely many type models for $\Phi_F$, and moreover, the property of being a type model for a set $F$ of bounded size $N$ is decidable.

\subsection{\textbf{Representation of type models as dependence models}}

First, it is easy to see that every dependence model induces a type model.
\begin{defi}
Given a dependence model $\bM=(M, A)$ and a set $\Phi$ as in the previous section, the \emph{$\Phi$-type} of an assignment $s\in A$ is defined as

\vspace{-2mm}

$$type(s) \,\, =\,\, \{ \psi \in \Phi \mid \bM, s \models \psi\}.$$

\vspace{-2mm}
\end{defi}

\vspace{-2mm}

\begin{fact}\label{Types}
For every assignment $s\in A$ in a dependence model $\bM=(M, A)$, its $\Phi$-type $type(s)$ is a Hintikka set. Moreover, the set $type(A):=\{type(s): s\in A\}$ of all $\Phi$-types occurring in $\bM$ is a type model.
\end{fact}

\begin{proof}\quad \,\,
Checking conditions (a)--(e) on Hintikka sets is straightforward. For the witness condition (f) in the type model, let $\Sigma = type(s)$ for $s \in A$, and let $\edual_X\psi \in \Sigma$, i.e. $s\models \edual_X\psi$. By the semantics of LFD, there exists $t\in A$ with $s=_X t$ and $t\models \psi$, i.e. $\psi\in type(t)$. By the Locality Lemma \ref{locality}, $s, t$ make the same formulas true whose free variables are among the $X$, which includes all dependence atoms $D_Xy$. Therefore, $s, t$ agree on all variables in the set $D_X^\Sigma$, and so, once more by Locality, we have that $type(s) \sim_X type(t)$ in the sense of Definition \ref{Xequivalence}. Finally, condition (g) reflecting  the uniform behavior of constants again follows from Locality in dependence models.

%Let $Y:=D_X^{type(s)}$. By Fact \ref{DepClosure}, the set $Y$ is dependence-closed wrt $type(s)$.
%Next, it is easy to check that  $s=_Y t$. For, let $y\in Y=D_X^{type(s)}$, i.e.,  $s\models D_X y$: together with
%$s=_X t$ this gives  $s=_y t$, as desired. Finally, by the Locality property of LFD, $s=_Y t$ implies that: $s\models \psi$ iff $t\models\psi$ for all formulas $\psi$ with $Free(\psi)\subseteq Y$. This means that $type(s)\sim_Y type(t)$, again as desired.
\end{proof}

The more challenging direction is now the converse: that every type model can be represented as the set of types of some dependence model.

\begin{thm}\label{Rep} Given a type model $\mathfrak{M}$, there exists a dependence model $\bM=(M, A)$ with

\vspace{-2mm}

$$\mathfrak{M}=\{type(s): s\in A\}.$$
\end{thm}

\begin{proof}\quad \,\,  \, First fix any Hintikka set $\Sigma_0 \in \mathfrak{M}$.  Define a \emph{good path} to be a finite sequence $\pi=\langle\Sigma_0, X^1,\Sigma_1, \ldots, X^n, \Sigma_n\rangle$ of any length $n+1\geq 1$ such that (i) $\Sigma_k\in \mathfrak{M}$ for each $k$ (hence each $\Sigma_k$ is a Hintikka set in  $\mathfrak{M}$), and (ii) each $X^k\subseteq V_F$ satisfies $\Sigma_{k-1}\sim_{X^k} \Sigma_k$.
%and is dependence-closed with respect to $\Sigma_{k-1}$ (and hence also with respect to $\Sigma_k$, by Fact \ref{DependenceSet}).
Write $last(\pi)=\Sigma_n$ for the \emph{last element} of path $\pi$.\footnote{This definition creates infinitely many good paths, and as we shall see in a moment, infinitely many objects. Whether this can be restricted to a finite set of paths and values is at present an open problem.}

\medskip

In what follows, it is convenient to view good paths as consisting of successive \emph{good transitions} of the form $(\Sigma, X, \Delta)$. Here we think of the variables in $X$, and those depending on them according to $\Sigma$, as keeping their value in the transition.
More precisely, we say that

\medskip

\emph{the variables kept fixed} in a transition $(\Sigma, X, \Delta)$

\medskip

\noindent are all those in the extended set of variables $D_X^{\Sigma}$ introduced in Definition \ref{DepClosure}.\footnote{$D_X^{\Sigma}$ equals $D_X^\Delta$ by Fact \ref{propequiv}, so the reverse transition is also good.} Sets of variables kept fixed in good transitions underlie many of the definitions and proofs that follow.

The good paths are finite sequences that form a rooted branching tree in a standard manner, with the $1$-length path $\langle\Sigma_0\rangle$ as its root. It may help the reader to keep a tree picture in mind in what follows, cf. Figure 1 below for a visual aid.

\medskip

Next, \emph{objects} will be special pairs of good paths and variables. Instead of defining these objects separately, we introduce them simultaneously with the following inductive definition of \emph{path assignments} $v_\pi$ for good paths $\pi$, that send variables to objects:

\vspace{3mm}

$v_\pi(x) = (\pi, x)$ if $\pi$ has length 1, i.e. $\pi=\langle\Sigma_0\rangle$ is the root of our tree.

\vspace{2mm}

\indent $v_\pi(x) = v_{\pi'}(x)$ if $\pi=(\pi', X, \Sigma)$ with $x \in D_X^{last(\pi')}$.

\vspace{2mm}

\indent $v_\pi(x) = (\pi, x)$ if $\pi=(\pi', X, \Sigma)$ with $x\not\in D_X^{last(\pi')}$

\vspace{3mm}

\noindent The second clause leaves the same values for variables if the last transition keeps them `fixed'. The third clause creates fresh objects as soon as this fixing is not prescribed. In particular, note that \emph{constants} $x$, i.e. special variables with $D_\emptyset x$ present in \emph{all} Hintikka sets in $\mathfrak{M}$, will get the \emph{same value} $(\langle \Sigma_0\rangle, x)$ under all path assignments. By condition (g) on type models, that value never changes for longer paths.

\medskip

%result of the construction is a sort of a tree unraveling of the given type model, and drawing a concrete tree picture will help the reader in understanding the arguments to follow.

Now, we define a first-order model $M=(O,I)$ by letting
$$O=\{v_\pi (x): \pi \mbox{ good path and } x\in V_F\}$$ be the set of all objects $(\pi, x)$ assigned by the assignments $v_\pi$  in the above manner.  Next, an interpretation $I(P)$ is given to each predicate by means of the
following `coherence condition':

\vspace{3mm}

\indent $I(P)$ holds for a finite sequence of objects $\overline{(\pi, x)}$  in $O$ if all paths $\pi$ occurring\\
\indent in the sequence are linearly ordered by the relation of initial segment, and \\
\indent the formula $P{\bf x}$
occurs in $last(\pi^*)$ on the longest path $\pi^*$ among these.

\medskip

\noindent Finally, a dependence model $\bM=(M, A)$ is obtained over the first-order model $M$ by setting
$$A=\{ v_\pi: \pi \mbox{ is a good path}\}$$

 The crucial semantic notion of equality of values among assignments $v_\pi$, $v_{\pi'}$ in the dependence model $\bM$ wrt a given set $X$ of variables may be described concretely as follows. In general, the paths $\pi$, $\pi'$ fork beyond a shared initial segment $\pi''$, that includes at least $\langle \Sigma_0\rangle$. The semantic equality $v_\pi=_X v_{\pi'}$ means that the values assigned by $v_\pi$ and $v_{\pi'}$ to all variables in $X$ have been set already by the final stage of $\pi''$ (cf. Figure 1):

\bigskip\par\noindent
\bigskip\par\noindent
\begin{figure}
\begin{center}
      \begin{tikzpicture}[node distance=2.3cm]
        \tikzstyle{zz}=[decorate,decoration={zigzag,post=lineto,post length=5pt}]

        \tikzstyle{w}=[draw=black,thick,circle,fill=black,inner sep=0pt,minimum size=3pt]

        \tikzstyle{every edge}=[draw,thick,font=\small]

        \tikzstyle{every label}=[font=\small]

        \tikzstyle{ev}=[anchor=center,node distance=3cm]

        \tikzstyle{wred}=[w,draw=black]

        %%
        %% nodes
        %%

        \node[w,label={below:$\Sigma_0$}] (w0) { };

        \node[above of=w0] (w7) { };

        \node[w, above right of=w0] (w1) { };

     \node[right of=w1] (w8) { };

     \node[right of=w8] (w13) { };

   \node[w, above right of = w0,  label={below:$\pi''$}] (w2) { };

  %    \node[w,above left of=w2] (w3) { };

   \node[above right of=w2]  (w4) { };

    \node[above of=w1] (w5) { };

     \node[w, above right of = w2,label={below:$\pi'$}] (wX) { };

      \node[w, above of=w2,label={right:$\pi$}] (w6) { };

 %           \node[w,above above right of =w8] (wQ) { };

 %  \node[above right of =wQ] (wY) { };

        %%
        %% edges
        %%

          \path (w0) edge[-] node[above]{$ $} (w1);

        %\path (w1) edge[-] node[above]{$ $} (w2);

        \path (w1) edge[-] node[below right]{$X$} (wX);

         \path (w1) edge[-] node[below left]{$X$} (w6);

        %\path (w1) edge[dashed] node[above]{ } (w5);

         \path (w0) edge[dashed]  (w7);
        \path (w0) edge[dashed]  (w13);

  %        \path (wQ) edge[dashed] (wY);

      \end{tikzpicture}
      \end{center}
\caption{Forking paths $\pi$ and $\pi'$ with $v_\pi=_X v_{\pi'}$: according to Fact \ref{Fork}, all variables in $X$ are kept fixed in all transitions (on these paths) beyond the shared path $\pi''$.}
\end{figure}
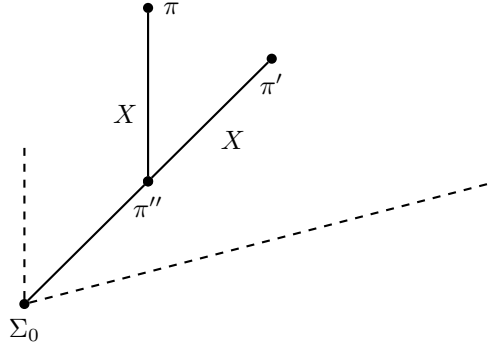
 % easychair.tex,v 3.5 2017/03/15

\vspace{-1.5cm}

\begin{fact}\label{Fork} For any two $v_\pi, v_{\pi'}$ in $\bM$ and any set of variables $X$, the following are equivalent:

\vspace{2mm}

(a) $v_\pi =_X v_{\pi'}$

\vspace{2mm}

(b) $\pi$ and $\pi'$ have the form $\pi = \pi'', X_1, \dots,  X_n ,last(\pi)$,
$\pi' = \pi'', X_1^{'}, \dots, X_m^{'}, last(\pi')$ \\
\indent with a shared path  $\pi''$,  where all variables in $X$ are kept fixed in the transitions\\
\indent  involving the displayed sets $X_1, \dots,  X_n$ and $X_1^{'}, \dots, X_m^{'}$.

\end{fact}

\begin{proof}\quad \,\,  This follows by inspection of the above definitions for values of assignments, noting that the identical objects assigned by $v_\pi$ and  $v_{\pi'}$ to any variables $x \in X$ must be of the form $(\pi^{\bullet}, x)$ for some initial segment $\pi^{\bullet}$ of the shared path $\pi''$, while no further changes have taken place.\footnote{Note that this description also covers the case when $X$ is empty: the `fork' can then be right after $\langle \Sigma_0\rangle$.}
\end{proof}

To complete the proof of the main theorem, we must show that our initially given type model coincides with the set of $\Phi$-types of all assignments in $\bM$, i.e. that we have: $\mathfrak{M}=\{type(v_\pi): v_\pi\in A\}$. And in order to establish this identity, it suffices to prove that

\medskip

\quad \quad $type(v_\pi) \, = \, last(\pi)$ \quad for all good paths $\pi$.

\medskip
\smallskip

\noindent Once we proved this claim, the desired identity $\mathfrak{M}=\{type(v_\pi): v_\pi\in A\}$ is immediate.\footnote{To see this, in one direction, each set $last(\pi)$ for a path  $\pi$ is by definition a Hintikka set in $\mathfrak{M}$, In the  opposite direction, each set $\Gamma\in \mathfrak{M}$ immediately gives a good path $\pi = (\Sigma_0, \emptyset, \Gamma)$ of length $2$, since $\sim_\emptyset$ was the universal relation on $\mathfrak{M}$, and hence there is a matching  assignment $v_\pi$ with $type(v_\pi) = last(\pi) =  \Gamma$.} %In other words, the final task is to prove the following:
Unfolding now the claim  $type(v_\pi) \, = \, last(\pi)$, we can see that our remaining task is to prove the following result.

\begin{fact}[Truth Lemma] For all formulas $\varphi \in \Phi$ and good paths $\pi$, the following  holds:

\medskip{
}$\bM, v_\pi \models \varphi$ iff $\varphi \in last(\pi)$.
\end{fact}

\begin{proof}\quad \,\,  The proof is by induction on the formula $\varphi$.

\medskip

Case 1: \emph{Atomic formulas}. By the truth definition for LFD, $\bM, v_\pi \models P{\bf x}$ iff $I(P)(v_\pi ({\bf x}))$. By the above definition of the atomic predicates in the first-order model $M$,  the objects $v_\pi ({\bf x})$ are pairs ($\pi', x)$ whose paths $\pi'$ are all initial subpaths of the longest path $\pi^*$ among them. Moreover, the formula
$P{\bf x}$ belongs to $last(\pi^*)$. Now, given the above inductive definition of assignments, all objects assigned by $v_\pi$ to variables have a path component which is an initial segment of $\pi$. In particular, $\pi^*$ is an initial segment of $\pi$, and also, again by the inductive definition of the assignments, no values of variables $x \in \bf{x}$ have changed along the remaining path from $\pi^*$ to $\pi$. This means, by Definition \ref{Xequivalence} for the relations  $\sim_X$ that the formula $P{\bf x}$ itself occurs in every Hintikka set in $\pi$ after $\pi^*$, and in particular, that $P{\bf x}$ occurs in the set $last(\pi)$.

%all initial subpaths of some good path $\pi'$ such that $P{\bf x}\in last(\pi')$. Now the given path $\pi$ will have $\pi'$ as an initial segment, since $v_\pi$ assigns only objects marked by subpaths of $\pi$. But it may continue further. However, given the definition of  objects assigned by  $v_\pi$, none of the relevant variables $x$ have changed their value through the steps after $\pi'$, which means that $P{\bf x} \in last(\pi)$.

\medskip

Case 2: \emph{Boolean combinations}. The proof is a straightforward appeal to the truth definition, the  inductive hypothesis, and the definition of Hintikka sets.

\medskip

Case 3: \emph{Dependence modalities}. For ease of presentation, we consider the existential LFD dependence modality instead of the universal one.

\smallskip

\emph{From right to left.} Let
$\edual_X \varphi \in last(\pi)$. By the witness condition (f) on type models, there exists a set $\Delta\in \mathfrak{M}$ with  $\varphi\in \Delta$ and
$last(\pi) \sim_X \Delta$.
Let $\pi^{+}=(\pi, X, \Delta)$ be the good path consisting of $\pi$ with a $\sim_{X}$-transition to $\Delta$ added. By the inductive hypothesis, $\bM, v_{\pi^{+}} \models \varphi$, and hence also $\bM, v_{\pi^{+}} \models \edual_X \varphi$. Now consider the objects that $v_{\pi^{+}}$ assigns to the variables in $X$. By the above definition for $v_{\pi^{+}}$, none of the variables $x\in X$ changed their value in the last step $\sim_{X}$  -- and so, these objects are the same as those assigned by $v_\pi$. Thus, the assignments $v_{\pi^{+}}$, $v_\pi$ agree on the values of the free variables for $\edual_X \varphi$, and so, by the Locality Lemma \ref{locality}, the latter formula is also true at $\bM, v_\pi$.

\smallskip

\emph{From left to right.} Let $\bM, v_\pi \models \edual_X \varphi$. By the truth definition, there is an  assignment $v_{\pi' }=_X v_\pi$ with
$\bM, v_{\pi' }\models \varphi$, so, by the inductive hypothesis $\varphi \in last(\pi')$. By condition (c) on Hintikka sets (dualized to the existential dependence modality), we then have $\edual_X \varphi \in last(\pi')$. Now  compare the two good paths $\pi, \pi'$, keeping Fact \ref{Fork} in mind concerning their shape wrt some shared initial path $\pi''$, and the fact that $X$ is contained in the set of  variables kept fixed in each transition made on the paths extending beyond $\pi''$ toward $last(\pi)$ and toward $last(\pi')$.

%\indent \emph{Any two good paths on a tree are equal up to some (possibly empty) initial segment.}

%\noindent This easily visualized feature makes it possible to reason about the `forks' of the tree occurring in this way, and trace which variables have changed values along paths and across forks.
Given that $\edual_X \varphi \in last(\pi')$, with free variables $X$, it follows by Definition \ref{Xequivalence} that this formula is present in each Hintikka set on the path toward $last(\pi'')$ and then in each Hintikka set on the path from there toward $last(\pi)$.\footnote{What we use here is the earlier observation that good transitions are good in both directions.}  So, finally, $\edual_X\varphi \in last(\pi)$.

%it suffices to transfer this fact from $\pi$ to $\pi'$, to obtain $\edual_X \varphi \in last(\pi)$. This works by inspection of each relational step $\sim_U$, traveling backward in the tree from the end of $\pi'$ to the fork and then upward to the end of $\pi$. Crucially, since $v_\pi, v_{\pi'}$ assigned the same objects to the variables $X$, they cannot have changed their values after the fork. Hence, step by step along the route, every dependence-closed set $U$ marking the transition steps contains every $x\in X$. Therefore,
% $Free(\edual_X \varphi)=X\subseteq U$;
%and hence, by the definition of $\sim_U$ on Hintikka sets, the formula $\edual_X \varphi$ is passed along each such transition step.

\medskip

Case 4: \emph{Dependence atoms.} The case of dependence atoms is proved in a similar manner, but interestingly, it makes no appeal to a witness clause for non-dependence in type models.

\smallskip
\emph{From right to left.} Let $D_Xy \in last(\pi)$.  Local semantic dependence of $y$ on $X$ at the assignment $v_\pi$ is shown as follows. Consider any assignment $v_{\pi'}\in A$ assigning the same objects to the variables in $X$, i.e.,  $v_\pi(X)=v_{\pi'}(X)$. Just as in the preceding Case 3, $X$-values have not changed after the largest common initial segment $\pi''$ of $\pi$ and $\pi'$. But then, since $Free(D_Xy)=X$, the formula $D_Xy$ is shared by the Hintikka sets in each of  these later transitions. Now the above recursive definition of values $v_\pi(u)$ for variables $u$ under assignments $v_\pi$ worked with  extended sets of variables $D_X^{last(\pi')}$ for immediately preceding subpaths $\pi'$, and these sets all include $y$ in the present case. It follows  that $v_{\pi}(y)=v_{\pi'}(y)$, as desired,

%transition step $\sim_U$. From this, together with $X\subseteq U$ and the fact that each set $U$ is dependence-closed (with respect to the appropriate Hintikka set), it follows that $y\in U$, for all the $U$'s encountered along the path from $\pi$ to $\pi'$. So, by the above definition of objects, $y$ does not change its denotation between $\pi$ and $\pi'$, i.e.,

%$v_{\pi}(y)=v_{\pi'}(y)$, as desired.

\smallskip

\emph{From left to right.} Let $\bM, v_\pi \models D_Xy$. Consider the good path
$\pi^{+}= (\pi, X, last(\pi))$  extending $\pi$ with one good $\sim_X$ transition to the Hintikka set $last(\pi)$. By the earlier definitions for the values given by our assignments,
%Put $Y=D_X^{last(\pi)}=\{y\in V_F: D_X y\in last(\pi)\}$. By Fact \ref{DepClosure}, $Y$ is dependence-closed with respect to $last(\pi)$, and we obviously have $last(\pi)\sim_Y last(\pi)$.
%So consider the good path By Fact \ref{DepClosure} again, it holds that $X\subseteq Y$, so by the earlier definitions,
$v_\pi, v_{\pi^{+}}$ assign the same objects to all the variables $x\in X$, kept fixed in the final transition. Therefore, by the given local semantic dependence at $v_\pi$, we also have that $v_{\pi^{+}}(y) = v_{\pi}(y)$. But this can only happen if the variable $y$, too, was kept fixed in the last transition of $\pi^{+}$, which means by definition that $y\in Y=D_X^{last(\pi)}$: i.e., $D_Xy \in last(\pi)$.
\end{proof}
This concludes the proof of Theorem \ref{Rep}.
\end{proof}

\subsection{\textbf{Decidability}}

The decidability of LFD can now be established.

 \begin{thm} Validity for formulas of \emph{LFD} on dependence models is decidable.
 \end{thm}

 \begin{proof}\quad \,\,
 By Theorem \ref{Rep} and Fact \ref{Types}, satisfiability for a formula $\varphi$ in dependence models is equivalent to $\varphi$'s occurring in some Hintikka set of some type model for the set $\Phi$ generated by $F=\{\varphi\}$ and all its subformulas and dependence formulas in the manner described earlier. As there are only finitely many type models of this sort, the latter test is decidable.
\end{proof}

\noindent {\bf Open problems} \, \emph{Does LFD have the Finite Model Property?} \ \emph{What is the computational \\ complexity of satisfiability for LFD?}

\begin{rem}
\emph{As noted earlier, the proof of decidability for LFD presented here is an extension of that for the Guarded Fragment of first-order logic, \cite{ABN}. An open problem is whether we can reduce the decidability problem for LFD to that for the Guarded Fragment with identity, \cite{Graedel99}, though this seems unlikely given the syntax of dependence atoms. Another issue in this connection is whether known decidable extensions of the Guarded Fragment such as the `loosely guarded fragment', \cite{ELD}, \cite{Bent2005}, have counterparts in natural extensions of LFD.}
\end{rem}

Finally, it may be worth noting that the preceding style of decidability argument can also be applied to first-order logic itself. Hintikka sets and type models can be defined just like above,  and the representation result for type models as dependence models also goes through. Moreover, it is decidable whether a given first-order formula has a type model. Given the undecidability of FOL,  it must then be an undecidable problem whether a given type model can be represented as a standard first-order model, i.e., a \emph{full} assignment model.

\section{\large{\textbf{Axiomatizations}}}\label{Axiomatize}

\vspace{-2mm}

It was shown in Section 3.2 that the set of LFD validities is recursive. In this section the structure of this set will be explored in more depth, in the form of  two complete deductive systems.

\vspace{-2mm}

\subsection{\textbf{A Hilbert-style axiomatization}}
A Hilbert-style \emph{proof system} {\bf LFD} is given in Table \ref{tb2}, consisting of: (I) the classical axioms and rules of propositional logic; (II) axioms and rules for dependence modalities, that can be seen as restricted duals of the classical Hilbert axioms for quantifiers; (III) axioms governing the behavior of dependence atoms (namely, Projection and Transitivity, already known to be equivalent to the conjunction of Reflexivity, Monotonicity and Transitivity); (IV) the key Transfer axiom, describing the interaction between dependence modalities and dependence atoms. The notions of formal derivation and provability are defined as usual.

\begin{table}[h!]
\begin{center}
\begin{tabularx}{\textwidth}{>{\hsize=0.5\hsize}X>{\hsize=1.5\hsize}X>{\hsize=0.8\hsize}X}

\toprule

\vspace{-0.7cm} \\
\textbf{(I)} &\textbf{Axioms and rules of classical propositional logic}
 \vspace{2mm} \ \\
 \textbf{(II)} &\textbf{Axioms and rules for dependence modalities $\dual$}  \vspace{2mm} \ \\
($\dual$-Necessitation)& From $\varphi$, infer $\dual_X\varphi$ \vspace{1mm} \ \\
($\dual$-Distribution) & $\dual_X (\varphi \to \psi) \to (\dual_X\varphi \to \dual_X\psi)$ \vspace{1mm} \ \\
($\dual$-Introduction) & $\varphi \to \dual_X\varphi$, provided that $Free(\varphi)\subseteq X$ \vspace{1mm}\ \\
($\dual$-Elimination) & $\dual_X\varphi \to \varphi$ \vspace{2mm} \ \\
\textbf{(III)} & \textbf{Axioms for dependence atoms $D$}  \vspace{2mm}  \ \\
(Projection) & $D_X x$, provided that $x\in X$ \vspace{1mm}\ \\
%(Monotonicity) & $D_X y \to D_Z y$, provided that $X\subseteq Z$ \vspace{1mm}\ \\
(Transitivity) & $\left( D_X Y \wedge D_Y Z \right)\to D_X Z$ \vspace{2mm} \ \\
%(Monotonicity) & $D_X y \to D_Z y$, provided that $X\subseteq Z$ \vspace{2mm}\ \\
 \textbf{(IV)} & \textbf{Axiom for $\dual$-$D$ interaction} \vspace{2mm}\ \\
(Transfer) &  $\left(D_X Y\wedge \dual_Y\varphi \right)\to \dual_X \varphi$ \hspace{2cm}
\vspace{1.5mm}\\
\bottomrule
\end{tabularx}
\end{center}
\vspace{-0.5cm}
\caption{The proof system $\mathbf{LFD}$.}\label{tb2}
\end{table}

%\vspace{-2mm}

\begin{fact}\label{K-Intro} In the context of the other axioms and rules presented in Table \ref{tb2}, the axiom schema ($\dual$-Introduction) can be replaced by its instances listed in Table \ref{tb3}.
\end{fact}

%\vspace{-2mm}

\begin{table}[h!]
\begin{center}
\begin{tabularx}{\textwidth}{>{\hsize=0.5\hsize}X>{\hsize=1.5\hsize}X>{\hsize=0.8\hsize}X}
\toprule
\vspace{-0.7cm}
\\
($\dual$-Intro$_1$) & $P x_1\ldots x_n \to \dual_{\{x_1, \ldots, x_n\}} P x_1\ldots x_n$ \vspace{1mm} \ \\
($\dual$-Intro$_2$) & $\dual_X\varphi \to \dual_X \dual_X \varphi$ \vspace{1mm}\ \\
($\dual$-Intro$_3$) & $\neg \dual_X\varphi \to \dual_X \neg \dual_X\varphi$ \vspace{1mm} \ \\
($\dual$-Intro$_4$) & $D_X\mb{y} \to \dual_X D_X\mb{y}$
\hspace{2cm}
\vspace{-5mm} \\
\bottomrule
\end{tabularx}
\end{center}
\vspace{-0.5cm}
\caption{
The relevant instances of dependence modality introduction.}\label{tb3}
\end{table}%

%\vspace{-0.3cm}

Note that, unlike with CRS, the provable principles for $\mathbf{LFD}$ are closed under substitution for predicate letters. Note also the analogy between $\dual$-Necessitation, $\dual$-Distribution, $\dual$-Elimination, $\dual$-Intro$_2$ and $\dual$-Intro$_3$ with the usual axioms and rules of the modal system $S5$. This is unsurprising and it is more than an analogy: as seen in Section
\ref{StandardRelationalSemantics}, our dependence modalities $\dual_X$ are in fact relational modalities for equivalence relations $\sim_X$, and so they automatically validate all the $S5$ laws (by known results in classical modal  correspondence theory, \cite{BRV}).

\begin{exam}\label{LFD theorems} The following formulas are derivable as theorems in  $\mathbf{LFD}$:

\vspace{1mm}

(a) $D_X Y$, \, for $Y\subseteq X$ \,\, \quad\quad\quad\quad\quad\quad (Inclusion)

(b) $(D_X Y \land D_Z U) \rightarrow D_{X\cup Z} (Y\cup U)$ \,\, \quad (Additivity of Dependence)

(c) $D_X Y\to D_Z Y$, \, for $X\subseteq Z$ \,\, \quad\quad\quad\quad (Monotonicity of Dependence)

(d) $\dual_X\varphi \rightarrow \dual_Y\varphi$, \, for $X \subseteq Y$ \,\, \quad\quad\quad\quad\quad (Monotonicity of Dependence Quantifiers)

%(d) the laws for universal quantifiers: \,
%$\forall_X \varphi \to \varphi$;  \, \,\,
%$\, \varphi \to \forall_X \varphi \, \, \mbox{ if } Free(\varphi)\cap X=\emptyset.$\end{exam}

(e)
$\forall_X \varphi \to \varphi$;  \, \,\,
$\, \varphi \to \forall_X \varphi \, \, \mbox{ if } Free(\varphi)\cap X=\emptyset$ \, \,\, (Universal Quantifier Laws).
\end{exam}

Note that the more general quantifier elimination rule via substitution $\forall_x\varphi\to [y/x]\varphi$ (as in classical FOL) is \emph{not} a theorem or axiom of ${\bf LFD}$: indeed, as we saw in Example \ref{valid}, this rule is not sound in our semantics.
%\medskip

\begin{thm}\label{Completeness} (\emph{Completeness})
The system {\bf LFD} is sound and complete for dependence models.
\end{thm}

\begin{proof}\quad \,\,  Given a consistent formula $\varphi$, consider the set $\Phi=\Phi_F$ generated by $F=\{\varphi\}$ as in Section 4.1. Fix some maximally consistent subset $\Sigma^{\ast}$  of $\Phi$ that contains $\varphi$, and let  $\mathfrak{M}$ be the family of all maximally consistent subsets of $\Phi$ that are connected to $\Sigma^{\ast}$ via a finite sequence of relations $\sim_X$ as introduced in Definition \ref{Xequivalence}.\footnote{This corresponds to taking a `generated submodel', a standard technique in modal logic.}

\begin{fact}\label{TypeFact}
The family $\mathfrak{M}$ is a type model.
\end{fact}

\noindent\emph{Proof.} Maximally consistent subsets are Hintikka sets:
they obviously satisfy the Boolean clauses, and the other closure conditions follow from their closure under deduction. To prove that $\mathfrak{M}$ satisfies the witness condition (f) on type models, let $\edual_X\psi\in \Sigma\in \mathfrak{M}$. Take $Y:=D_X^\Sigma$ (the dependence closure of $X$ with respect to $\Sigma$), and consider the set
\vspace{-1mm}
$$\Delta_0\,\, :=\,\, \{\psi\}\cup \{\theta\in \Sigma: Free(\theta)\subseteq Y\}.$$
This set of formulas is consistent by a standard modal argument using the $S5$ axioms\footnote{These are $\dual$-Necessitation, $\dual$-Distribution, $\dual$-Elimination, $\dual$-Intro$_2$ and $\dual$-Intro$_3$, all available in {\bf LFD}.} for $\dual$, the
presence of the formulas $D_Xy$ in $\Sigma$, and the Transfer Axiom of {\bf LFD}. The required Hintikka set $\Delta$ can be taken to be any maximally consistent set in $   \mathfrak{M}$ that includes $\Delta_0$. Finally, condition (g) on type models is satisfied because all sets in $\mathfrak{M}$ are connected by $\sim_X$ transitions, which are also $\sim_{\emptyset}$ transitions by the Monotonicity property provable in {\bf LFD}.

This concludes the proof of Fact \ref{TypeFact}, and of the completeness theorem.\end{proof}

%\medskip

Theorem 5.3 states `weak completeness' only. `Strong completeness' says that provability also matches semantic consequence
 from possibly infinite sets of formulas.

\begin{thm}
The proof calculus {\bf LFD} is strongly complete.
\end{thm}

\begin{proof}\quad \,\,
First, the Compactness Theorem holds for LFD.
This follows from the first-order translation in Fact 3.8, plus
compactness for first-order logic.
Given this, given any valid semantic consequence $\Psi\models \varphi$, we also have a valid consequence $\Psi_0\models \varphi$ from some \emph{finite subset} $\Psi_0\subseteq \Psi$ of the premises -- and this amounts to the validity
of a single formula $\bigwedge\Psi_0\to\varphi$. By the weak completeness
theorem, there is a formal proof of this formula, hence $\varphi$ is provable from $\Psi$.
\end{proof}

\medskip

In Appendix A, we give another  proof of strong completeness, that proceeds along more standard lines using modal logic techniques.

%\vspace{-2mm}

\subsection{\textbf{Sequent calculus, cut elimination and strong interpolation}}
An alternative formulation of the proof system is as a sequent calculus. To avoid the use of the rules of Contraction and Permutation, we take a Gentzen calculus using \emph{sets} of formulas rather than sequences. In the following, $\Gamma$, $\Delta$  denote sets of formulas,  $\Gamma, \varphi$ denotes  $\Gamma\cup\{\varphi\}$, etc. $Var(\Gamma)$ is the set of all variables occurring in $\Gamma$, and $Free(\Gamma)$ is the set of free variables in $\Gamma$.

%Once again, an equivalent presentation can be given in terms of a Gentzen-style sequent calculus, consisting of the standard sequent axioms and rules for Propositional Logic (including Cut, Weakening and Contraction), together with the following axioms and rules:

\smallskip

\begin{defi} The \emph{sequent calculus} for \emph{LFD} has the standard Gentzen axioms and rules for classical propositional logic (including structural rules of Identity, Weakening and Cut), together with the following additional axioms and rules:

%\vspace{-1mm}

\begin{prooftree}
%\small
\AxiomC{}
\LeftLabel{(Projection)}
\RightLabel{\, where $x\in X$}
\UnaryInfC{$\vdash D_X x$}
\end{prooftree}

%\vspace{-3mm}

\begin{prooftree}
%\small
\def\fCenter{\mbox{\ $\vdash$\ }}
\AxiomC{$\Gamma \vdash \Delta, D_X Y$}
\AxiomC{$\Gamma \vdash \Delta, D_Y Z$}
\LeftLabel{(Transitivity)}
\BinaryInf$\Gamma \fCenter \Delta, D_X Z$
\end{prooftree}

%\vspace{-3mm}

\begin{prooftree}
%\small
\def\fCenter{\ \vdash\ }
\Axiom$\varphi,\Gamma \fCenter \Delta$
\LeftLabel{($\dual_L$)}
\UnaryInf$\dual_X\varphi, \Gamma \fCenter \Delta$
\end{prooftree}

%\vspace{-3mm}

\begin{prooftree}
%\small
\def\fCenter{\ \vdash\ }
\Axiom$\Gamma \fCenter \Delta,\varphi$
\LeftLabel{($\dual_R$)}
\RightLabel{\, where $Free(\Gamma\cup\Delta)\subseteq Y$}
\UnaryInf$D_X Y,\Gamma \fCenter \Delta, \dual_X \varphi $
\end{prooftree}

\end{defi}

\smallskip

\noindent Note that, compared with the classical sequent calculus for FOL, there are now extra structural rules for $D$-Projection and $D$-Transitivity. Next, the left-introduction rule ($\dual_L$) is weaker than (the dual version of) the classical left-introduction rule for the universal first-order quantifier $\forall$, as it does not allow for variable or term substitutions. Also, the right-introduction rule ($\dual_R$) is different from, and in fact stronger then, the (dual version of the)
classical rule for $\forall$: note it involves a dependence-atom premise (incorporating the Hilbert-style Transfer axiom). But also note that ($\dual_R$) implies the weaker rule
%\vspace{-1mm}
\begin{prooftree}
%\small
\def\fCenter{\ \vdash\ }
\Axiom$\Gamma \fCenter \Delta,\varphi$
\RightLabel{\, \emph{where} $Free(\Gamma\cup\Delta)\subseteq X$,}
\UnaryInf$\Gamma \fCenter \Delta, \dual_X \varphi,$
\end{prooftree}
%\vspace{-1mm}
which can indeed be seen as a dualization of the classical right-introduction rule for the universal quantifier of FOL.
%corresponding dual-quantifier rule in (the dual version of) $CRS$. The reason is that this stronger rule incorporates the Transfer axiom from the Hilbert-style system.

\medskip

It is easy to show that \emph{the two proof calculi are equivalent in terms of their output}:

\begin{fact} The provable sequents $\Gamma \vdash \Delta$ in the above calculus match exactly the provable implications $\bigwedge \Gamma \rightarrow \bigvee \Delta$ in the axiomatic system {\bf LFD}.
\end{fact}

Although our sequent calculus lacks standard cut elimination in its full generality, it does have it in a restricted form. Namely, Cut is eliminable in favor of `DA Cut': this version of the Cut Rule allows cutting only \emph{dependence atoms that involve variables actually occurring in the conclusion}. To ensure the subformula/subterm property, it is also convenient to absorb Weakening into the logical rules (cf. \cite{T&S}, or the explanation in Appendix B), while simultaneously restricting Projection and Transitivity to the variables that actually occur in the sequent to be proven. A \emph{restricted-cut proof} uses only these modified rules and the DA Cut rule. We obtain a limited, but very useful, form of the Cut Elimination Theorem:

\begin{thm}\label{CutElim} (\emph{Restricted Cut Elimination}) Every provable sequent $\Gamma\vdash \Delta$ has a restricted-cut proof. Such a proof involves only subformulas of the sequent formulas, or dependence atoms for variables occurring in the final sequent proved.
\end{thm}

The details, as well as a sketch of the proof, are in Appendix B.

\begin{rem}[Decidability revisited] \emph{These results yield a purely proof-theoretic proof of \emph{decidability} for $LFD$. For a given sequent $\Gamma \vdash\Delta$,  proof search in the above system with no other structural rule than  DA Cut is finite. The search produces a tree whose nodes are sequents $\Gamma'\vdash\Delta'$ consisting only of subformulas of the original sequent or formulas $\dual_{\mb{x}}\mb{y}$ with all $x_i, y_j\in Var(\Gamma\cup\Delta)$. There are only finitely many such formulas, and thus only finitely many such sequents $\Gamma'\vdash\Delta'$ (since $\Gamma', \, \Delta'$ are sets, there are no repetitions). The pruned tree will be finite, and it contains a proof of the original sequent iff such a proof exists.}
\end{rem}

\smallskip

\noindent Another spin-off is a strong version of Craig Interpolation for LFD. A formula $\theta$ is a \emph{strong interpolant} for a sequent $\Gamma\vdash\Delta$ if we have: (1) $\Gamma\vdash \theta$ and $\theta\vdash\Delta$ are valid, (2) all predicate symbols in $\theta$ occur both in $\Gamma$ and in $\Delta$, and (3) all \emph{variables} in $\theta$ occur in both $\Gamma$ and in $\Delta$, i.e., we have $Var(\theta)\subseteq Var(\Gamma)\cap Var(\Delta)$.

\begin{thm}\label{Interpol} (\emph{Strong Interpolation}) If $\Gamma\vdash \Delta$ is valid, then there exists a strong interpolant for this sequent.\end{thm}
%\footnote{\emph{Proof}.}
\begin{proof}\quad \,\,  By Completeness and Restricted Cut Elimination, $\Gamma\vdash \Delta$ has a restricted-cut proof. So, it is enough to find strong interpolants for all sequents that are restricted-cut-provable. For this, it suffices to provide strong interpolants for the axioms, and then show how to turn strong interpolants for the premises of each of the above modified rules (including DA Cut) into a strong interpolant for the conclusion. This can be  done in the usual way. The strong version of the above interpolation result arises thanks to the tighter variable management provided by DA Cut and restricted Projection and Transitivity.
\end{proof}

As usual,  interpolation implies a version of the Beth Definability Theorem. Given a sequent $\Gamma$, an $n$-ary relation symbol $P$ and a tuple of $n$ fresh variables $\mb{x}=(x_1, \ldots, x_n)$ with $x_i\not\in Var(\Gamma)$,  say that $\Gamma$ \emph{implicitly defines $P$ in variables $\mb{x}$} if the sequent
$$\Gamma, \Gamma' \, \vdash \, P\mb{x}\leftrightarrow P'\mb{x}$$
is valid, where $P'$ is any fresh relation symbol of the same arity as $P$, and $\Gamma'$ is the sequent obtained from $\Gamma$ by replacing every occurrence of $P$ with $P'$.

\begin{thm}\label{Beth} (\emph{Strong Beth Definability})
If $\Gamma$ implicitly defines $P$, then there is a formula $\theta$ with $Var(\theta)\subseteq Var(\Gamma)\cup \{x_1, \ldots, x_n\}$, such that the sequent  $\Gamma \, \vdash \, P\mb{x}\leftrightarrow \theta$ is provable.
\end{thm}

\subsection{\textbf{Adding special axioms}}
Further axioms beyond the logic {\bf LFD} may hold on special classes of dependence models. We give just a few illustrations here, relying heavily on known notions and results from modal logic. For convenience, we will mostly use the existential version of the dependence modality.

\begin{exam} Consider the following operator interchange principle:

\medskip

 \quad $\edual_X\edual_Y\varphi \rightarrow \edual_Y\edual_X\varphi$ \quad \quad\quad \emph{(Commutation)}

 \medskip

 \noindent The following dependence model $\bM$ is a counterexample. Take two variables $x, y$ and three assignments $s, t, u$ with $s(x) = s(y) = 0, t(x) = 0, t(y) = 1, u(x) = u(y) = 1$. Let $R$ be a binary predicate holding only of the tuple of objects $(1, 1)$. Then $\bM, s \models \edual_x\edual_y Rxy$, as one can reach $u$ by first keeping the value of $x$ fixed, and then that of $y$. But $\bM, s \models \edual_y\edual_x Rxy$ is false: there is no way of getting from $s$ to $u$ by first keeping the value of $y$ fixed, and then that of $x$.
\end{exam}

On the other hand, it is easy to see that $\edual_X\edual_Y\varphi \rightarrow \edual_Y\edual_X\varphi$ holds on \emph{full} dependence models (with all functions from $V$ to $O$ as assignments). The crucial property here is the following:

\begin{fact} The Commutation axiom $\edual_X\edual_Y\varphi \rightarrow \edual_Y\edual_X\varphi$ is valid on the class of all dependence models $\bM=(M,A)$ satisfying the following closure property for available assignments:

\vspace{2mm}

for every three assignments $s, t, u\in A$, if $s =_X t =_Y  u$, then \\
\indent there also exists an assignment
$v\in A$ in $\bM$ with $s =_Y v =_X  u$.
\end{fact}

This technical condition is a  Church-Rosser principle requiring the set of available assignments to be rich in alternative pathways. It is in fact the exact semantic content of Commutation, but  formulating this precisely requires the  modal notion of \emph{frame correspondence}, \cite{BRV}, that we will demonstrate with a different example below. The result of the above Church-Rosser restriction on dependence models is striking:

\begin{fact} The logic {\bf LFD} plus the Commutation axiom is undecidable.
\end{fact}

\begin{proof}\quad \,\,  It is known  that the  modal CRS-type logic of generalized assignment models plus the commutation axiom $\exists x \exists y \varphi \rightarrow \exists y \exists x \varphi$ is undecidable,  \cite{Nem85}. Given that this logic can be translated effectively into {\bf LFD} plus the Commutation axiom, the latter logic is undecidable too.\footnote{The undecidability can be understood as follows.  Commutation is a modal `Sahlqvist'-type axiom supporting a  completeness theorem for its frame-corresponding property, cf. \cite{BRV} for details. In particular, the cited CRS-type logic is complete for dependence models satisfying the Church-Rosser constraint. Given the grid-like structure of such models, one can then express standard undecidable tiling problems on geometrical grids as satisfiability problems for the logic. Cf.  \cite{Marx2006} for details of this widely used reduction technique for proving undecidability.}
\end{proof}

To show a bit more detail of how frame correspondence analysis works, we give an illustration for a related special dependence axiom. Recall the invalid principle $\dual_X \dual_Y\varphi \rightarrow \dual_{X\cap Y}\varphi$ mentioned in Example \ref{valid}, perhaps better understood in its existential form

\medskip

\indent $\edual_{X\cap Y}\varphi \rightarrow \edual_X\edual_Y \varphi$ \quad\quad\quad\quad (`Stepwise')

\medskip

\noindent  Like Commutation, Stepwise expresses an existence constraint on available assignments that holds in full dependence models, but not in all of them. We now give a semantic correspondence analysis, for convenience, in terms of only \emph{three variables} $x, y, z$. Call an LFD formula $\varphi$ \emph{true} in a \emph{dependence frame} (a dependence model without an added interpretation for atomic predicates) if, for every interpretation of the predicate letters on the frame (where dependence atoms always keep their fixed interpretation), $\varphi$ is true at every assignment.

\begin{fact}\label{stepwise}
The  \emph{Stepwise} axiom is true in a dependence frame  iff that frame is a full Cartesian product with all possible combinations of values for the values.
\end{fact}

\begin{proof}\quad \,\,  We show that,  with three variables, frame truth of Stepwise expesses that the  admissible assignments $A$  include all functions from  $x, y, z$ to the Cartesian product $Val(x) \times Val (y) \times Val(z)$, where $Val(x):=\{s(x):s\in A\}$ and similarly for $Val(y)$, $Val(z)$.

In one direction, this is straightforward. If the frame has the stated Cartesian structure, then it is easily verified that Stepwise will hold everywhere under every interpretation of the atomic predicates. In the opposite direction, starting from the frame truth of Stepwise, the quantification over all interpretations of atomic predicates allows us to  assume that for each assignment $s$, there exists some predicate $Pxyz$ that holds uniquely for the values $s(x), s(y), s(z)$.\footnote{See again \cite{BRV} for details of this standard move in a modal frame correspondence argument.}

Now, suppose  some value $d$ occurs for $x$ at some available assignment $s$. Suppose also that  value $e$ occurs for $y$ at some assignment $t$, uniquely defined by an atomic formula  $Pxyz$. One can reach $t$ from $s$ via the universal relation $=_{\emptyset}$, so $s$ satisfies $\edual_{\emptyset}Pxyz$. Now   write $\emptyset = \{x\} \cap \{y, z\}$. Then by Stepwise,  we also have $\edual_{\{x\}}\edual_{\{y, z\}} Pxyz$ true at $s$. But that means one can go from $s$ to some assignment $u$ keeping the value of $x$ fixed, and then from $u$ to $t$ keeping the values of $y, z$ fixed. It follows that $u(x) = d, u(y) = e$.  Next assume that $z$ takes on value $f$ at some assignment $v$. Repeating the preceding argument for $u$ and $v$, now making the split $\emptyset = \{x, y\} \cap \{z\}$, we find an assignment $w$ with $w(x) = d, w(y) = e, w(z) = f$.

%Next suppose that some value $e$ occurs for $y$ at some assignment $u\in A$. Writing the empty set as $\{x\} \cap \{y, z \}$ this time, the value for $s_1(x)$ can be kept the same to now arrive at an assignment $s_2$ where $y$ already has the value $e$. Repeating this argument with an arbitrary value $f$ for $z$, one finds an assignment $s_3$ with $s_3(x) = d, s_3(y) = e, s_3(z) = f$.

%The Cartesian product obtained here is not a full model in our sense, since it is not a full cube: each variable can have its own range of objects. So this does not correspond to any standard model for single-sorted FOL. However, product models of this kind are standard models for the \emph{three-variable fragment of} \emph{many-sorted first-order logic}, which is known to be undecidable, \cite{HMT}. Moreover, given that CRS quantifiers are definable by LFD dependence modalities (cf. Section 3.1), while CRS quantifiers coincide with first-order quantifiers on standard models,  satisfiability of first-order formulas in the above three-variable fragment reduces to satisfiability of LFD formulas in the preceding full Cartesian models. One just  replaces first-order quantifiers $\exists u$ by their obvious LFD-counterparts $\edual_{\{x, y, z\} - \{u\}}$.
\end{proof}
Again, there is a consequence in terms of logics extending {\bf LFD}.

\begin{fact}\label{stepwise}
The logic {\bf LFD} plus the  \emph{Stepwise} axiom is undecidable.
\end{fact}

\begin{proof}\quad \,\,
The Stepwise axiom has the  modal Sahlqvist form mentioned in Footnote 40, and hence, by general results, \cite{BRV}, this logic is complete for dependence frames satisfying the corresponding condition identified above.
%Next suppose that some value $e$ occurs for $y$ at some assignment $u\in A$. Writing the empty set as $\{x\} \cap \{y, z \}$ this time, the value for $s_1(x)$ can be kept the same to now arrive at an assignment $s_2$ where $y$ already has the value $e$. Repeating this argument with an arbitrary value $f$ for $z$, one finds an assignment $s_3$ with $s_3(x) = d, s_3(y) = e, s_3(z) = f$.
Now, the Cartesian product structure obtained here is not a full dependence model in our sense, since  each variable can have its own range of objects. But this is no obstacle to the following  analysis combining two known facts.

Dependence models with the preceding structure are standard models for the \emph{three-variable fragment of} \emph{many-sorted first-order logic}, whose satisfiability problem is known to be undecidable, \cite{HMT}.  Moreover, CRS quantifiers are definable by LFD dependence modalities (cf. Section 3.1), while CRS quantifiers just are the first-order quantifiers on standard models.

It follows that  satisfiability of first-order formulas in the many-sorted three-variable fragment reduces to satisfiability of LFD formulas in the preceding Cartesian models. In particular, one just  replaces first-order quantifiers $\exists u$ by their obvious LFD-counterparts $\edual_{\{x, y, z\} - \{u\}}$.
\end{proof}

While the above examples concern semantic restrictions in the spirit of modal logic, the dependence setting also suggests new questions of axiomatization. Recall the three representation results for abstract dependence relations listed in Proposition \ref{Armstrong completeness}. The pivotal second result there concerned \emph{uniform dependence models} where all local dependence relations between variables are the same, and hence also equal the global dependence relation. Uniform dependence models validate the following principles, where $\dual_\emptyset$ is the universal modality:

\vspace{2mm}
\indent $D_X y \rightarrow  \dual_\emptyset D_X y, \quad \neg D_X y \rightarrow  \dual_\emptyset \neg D_X y$, \quad for arbitrary variables $X, y$

\vspace{2mm}

\noindent It is easy to find counter-examples to these implications in arbitrary LFD models.

\medskip

\noindent {\bf Open problem} \quad \emph{Axiomatize LFD over uniform dependence models.}\footnote{The disjoint unions of uniform dependence models in the proof for Proposition \ref{Armstrong completeness}, Clause 3,  do not validate the above implications. Still, since the components used disjoint sets of values, except for the common constants, these models validate modified uniformity principles. Again a question of axiomatization arises.}

\medskip

%Perhaps more interesting special axioms arise in the study of concrete notions of dependence. One example, found in Section 8, is the Steinitz Exchange Principle for linear dependence of vectors, which constrains dependence models in new ways.

This concludes the analysis of properties of the system LFD. The remaining part of this article explores what lies beyond the base system LFD: extensions of the language, enrichments of the framework, and concrete dependence notions in a number of areas.

\section{\large{\textbf{Richer dependence languages}}}

The modal language of LFD can be extended to describe other natural features of dependence. This section contains a few examples, all  with first-order truth conditions, thus making it possible to extend the translation of Section 3.3 making all logics effectively axiomatizable. Some of these extensions are straightforward, and do not affect the decidability of the logic, others do.

\vspace{-2mm}

\subsection{\textbf{Function symbols and constants}}

Recall the functional perspective of Section 2.3. It makes sense to add to LFD function terms, built from variables $x$ using a given family of operation symbols $f$ with arities marked. 0-ary function symbols are individual constants $c$ denoting objects. Terms are constructed by the rule

\medskip
\indent $t ::= x \mid f {\bf t}$, \quad \quad with {\bf t} a tuple of terms of the arity of $f$.
\medskip

\noindent In the syntax of formulas, the earlier sets of variables $X$ now become sets $T$ of terms,  and one can correspondingly extend the LFD syntax with operators $D_T t$ and $\dual_T\varphi$ for such sets of terms $T$ and single terms $t$. This allows for new sorts of dependence statements, such as

\medskip

$D_{fxy} {gyz}$ \quad the value of $gyz$ depends on that of $fxy$
\smallskip

\indent $\dual_{fx} \varphi$ \quad\quad\, the current value of $fx$ fixes the truth of $\varphi$.

\vspace{2mm}

 Models $\bM$ for this extended language come with an interpretation map $I$ for operation symbols, where the semantic clauses for term values read

 \medskip
(a) $val_{s}(x) = s(x)$
\smallskip

\indent (b) $val_{s} (f{\bf t}) = I(f)(val_{s}({\bf t}))$.

\medskip

\noindent In this setting, it is straightforward to define agreement $s =_T s'$ on the values of all terms in a set $T$, and use it to give the corresponding semantic clauses for $D_T y$ and $\dual_T\varphi$.

This logic is still decidable, but to show this the following notion is needed.

\begin{defi} A dependence model is \emph{distinguished} if distinct variables can only take distinct values. For every two distinct variables $x\not=y$ and every assignment $s\in A$:  $s(x)\not= s(y)$.
\end{defi}

\begin{fact}\label{Distinguish}
Every dependence model $\bM=(O,I,A)$ induces a \emph{distinguished} model $\bM^d$ of the form $(O^d, I^d,A^d)$ with:
$O^d=V\times O$; $I^d(P)((x_1,o_1), \ldots (x_n, o_n))$ iff $I(P)(o_1, \ldots, o_n)$ holds; and $A^d=\{s^d: s\in A\}$, where each assignment $s\in A$ has an associated assignment $s^d:V\to O^d$, given by
$s^d(x)=(x,s(x))$. Moreover, the two models are LFD-equivalent: for all assignments $s\in A$ and formulas $\varphi$ of \emph{LFD}:

\vspace{-2mm}

$${\bM},s\models \varphi \,\, \mbox{ iff }\,\, {\bM^d},s^d\models \varphi.$$\end{fact}

\smallskip

\begin{fact}\label{Function Translation}
The logic \emph{LFD} extended with function terms is decidable.
\end{fact}

\begin{proof}\quad \,\,
One can  translate formulas $\varphi$ in the extended language to formulas $\tau(\varphi)$ in the original LFD language so that $\varphi$ is satisfiable iff $\tau(\varphi)$ is satisfiable. First, associate to each complex term $t$ occurring in $\varphi$ some distinct new variable $v_t$, while keeping the old variables the same. Let $V'$ be the total extended set of variables, and let $\tau_0(\varphi)$ be the LFD formula obtained by replacing all terms $t$ in $\varphi$ by the matching variables $v_t$. The required functional dependencies between the variables are expressed as global dependence formulas, e.g., $\rotatebox[origin=c]{180}{$\forall$} D_{\{v_{t}, v_{t'}, v_{t''}\}} v_{f t t' t''}$. Let $\varphi_0$ be the conjunction of all these global dependence formulas, for all terms in $\varphi$. Then the translation $\tau(\varphi)$ is simply given by the conjunction $\varphi_0\wedge \tau_0(\varphi)$.\footnote{For example, the translation of the formula $Px f(x,g(y))$ is $\rotatebox[origin=c]{180}{$\forall$} D_y w \wedge \rotatebox[origin=c]{180}{$\forall$} D_{x,w}v \wedge Pxv$, where $w$ and $v$ are the fresh variables associated to terms $g(y)$, $f(x,g(y))$, respectively.}

To check that our translation preserves satisfiability, first assume that a formula $\varphi$ in the extended language holds for some assignment $s_0$ in a dependence model $\bM=(O,I,A)$. Now construct a model $\bM'$ for the extended set of variables $V'$, with the same  objects $O'=O$ and  interpretation $I'=I$. For this, we take $A'=\{s': s\in A\}$ as our new set of assignments, where we associated to each old assignment $s\in A$ a new \emph{extended assignment} $s': V'\to O$, defined by recursively putting: $s'(x)=s(x)$ for $x\in V$, and $s'(v_{f{\bf t}})=I(f)(s'({\bf t}))$. It is easy to see that $s'(v_{\bf t})=val_s({\bf t})$ holds for all tuples ${\bf t}$ of terms in $\varphi$, and moreover that
$\tau(\varphi)$ is satisfied by the assignment $s_0$ in the model $\bM'$.

For the converse, let the LFD formula $\tau(\varphi)$ hold for some assignment $s_0$ in a dependence model $\bM=(O,I,A)$. By Fact \ref{Distinguish},  $\bM$ may be taken to be distinguished. Now construct a model $\bM'$ for the language extended with function terms,  by enriching  $\bM$ with an interpretation $I(f)$ for each function symbol,  putting $I(f)(s(v_{\bf t}))=s(v_{f{\mb t}})$. Here, if any of the objects $o_1, \ldots, o_n\in O$ is not the value of any term for an assignment in $A$, just put $I(f)(o_1, \ldots, o_n)=o_0$ for some arbitrarily chosen object $o_0\in O$. These functions are well-defined because $\bM$ is distinguished, so there is no clash. It is easy to check that $\varphi$ is satisfied in $\bM'$ by the same assignment $s_0$.
\end{proof}

\begin{fact}\label{Function Translation2}
\emph{LFD} with function terms is axiomatized by the system {\bf LFD} plus
\begin{itemize}

\vspace{-1mm}
\item\quad The Functionality Axiom \, $D_{\bf x} f{\bf x}$ \, for all function symbols $f$.

\vspace{-1mm}
\item\quad The Substitution Rule \, ``from $\varphi$, infer $[{\bf t}/{\bf x}]\varphi$".
\end{itemize}
\end{fact}

\begin{proof}\quad \,\,  The proof is similar to the previous one, except that we now need a theorem-preserving translation $\tau'(\varphi)$ between the two systems. For any given formula $\varphi$ in the extended language, we associate new variables $v_t$ as in the proof of Fact \ref{Function Translation} to each of its terms $t$, and we construct the formulas $\tau_0(\varphi)$ and $\varphi_0$ as in that proof. Then our translation $\tau'(\varphi)$ is simply given by the implication $\varphi_0\to \tau_0(\varphi)$. It is now easy to check that $\varphi$ is a theorem in the above extended proof system iff $\tau'(\varphi)$ is a theorem in the basic system {\bf LFD}. The Substitution Rule, as well as the theorem $\rotatebox[origin=c]{180}{$\forall$} D_{\bf x} f{\bf x}$ (provable in the extended system by applying the Necessitation Rule to the Functionality Axiom) plays a key role in this verification.  \end{proof}

\noindent Note that the additional axiom and rule can be used to establish facts about complex terms. For instance, by the Functionality axiom we have $D_x g(x)$, and then by applying the Substitution rule we get $D_{f(x)} g(f(x))$. Combining this with $D_x f(x)$ (itself another instance of the Functionality axiom) and applying the Transitivity of dependence, we obtain that $D_x g(f(x))$. Applying the Necessitation Rule, we see that in fact this holds globally: $\rotatebox[origin=c]{180}{$\forall$} D_x g(f(x))$, i.e. we have $=(x; g(f(x)))$.

\smallskip

This extended logic can Skolemize implicit dependencies, in the spirit of Section \ref{Functions} on operational views of dependence, using function symbols as witnesses:

\begin{prop}
Let $\varphi(\mb{x}, \mb{y},\mb{z})$ be an LFD formula with free variables $\mb{x}, \mb{y}, \mb{z}.$ Let $X=\{x_1, \ldots, x_n\}$,  $Y=\{y_1, \ldots, y_m\}$, and  $f_1, \ldots, f_m$  fresh $n$-ary relation symbols not in $\varphi$. Then

\vspace{-2mm}

$$\vdash \varphi(\mb{x}, f_1\mb{x}, \ldots, f_m\mb{x}, \mb{z}) \,\, \mbox{ iff } \,\, \vdash \rotatebox[origin=c]{180}{$\forall$} D_X Y \to \varphi(\mb{x},\mb{y}, \mb{z}).$$
\end{prop}

\begin{proof}\quad \,\,  Apply the same construction as in the proof of Fact \ref{Function Translation2} to the formula on the left, associating fresh variables $y'_1, \ldots, y'_m$ to each of the terms $f_1\mb{x}, \ldots, f_m\mb{x}$. The same argument as in the preceding proof shows that: $\vdash \varphi(\mb{x}, f_1\mb{x}, \ldots, f_m\mb{x}, \mb{z})$ (a) iff $\vdash \rotatebox[origin=c]{180}{$\forall$} D_X Y'\to \varphi(\mb{x},\mb{y'}, \mb{z})$ (b) (where $Y'=\{y'_1, \ldots, y'_m\}$). To show that (b)  implies $\vdash \rotatebox[origin=c]{180}{$\forall$} D_X Y \to \varphi(\mb{x},\mb{y}, \mb{z})$, take a proof of $\vdash \rotatebox[origin=c]{180}{$\forall$} D_X Y' \to \varphi(\mb{x},\mb{y'}, \mb{z})$ and replace any occurrence of variables $y'_i$ by the corresponding variables $y_i$, obtaining a proof of $\vdash \rotatebox[origin=c]{180}{$\forall$} D_X Y \to \varphi(\mb{x},\mb{y}, \mb{z})$. The converse is proven by the inverse substitution (replacing every occurrence $y_i$ in the proof by the corresponding $y'_i$).\footnote{This syntactic variable replacement property for proofs in {\bf LFD} matches the semantic Renaming Lemma \ref{Renaming}.} \end{proof}

%\smallskip

However, this functional language still cannot talk about identity of term values, making it impossible to witness implicit dependencies by means of explicit statements  $\rotatebox[origin=c]{180}{$\forall$} (y = f(x))$.

\vspace{-2mm}

\subsection{\textbf{Explicit equality}}

We can easily extend our set of predicate symbols with an identity relation $=$ on objects, with the obvious semantics. It is convenient to work with a countably infinite set $C$ of constants, and allow complex terms (built from variables and constants using function symbols) as in the previous section. We denote by $c, d$, etc. arbitrary constants, and by $t, t'$ arbitrary terms. A \emph{ground term} is one that does not contain any variables (i.e., it is constructed only from constants using function symbols).
As before, it is useful to extend our dependence quantifiers and dependence atoms to terms, writing e.g. $\dual_T \varphi$ and $D_T t$, where $t$ is any arbitrary term and $T$ is any finite set of terms. As before, we use $\mb{x}$ and $\mb{t}$ for finite tuples of variables and terms.
%We also write $\dual_{\mb{t}} \varphi$ for $\dual_T \varphi$ whenever $\mb{t}$ is an enumeration of the set $T$, and similarly write $D_{\mb{t}} t'$ for $D_T t'$.

\medskip

Let us call this new logic $LFD^=$.
Our translation to FOL can be easily extended to $LFD^=$, so the logic is compact and its set of validities is recursively enumerable. But the new syntax has several advantages, such as supporting a more perspicuous axiomatization of the logic.

\medskip

A Hilbert-style \emph{proof system} $\mathbf{LFD^=}$ is given in Table \ref{tb4}, where the letters $P$ shown range over all predicate symbols, including equality.
%Essentially, the Value-Existence Rule asserts that each term always has a current value;
%the Dependence-Atom Axiom reduces local dependence to a universal implication when the current values of the %variables are explicitly given; and the Dependence-Quantifier Axiom does the same for dependence quantifiers.

\begin{table}[h!]
\begin{center}
\begin{tabularx}{\textwidth}{>{\hsize=0.8\hsize}X>{\hsize=1.9\hsize}X>{\hsize=0.8\hsize}X}
\toprule
%\vspace{0.1mm}
\\
\textbf{(I)} &\textbf{Axioms and rules of classical propositional logic}
 \vspace{1.3mm} \ \\
 \textbf{(II)} &\textbf{Special rules}  \vspace{1mm} \ \\
(Variable Substitution) & From $\varphi$, infer $[\mb{t}/\mb{x}]\varphi$. \vspace{1mm} \ \\
(Value Existence Rule) & From $t=c\to \varphi$, infer $\varphi$, provided that $c$ does not occur in $\varphi$. \vspace{0.7mm} \ \\
 \textbf{(III)} &\textbf{Axioms and rules for the universal modality}  \vspace{1mm} \ \\
($\rotatebox[origin=c]{180}{$\forall$}$-Necessitation)& From $\varphi$, infer $\rotatebox[origin=c]{180}{$\forall$}\varphi$. \vspace{1mm} \ \\
($\rotatebox[origin=c]{180}{$\forall$}$-Distribution) & $\rotatebox[origin=c]{180}{$\forall$} (\varphi \to \psi) \to (\rotatebox[origin=c]{180}{$\forall$}\varphi \to \rotatebox[origin=c]{180}{$\forall$}\psi)$ \vspace{1mm} \ \\
($\rotatebox[origin=c]{180}{$\forall$}$-Introduction$_1$) & $P\mb{t} \to \rotatebox[origin=c]{180}{$\forall$} P\mb{t}$, provided that $\mb{t}$ consists only of ground terms. \vspace{1mm}\ \\
($\rotatebox[origin=c]{180}{$\forall$}$-Introduction$_2$) &   $\rotatebox[origin=c]{180}{$\forall$}\varphi\to \rotatebox[origin=c]{180}{$\forall$}  \rotatebox[origin=c]{180}{$\forall$} \varphi$ \vspace{1mm} \ \\
($\rotatebox[origin=c]{180}{$\forall$}$-Introduction$_3$) &   $\neg \rotatebox[origin=c]{180}{$\forall$}\varphi\to \rotatebox[origin=c]{180}{$\forall$}  \neg \rotatebox[origin=c]{180}{$\forall$} \varphi$ \vspace{1mm}\ \\
($\rotatebox[origin=c]{180}{$\forall$}$-Elimination) & $\rotatebox[origin=c]{180}{$\forall$}\varphi \to \varphi$
\vspace{1.3mm} \ \\
\textbf{(IV)} & \textbf{Equality axioms}  \vspace{1mm}  \ \\
(Reflexivity)  & $x=x$ \vspace{1mm}\ \\
(Symmetry) & $x=y\to y=x$ \vspace{1mm}\ \\
(Transitivity) & $(x=y\wedge y=z)\to x=z$ \vspace{1mm}\ \\
(Functional Substitution) & $\mb{x}=\mb{y} \to f\mb{x}=f\mb{y}$ \vspace{1mm}\ \\
(Substitution of Equals)  & $\mb{x}=\mb{y} \to (P\mb{x} \leftrightarrow P\mb{y})$ \vspace{1.3mm} \ \\
\textbf{(V)} & \textbf{Axioms for dependence atoms and modalities} \vspace{1mm}  \ \\
(Dependence Atom) & $(\mb{x}=\mb{c}\wedge y=d) \, \to \, \left(D_X y\leftrightarrow \rotatebox[origin=c]{180}{$\forall$} (\mb{x}=\mb{c}\to y=d)\right)$  \vspace{1mm}\ \\
(Dependence Modality) & $\mb{x}=\mb{c} \, \to \, \left(\dual_X\varphi
\leftrightarrow \rotatebox[origin=c]{180}{$\forall$} (\mb{x}=\mb{c}\to \varphi)\right)$,
\hspace{2mm}
\vspace{1mm}\ \\
& where in both cases, $X$ is the set of variables occurring in $\mb{x}$.
\hspace{0.1mm}
\vspace{-0.5mm}
\\
\bottomrule
\end{tabularx}
\end{center}
\vspace{-0.5cm}
\caption{The proof system $\mathbf{LFD^=}$.}\label{tb4}
\end{table}%

%\vspace{-0.2cm}

Substitution of Equals is a special case of Leibniz' Law of `indiscernability of identicals', allowing  substitution of equal variables in atomic formulas.\footnote{The general Leibniz Law, that allows substitution of equal variables in \emph{arbitrary} formulas, is not valid, due to the modal character of our semantics: the equality may hold locally at a given world/assignment, while the truth of the relevant formula may depend on the values of the variable at other worlds/assignments.}
Essentially, the Value-Existence Rule asserts that each term always has a current value.
The Dependence-Atom Axiom `reduces' local dependence to a universal implication when the current values of the variables are explicitly given, and the Dependence-Modality Axiom does the same for dependence modalities. As a consequence of these context-dependent `reductions', all the $\mathbf{LFD}$ principles characterizing the dependence modalities and dependence atoms (i.e. the axioms and rules II, III, IV in Table 1 for $\dual$ and $D$)
are missing from from the definition of the new system: indeed, they now become provable theorems in $\mathbf{LFD^=}$.

\begin{thm}\label{CompletenessEquality} (\emph{Completeness.})
The calculus $\mathbf{LFD^=}$ is sound and complete for validity in the dependence language with equality w.r.t. dependence models.
\end{thm}

The completeness proof follows exactly the lines of that for the logic LED in \cite{Baltag2016}. The proof uses a Henkin-style canonical model, with the additional twist that the maximally consistent theories are also required to be `witnessed': for every term $t$ there exists some constant $c$ such that $t=c$ is in the theory. Beyond completeness, however,
 there is  a  complication. The type model proof used in Section 4 to show the decidability of LFD does not seem to work in the presence of the Value-Existence Rule.
%\footnote{A claim to the contrary was made in
%\cite{Baltag2016}, but subsequently the author discovered a gap in the proof.}

\medskip

\noindent {\bf Open problem} \, \emph{Is LFD with equality decidable?}\footnote{In response to a preprint version of this paper, \cite{Aachen1} has announced a negative answer, proved by reducing satisfiability for the undecidable Kahr-Class of first-order formulas to satisfiability for $LFD^=$-formulas.}

%\subsection{From functional to relational dependence} Another extension relaxes the meaning of dependence atoms $D_X y$ to capture \emph{non-functional} dependence. In Section 3, it was shown how LFD can express some non-functional dependencies by using formulas involving the dual quantifier $\dual_X$, sometimes in combination with the functional dependence atom $D_X y$. However, on the analogy noted in Section 2 between dependence $D_X y$ and single-conclusionve sequents $\Phi \vdash \psi$, the language may be extended with a dependence-analogue of multiple-conclusion sequents $\Phi\vdash\Psi$: $X\multimap Y$ is true at $s$ iff fixing the values of all variables in $X$ to the their current values under $s$ fixes the values of at least one of the variables in $Y$ to its current value under $s$. Again the techniques used for LFD do not apply straightforwardly here. \medskip \noindent {\bf Open problem} \, Is the logic of relational dependence decidable? And regardless of decidability, what would be its complete axiomatization?

 \subsection{\textbf{Independence}}

 A major natural  extension for LFD concerns the notion of \emph{independence}. Intuitively, saying that $y$ is independent from $x$ is by no means the same as the statement $\neg D_xy$ which just expresses that $x$ does not fix the value of $u$ at the current assignment, and even the universal quantification $\rotatebox[origin=c]{180}{$\forall$}\neg D_xy$ is too weak. What independence of $y$ from $x$ should mean in the present setting is that \emph{the current value of $x$ does not constrain the values that $y$ can take}.\footnote{Varieties of independence have been studied extensively in IF-logic, \cite{HintSand}, and in Dependence Logic, \cite{GraVa}. Our central concern here is how dependence and independence behave on our simple modal basis.}

 In epistemic terms, this amounts to saying that knowing the current values of $x$ tells us nothing about the current value of $y$. To make this precise, we need the following notion.

\begin{defi} In any dependence model $\bM=(M,A)$, the \emph{information carried by $X$ about $Y$ at an assignment $s\in A$} is given by the range of $Y$-values that are compatible with the current values of $X$ (at $s$):
$$Inf_s(X, Y) \,\,\, :=\,\,\, \{ t\upharpoonright Y : t\in A, t =_X s\},$$
where   $t\upharpoonright Y$ is the restriction to $Y$ of the assignment $t:V\to O$.
\end{defi}

We  now introduce \emph{independence atoms} $I_X Y$, saying that $Y$ is independent from $X$ at $s$.

 \begin{defi} For any model $\bM$, sets of variables $X, Y$, and assignment $s\in A$, we put:
$$\bM, s \models I_X Y \,\, \, \mbox{ iff } \,\, \, Inf_s(X,Y)=Inf_s(\emptyset,Y).$$
One can also define more general  \emph{conditional independence} atoms $I_X Y|Z$, saying that given $Z$, $X$ gives no further information about $Y$:
$$\bM, s \models I_X Y|Z \,\, \, \mbox{ iff } \,\, \, Inf_s(X\cup Z,Y)=Inf_s(Z,Y).$$
\emph{Global independence} (both conditional and unconditional)\footnote{Global conditional independence as defined here may be viewed as a qualitative counterpart to the notion of conditional independence found in Probability Theory.} can then be defined from the local versions in the obvious manner, using the universal modality available in LFD:
$$X\rotatebox[origin=c]{90}{$\models$} Y \,\,\, :=\,\,\,  \rotatebox[origin=c]{180}{$\forall$} I_X Y.$$
$$(X\rotatebox[origin=c]{90}{$\models$} Y)|Z \,\,\, :=\,\,\,  \rotatebox[origin=c]{180}{$\forall$} (I_X Y|Z).$$
As usual, when either $X$ or $Y$ are singletons,  notation simplifies to $I_X y$, $I_x y$, etc.
\end{defi}

 %\footnote{Since independence has this natural interpretation, our terminology `dependence models' may be unfortunate, and a neutral term like `state spaces' might be preferable eventually.}

Reasoning with independence atoms has some interesting features. For instance, it is easy to see that $I_XY$ does not imply $I_YX$ locally (the current value of $X$ might carry no information about $Y$, but that of $Y$ might carry information about $X$). In contrast with this, however, global independence $X \rotatebox[origin=c]{90}{$\models$} Y$ is \emph{commutative}, and the conditional version $(X \rotatebox[origin=c]{90}{$\models$} Y)|Z$ is commutative in $X,Y$. If in a model, every value of $y$ can be taken at every value of $x$, then the set of joint values $(x,y)$ must be a full Cartesian product of the ranges of $x$ and $y$.

There are also interesting valid principles connecting the modalities $I$ and $D$. As an illustration, $(D_xz \wedge I_xy) \to I_zy$ is valid. If the current value of $x$ gives full information about $z$ but no information about $y$, then the current value of $z$ does not yield any information about $y$ either.

\begin{fact}
The \emph{dependence atoms can be defined in terms of conditional independence}, via the equivalence $D_X Y \leftrightarrow I_Y Y|X$,
and the same goes for the global versions.\end{fact}

These principles are part of a new logic LFDI extending the purely structural rules for independence in \cite{PearleInd}. It consists of LFD extended with the basic independence modalities $I_X y$.

 \medskip

 \noindent {\bf Open problem} \quad \emph{Axiomatize the logic LFDI.}

 \medskip

 Interestingly, the core logic of independence differs essentially from that of dependence: LFDI is more complex than LFD. The reason is explained in the proof to follow.

\begin{thm} The modal logic \emph{LFDI} is undecidable.
\end{thm}
\begin{proof}\quad \,\,  The proof is reminiscent of that for Fact \ref{stepwise}, and uses the undecidability of the three-variable fragment of many-sorted first-order logic. Formulas $\varphi$ of the  latter language with variables $x, y, z$ can be translated into formulas $\tau(\varphi)$ of LFD as indicated at several places earlier on, replacing  first-order quantifiers $\exists u \psi$ by existential LFD modalities $\edual_{\{x, y, z\} - \{u\}}\tau(\psi)$. Now, crucially, the independence modalities
 can be used as follows to force the values of $x, y, z$ to be a full Cartesian product, via the formula:

 \vspace{1ex}

 $\rotatebox[origin=c]{180}{$\forall$} (I_xy \land I_{\{x, y\}}z)$, \quad where $\rotatebox[origin=c]{180}{$\forall$}$ is the universal modality, available in LFD.

 \vspace{1ex}

Let us show that every model of this formula is a full Cartesian product. Suppose that $x$ takes value $d$ at some assignment in $\bM$, $y$ can take $e$, and $z$ can take $f$. Consider an assignment $s$ with $s(x) = d$. Since $I_xy$ holds everywhere, the value $e$ for $y$ occurs with this value for $x$, at some assignment $t$ with $t(x) = d, t(y) = e$. But since also $I_{\{x, y\}}z$ holds everywhere, there must also be an assignment $u$ with $u(x) = d, u(y) = e, u(z) = f$.

% \smallskip

 Now it is immediate that any three-variable first-order formula $\varphi$ is satisfiable iff the matching LFDI formula $\tau(\varphi) \land \rotatebox[origin=c]{180}{$\forall$} (I_xy \land I_{\{x, y\}}z)$ is satisfiable. \end{proof}

 \begin{rem}\emph{Moving beyond the  contrast between  independence and  functional dependence, the more general perspective for this section  is the notion of \emph{correlated behavior}. Local independence $I_xy$ in our sense  says that the current value of $x$ does not place any constraint on the values that $y$ can take at the present location. This is at the opposite extreme from  functional dependence $D_xy$ which restricts the range of $y$ to just one value. Clearly, there are other natural notions here. For instance, the mere negation of independence, $\neg I_xy$, says that the current value of $x$ excludes at least one value for $y$, which can be seen as a weak form of \emph{correlation}. This gives a very minimalistic notion of `dependence', much weaker than our functional dependence, but one that is of interest on its own. The examples of failure of classical laws for CRS quantifiers presented in the Introduction do not necessitate functional dependence, but only weak correlations: any breach of full independence between any two variables can lead to such a failure of a classical validity. The complete logic of weak correlation and independence would be the \emph{pure modal logic of $I_X y$}. Stronger notions of correlation or partial dependence between $x$ and $y$ arise when we put further constraints on how far the local value of $x$ constrains that of $y$, with full functional dependence in the limit.}
 \end{rem}

 \vspace{2ex}

 \noindent {\bf Open problem} \quad \emph{Axiomatize the pure modal logic of independence. Is it  decidable?.}

\begin{rem}\emph{
The preceding analysis also suggests   more general  \emph{comparative informational assertions} $X\geq_Y Z$ (`$X$ carries at least as much information about $Y$ as $Z$ does'):
$$s\models X\geq_Y Z \,\,\, \mbox{ iff } \,\,\, Inf_s(X,Y)\subseteq Inf_s(Z,Y).$$
All the above notions of dependence and independence are definable in terms of these comparative assertions. E.g., the conditional independence statement $I_XY|Z$ says that $Z$ is at least (in fact, just) as informative about $Y$ as  $X\cup Z$ is, which can be expressed formally as the equivalence $I_X Y|Z \leftrightarrow Z\geq_Y (X\cup Z)$.
}\end{rem} \smallskip

 \noindent {\bf Open problem} \quad \emph{Axiomatize the logic of comparative informational assertions.}

\subsection{\textbf{Dynamics and model change}}\label{DepDynamics}

Typically, epistemic events carrying new information can change a current model. One may learn the current value of some variable, or more general facts. There can also be non-informational reasons for changing a current model, say, with a shift of a current dynamical system. A few instances of the dynamics of  dependence models will be discussed here, using methods from dynamic-epistemic logic, \cite{BMS}, \cite{DHK}, %\cite{DHK},
\cite{LDII}.
%Even so, the topic of dependence dynamics is vast, and a serious treatment would require a separate paper of its own.

\medskip

\noindent {\bf Learning current values.} One can update a knowledge base after learning the true values of a set of variables $X$ at (the current assignment) $s\in A$. This changes the model $\bM=(M,A)$ to the submodel $\bM\hspace{-0.1cm}\mid\hspace{-0.1cm}X= M\hspace{-0.1cm}\mid\hspace{-0.1cm}\{t\in A: s=_X t\}$ that retains only the assignments that agree with $s$ on all $X$-values. Now interpret the dynamic modality $[X]\varphi$ as follows:
$$\bM, s \models [X]\varphi \, \, \mbox{ iff } \,\, \bM\hspace{-0.1cm}\mid\hspace{-0.1cm}X, s \models \varphi$$

\noindent This  modality occurs in epistemic logic under the name of ``public inspection of a value",  \cite{vEGWang}.
%In the present setting, the following can be observed.

\begin{fact}
The logic \emph{LFD} with the modalities $[X]\varphi$ is completely axiomatizable and decidable.
\end{fact}

\begin{proof}\quad \,\,  It suffices to observe that the following recursion axioms are valid:

\medskip

(a) $[X]P{\bf y} \leftrightarrow P{\bf y}$, (b) $[X]\neg\psi \leftrightarrow \neg[X]\psi$, (c) $[X](\alpha \land \beta) \leftrightarrow ([X]\alpha \land [X]\beta)$,

\medskip

 (d) $[X]\dual_Y\psi \leftrightarrow \dual_{X \cup Y}[X]\psi$, \, (e) $[X]D_Yz \leftrightarrow D_{X \cup Y}z$
\medskip
\smallskip

\noindent Used iteratively in a standard dynamic-epistemic style, these reduce each formula in the extended dynamic language to an equivalent base formula of LFD.
\end{proof}

%Simple as it is, this  style of analysis by recursion axioms for dynamic modalities is significant, since failure to achieve it usually signals higher complexity of a logical system.

\noindent {\bf Learning new facts.}  Another form of information update happens when  learning a new \emph{true fact} $\varphi$ about the current assignment $s$. Semantically, an update with a formula $\varphi$ transforms the model $\bM=(M,A)$ into the relativized submodel $\bM\hspace{-0.1cm}\mid\hspace{-0.1cm}\varphi = M\hspace{-0.1cm}\mid\hspace{-0.1cm}\{s\in A: s\models\varphi\}$ retaining only the assignments satisfying $\varphi$. In the syntax, this is reflected by dynamic modalities $[\varphi]\psi$, with a semantic truth condition given by:
$$\bM, s \models [\varphi]\psi \,\, \mbox{ iff } \,\, \bM, s \models \varphi \mbox{ implies } \bM\hspace{-0.1cm}\mid\hspace{-0.1cm}\varphi, s \models \psi$$
\begin{rem}  \emph{The logic of this type of update modality (`public announcement logic') is a well-known pilot system of information update. But updating a dependence model can mean different things. Going to a submodel with fewer assignments typically adds to the existing dependencies. In  epistemic scenarios, this increase is fine, and in fact useful.\footnote{Suppose the truth values of $p$ and $q$ are unknown, whence observing one of them says nothing about the other. But after learning a new fact $p \leftrightarrow q$ imposing a dependency, observing the truth value of one of $p, q$ automatically gives the other. Thus, new dependencies speed up information flow.} However, if  the dependence model is a  state space for some known current process, changes in the dependence structure create a new process, and this needs to be motivated by other considerations.\footnote{A typical model change different from information update is changing the space of relevant variables. This happens, e.g., when analyzing two correlated variables by introducing a new variable on which both depend.}}
\end{rem}

A dynamic-epistemic analysis still works for the new extended setting, but there is no longer any reduction to the base language of LFD. Admittedly, the dependence modalities after an update can be reduced to the original ones in a similar way to the well-known recursion law for epistemic modalities:
\vspace{-1mm}
$$[\varphi] \dual_X \psi \, \leftrightarrow \, (\varphi \rightarrow \dual_X (\varphi \rightarrow [\varphi]\psi))$$
But the new dependencies in the updated model $\bM\hspace{-0.1cm}\mid\hspace{-0.1cm}\varphi$ can only be `pre-encoded' in the original model $\bM$ by means of a \emph{conditional dependence operator} $D_X^{\varphi} y$, with a semantics given by:
$$s\models D^{\varphi}_X y\,\,\, \mbox{ iff } \,\,\, \forall t\in A \left(\, t\models\varphi \mbox{ and } s=_X t \mbox{ imply } s=_y t \,\right).$$
This is illustrated  in the following recursion equivalence, whose validity is easy to check:
\vspace{-1mm}
$$[\varphi]D_Xy \, \leftrightarrow \, (\varphi \rightarrow D_X^{\varphi}y)$$
Of course, conditional dependence needs a recursion law in its turn, and the following  is valid:
\vspace{-1mm}
$$[\varphi]D_X^{\alpha}y \, \leftrightarrow \, (\varphi \rightarrow D_X^{\varphi \land [\varphi]\alpha}y)$$
The logic with this update modality can be reduced to its static base logic (with conditional dependence operators) via such recursion laws.

But in this case, the static base logic itself is no longer a routine extension of LFD. The difficulty lies in the following result.

\begin{fact}
The conditional dependence atom is not definable in \emph{LFD}.
\end{fact}

\begin{proof}\quad \,\,  For simplicity, consider a language with two variables $x, y$ and one atom $Pxy$. Take a dependence model $\bM$ with just two admissible assignments $s, t$ where $s(x) = s(y) = 0, \, t(x) = 0, t(y) = 1$, while the binary predicate $P$ holds only of (0, 0) in the underlying first-order model. As for non-trivial dependence atoms in this language, at both assignments, the formulas $D_yx, \neg D_xy$ are true. Now extend $\bM$ to a model  $\bM'$  with a third assignment $u$ such that $u( x) = 0, \, u(y) = 2$, while $P$ now also holds of (0, 2). In $\bM'$, $D_yx, \neg D_xy$ are true at all of $s, t, u$.

Now, it is easy to prove by induction  that the map $F$ sending the two assignments $s$ and $u$ in $\bM'$ to $s$ in $\bM$, and the assignment $t$ in $\bM'$ to $t$ in $\bM$ has the following property. For any LFD formula $\varphi$ and any assignment $v$, $\bM', v \models \varphi$  iff $\bM, F(v) \models \varphi$.\footnote{The  general fact here is that $F$ is a modal `$p$-morphism', cf. \cite{BRV}, in a sense appropriate to  dependence models.}

But, conditional dependence sees a difference here: the formula $D_x^{Pxy}y$ is true at $s$ in the model $\bM$ (the restriction leaves only the assignment $s$), but not in $\bM'$, since both $s, u$ remain after the restriction. So $D_x^{Pxy}y$ cannot be definable in terms of LFD formulas.
\end{proof}
\noindent{\bf Open problem} \quad \emph{Axiomatize the modal logic of conditional dependence. Is it decidable?}

%\begin{rem} \emph{A special case of public announcement arises when some variable $x$ is set to one of its %possible values, as  happens with `interventions' in causal models, \cite{Pearl}.
%\footnote{Interestingly, this is often described as releasing $x$ from its causal dependence on other %variables, whereas the opposite is true in LFD: a constant function is dependent on everything.}
%Describing this move as a modality $[x := a]\varphi$ leads to the special case of public announcement of an %atomic formula $x = a$, assuming that the language has constants $a$ for the desired value, as well as %explicit identity symbols. This case might seem easier than the preceding, because of the following validity:} %\medskip

%$[x := a]D_Yz$ \, \emph{is equivalent to} \, $\rotatebox[origin=c]{180}{$\forall$}(x = a \rightarrow D_{Y \cup %\{x\}} z)$

%\medskip \noindent \emph{Still, this reduction requires the presence of explicit identity in the language, cf. %Section 6.2.} \end{rem}

\noindent \begin{rem}[Enlarging models] \, \emph{Natural updates can just as well extend current dependence models with new assignments, thereby possibly giving up dependencies that used to hold.} \end{rem}

%An example occurs with the standard FOL semantics of a first-order quantifier $\exists x \varphi$. One lets %the value of $x$ range through the model to find a witness making $\varphi$ true, while keeping the values for %the free variables fixed. In the CRS-style semantics of Section 1, this running is done only over the %available assignments $s$ in the model. But one can also think of the process as surveying arbitrary %assignments of the form $s[x := d]$, \emph{freeing} $x$ from any significant dependencies it might have had in %the original model. Such model-enlarging updates need to be understood better.

%\begin{rem}
%\emph{The issues discussed in Section 4.3 are also relevant here. %When dependence and epistemic information %come apart, richer two-dimensional models may be needed as `update universes', of a sort we will not pursue %here.}
%\end{rem}

\medskip

\noindent {\bf Broader dynamic perspectives.} \, An update perspective suggests extending the semantics of LFD from considering just single dependence models to \emph{families} of these.

 %The discussion in Section 6.5 does not exhaust all plausible intuitions linking dependence and information. Consider the initial database in Example 1.1. A database is often seen as an information state recording all currently known atomic facts, and in richer views of databases, also the known general rules, say, in the form of `integrity constraints'. In that setting, adding a new database entry represents update with new information, while removing an entry represents culling perhaps unreliable earlier information. But then, there is a mismatch with the earlier epistemic view of dependence models. Adding facts extends the model with a new assignment, possibly disrupting earlier dependencies, and in general, losing information. Removing facts restricts the model, adding information. Things go  the wrong way around.
 %A solution requires disentangling two views of information: the syntactic view of a database as a set of atomic statements, and the semantic one of a range of complete options, as found in epistemic logic. In the latter perspective, one wants to learn the exact extent of some relation $\mathfrak{R}$ encoded in the database. The current relation $R$ is a subset of this true relation, and so the universe of epistemic options might be taken to be the family of all relations in between $R$ and the universal relation. \footnote{Of course, the database content can be refined with negative facts, rules, and so on.} Now, adding a fact, i.e., some tuple in the true relation $\mathfrak{R}$,  amounts to eliminating all candidate relations lacking that fact.
%\medskip

 \begin{defi} A \emph{dependence universe} \begin{large}$\mathfrak{U}$\end{large} is a family of dependence models.

 \end{defi}

 \smallskip

Epistemically, each  model in \begin{large}$\mathfrak{U}$\end{large} can be seen as a candidate for the true structure of the world, and the  dependence universe then represents a `space of inquiry'. But  a dependence universe might also be a family of available processes one can switch between. Either way, just as a dependence model, a dependence universe need not be a full power set of some sort: many possible structures may be missing from the space of inquiry or process repertoire.\footnote{In an epistemic perspective, gaps encode information about how the process of inquiry may go. Related structures occur in dynamic-epistemic logic under the heading of `protocol models', \cite{LDII}. One can think of protocols with gaps as higher-order dependencies or correlations on how information can be acquired.}
 %For a more temporally flavored intuitionistic take on these, cf. van Benthem  2007.

\bigskip

A natural extension of LFD  describes triples $(\mathfrak{U},\bM, S)$  of a dependence universe  $\mathfrak{U}$, a dependence model $\bM\in \mathfrak{U}$, and one or more binary relations on dependence models $S\subseteq \mathfrak{U}\times \mathfrak{U}$, expressing relevant changes of the dependence models. It  contains the dependence modalities interpreted as before, but also  modalities accessing alternative models via the relations $S$. %The following two clauses will illustrate the format.

 \begin{exam} \emph{Truth conditions for a bimodal epistemic dependence language:}
 %for any assignment $s\in A$, we put

 \vspace{2mm}

 $\mathfrak{U}, \bM, S, s \models \dual_X\varphi \, \, \mbox{  \emph{iff}  } \,\,
\mathfrak{U}, \bM, S, s' \models \varphi \mbox{ holds
 for all $s'\in A$ with $s =_X s'$}$

 \medskip

$\mathfrak{U}, \bM, S, s \models \langle S\rangle\varphi \,\, \mbox{ \emph{iff } } \,\, \mathfrak{U}, \bM', S, s \models \varphi \mbox{ holds for some $\bM'\in \mathfrak{U}$ with $S \, \bM \,\bM'$ and $s\in A'$ }$
 \end{exam}

A natural relation $S$ for exploring dependence universes is that of \emph{submodel}. Call its downward-looking modality $[\downarrow]\varphi$. Now part of the semantics of LFD can be internalized in this richer logic.

\begin{exam} The following two principles are valid in dependence universes:

\medskip

(a) $D_Xy \rightarrow [\downarrow]D_Xy$, \quad (b) $\dual_X[\downarrow]\varphi \rightarrow [\downarrow]\dual_X\varphi$.

\end{exam}

\noindent{\bf Open problem}\, \, \emph{What is the complete logic of LFD plus the downward and upward submodel \\ \indent modalities on dependence universes?}

\medskip

This concludes our exploration of logical operators that extend the basic language of LFD.

 %This two-level framework on top of LFD raises intriguing further  questions. For instance, what about higher-level dependencies triggered by the gaps in the family of models \begin{large}$\mathfrak{U}$\end{large}?

%This concludes our discussion of extensions for the basic logic LFD. Many semantic open problems were raised -- and of course, there is the proof-theoretic issue of what logics  will emerge.

%\newpage

\section{\large{\textbf{Dependence in concrete settings}}}\label{Other}
The dependence semantics and logic of this paper are simple, and many notions of dependence in actual use add further features. This section presents a few cases, mainly to show that they fit with the basic LFD  perspective, while also highlighting their interesting more specialized structures that call for further logical investigation.
%All topics to be raised are mainly just pointers to further work to be done in bringing the framework of this logic paper to the realities of dependence.

\subsection{\textbf{Databases}}

\vspace{-2mm}

This paper started with a simple database example, which nevertheless does not do justice to the more sophisticated structures studied in database theory, \cite{AbitHV}. Much of this theory is in terms of first-order logic and its low-complexity fragments, and in this light, LFD can be seen as an attempt at capturing some high-level features of databases in a modal style. Indeed, various kinds of dependence and independence in databases can be represented in LFD-style languages, especially with the extensions introduced in Section 6.
%\footnote{Databases also posed a challenge:  viewed as syntactic information states to be updated toward the true relation,  Section 6.6 showed that their match with epistemic models is not as straightforward as Section 1 may have suggested.}

As a further point, databases consist of `facts' and `rules'. Rules are hard-wired regularities, telling us how to close the database under inferences. For semantic dependencies in a model, this suggests a natural distinction: some are `accidental',  others are `essential'. This distinction cannot be seen inside dependence models, it requires an additional external decision which regularities are important  and which ones are not.
%\footnote{A bit of the flavor can be found in Section 2.3 on operational views. Important dependencies can be defined explicitly some initial repertoire of operations on objects, while less important dependencies might be nameless.}
A semantic setting for getting at the distinction are the dependence universes of Section 6.6. Accidental dependencies $D_Xy$ just hold in the current model, while essential ones continue to hold even under  relevant updates of that model, which can be expressed using modalities such as $[\downarrow]\varphi$ and $[\uparrow]\varphi$.

%\vspace{-4mm}

\subsection{\textbf{Vector spaces}}\label{Vectors}

\vspace{-2mm}

The next example comes from linear algebra where dependence is the fundamental notion behind a wide range of applications to computation, defining geometrical dimension, and much more. A vector $y$ depends on a set of vectors ${\bf x}$ if $y$ can be written as a linear combination of the $z \in {\bf x}$. This notion is not primarily semantic, but it rather ties in with the equivalent functional definability perspective on dependence of Section \ref{Functions}. However, given the special operations used in linear algebra, there are interesting valid properties beyond those provided by LFD.

\begin{exam} \emph{The Steinitz Exchange Principle, \cite{Lambal}, reads as follows in the LFD language:}

\medskip

$D_{(X\cup\{y\})} z \rightarrow (D_{X} z  \lor D_{(X\cup\{z\})} y)$

\medskip

\noindent \emph{The reason for its validity in linear algebra is that, if $z$ is a linear combination of $X, y$, then either the coefficient for $y$ is 0, and the first disjunct holds, or that coefficient is not 0, and then one can divide by it, obtaining a formula expressing $y$ as a linear combination of $X, z$.}
\end{exam}

The Steinitz principle is not valid in LFD: a counter-example on dependence models occurs in Example \ref{Numerical}, where we have $D_{\{z, x\}}y, \neg D_zy$ and $\neg D_{\{z, y\}}x$. However, in the spirit of Section 5.3, one can ask for a modal \emph{correspondence} result: which constraint on assignments in dependence models ensures the validity of the Steinitz principle? What comes to mind is a principle about existence of inverses for implicit functions that truly depend on their arguments, but a precise solution remains to be found. In addition, there is a natural question of axiomatization.

\medskip{
\noindent{\bf Open problem} \emph{Axiomatize the complete theory of LFD-style assertions about dependence between vectors. Is it just the basic proof system {\bf LFD} plus Steinitz Exchange?}

%\vspace{-2mm}

%This difficulty with frame correspondence analysis may reflect a  mismatch.
\begin{rem} Matroid Theory \emph{ studies  abstract linear dependence and independence.
%\cite{Oxley}.
Matroids are finite families of sets of vectors satisfying  conditions implying the uniqueness of finite dimension. Matroids can be represented as dependence frames for LFD, \cite{Gonza}, but there is an issue of the best  logical framework. In the matroid setting, \emph{sets} of variables are the central notion, and LFD does not describe such sets in an abstract algebraic way, except by brute enumeration. It would be of interest to develop a modal perspective on Matroid Theory.}
\end{rem}

\vspace{-4mm}

\subsection{\textbf{Topologizing LFD: the logic of continuous dependence}}

\vspace{-2mm}

%The semantics of LFD may make unrealistic requirements in concrete settings.
In empirical contexts, the exact values of most variables are never accessible. Then, the existence of a functional dependence in the sense of LFD is a moot point, of only theoretical importance. What matters is  whether there is a \emph{knowable} dependence: given what can be known in principle, by measurements of any precision, about the value of $x$, can  the value of $y$ be found with any desired degree of precision?

Making sense of this intuition calls for a \emph{topological} setting, with its intuitions of approximation and continuity. This section outlines such a \emph{logic of continuous dependence} LCD, though a full presentation and development is postponed to our forthcoming paper \cite{BvB2020}.
%Indeed, in the present article, this section can be skipped without loss of continuity.
% \subsection{Continuity as knowable dependence}

\medskip

A variable \emph{$y$ depends continuously on $x$} at an assignment $s$ if the value of $y$ at $s$ is determined to any desired degree of approximation by some (possibly better) degree of approximation of the value of $x$ at $s$. Epistemically, this means that one can know the value of $y$ with any desired accuracy, if given a sufficiently accurate estimate of the value of $x$.\footnote{A stronger notion, perhaps closer to the stated epistemic intuition about approximation, would be continuous dependence at some open neighborhood of the current point. This can be defined in the language to follow.}

This suggests having a topology $\tau$ on the set of objects $O$, to capture approximations of values $s(x)$ as open neighborhoods $U\in \tau$ with $s(x)\in U$. Global dependence $D^{\bM}_x y$ in such a \emph{topo-dependence model} $\bM=(M,A,\tau)$ requires existence of a continuous map from $x$-values to $y$-values, while local dependence $D^s_x y$ is given by:
$$s\models D_x y  \,\,\, \mbox{ iff } \,\,\, \forall V\in \tau(s(y)) \, \exists U\in \tau(s(x))\, \forall t\in A \left( t(x)\in U \to t(y)\in V \right),$$
where $\tau(o)=\{U\in \tau: o\in U\}$ is the family of open neighborhoods of an object $o\in O$. This can be generalized to dependence $D_X y$ on a \emph{set} $X$ of variables, by using the \emph{product topology} on $O^{|X|}$. Intuitively, $D_X y$ holds at an assignment $s$ if all assignments that assign to $X$ values that are close enough to their current ones also assign to $y$ a close enough value to its current one.

The natural analogue semantic clause for simple dependence modalities is:
$$s\models \dual_x \varphi \,\,\, \mbox{ iff } \,\,\, \exists U\in \tau(s(x))\,  \forall t\in A \, \left(t(x)\in U\to t\models\varphi\right).$$
The definition can  be generalized to set-based dependence modalities $\dual_X \varphi$ using a product topology, but we skip details here. Intuitively, $\dual_x \varphi$ holds at an assignment $s$ if $\varphi$ holds at all admissible assignments that assign to $x$ values that are `close enough' to their current value $s(x)$.
This connects to the well-known topological-interior semantics for the modal logic $S4$: an assignment $s$ satisfies $\dual_x \varphi$ iff the current $x$-value $s(x)$ is in the \emph{interior} of the set $\{t(x):t\models \varphi\}$ of all $x$-values of $\varphi$-assignments.
Philosophically, the interior semantics points to an \emph{evidential conception of knowledge}: $\varphi$ is knowable from $X$ if there exist some potential pieces of evidence about $X$  that entails $\varphi$.

As for defined notions, $\rotatebox[origin=c]{180}{$\forall$}\varphi:= \dual_\emptyset\varphi$ is still the universal modality over all assignments in $A$. Global dependence, defined as before by $\rotatebox[origin=c]{180}{$\forall$}D_x y$, now expresses the existence of a \emph{continuous} map $f:O\to O$ with $s(y)=f(s(x))$ for all $s\in A$. Other defined operators acquire a different meaning. The formula $D_\emptyset y$ used to mean in LFD that $y$ is constant, taking only one value. But in a topological setting it expresses a more complex condition on the `specialization pre-order', which only reduces to constancy in the presence of the separation axiom $T_1$.

\medskip

More details, including decidability and a complete axiomatization, as well as further extensions to include uniform continuity and links with Domain Theory, will be presented in \cite{BvB2020}. For now, we note that the proof calculus for LCD involves modal logic S4 rather than S5 for its dependence modalities.  Moreover,
 %by the inference rules of Modus Ponens and Necessitation, as well as the following principles:
%\vspace{-1mm}
%\begin{itemize}
%\vspace{-1mm}
%\item The axioms of modal \emph{S4} for each separate topo-modality $\dual_X\varphi$
%%\item Monotonicity for sets: $D_X\varphi \rightarrow D_Y\varphi$, if $X \subseteq Y$.
%\vspace{-1mm}
%\item \emph{S5} for $\dual_\emptyset$: \,
%$\neg \rotatebox[origin=c]{180}{$\forall$} \varphi \to  \rotatebox[origin=c]{180}{$\forall$}\neg \rotatebox[origin=c]{180}{$\forall$}\varphi$
%%\item Persistence of atoms: $P{\bf x} \rightarrow D_YP{\bf x}$, if all $x \in {\bf x}$ are in $Y$.
%\vspace{-1mm}
%\item Reflexivity, Monotonicity and Transitivity for dependence atoms.
%%(a) $D_Xx$ for all $x \in X$, (b) $D_Xy \rightarrow D_Zy$, if $X \subseteq Z$, (c) $(D_XY \land D_YZ) \rightarrow D_XZ$
%\vspace{-1mm}
%\item The Transfer Axiom: \, $(D_XY \land \dual_Y\varphi) \rightarrow \dual_X\varphi$
%\vspace{-1mm}
%    \item Dependence on  the empty set: \, $\edual_y D_\emptyset y \to D_\emptyset y$.
%\end{itemize}
%\begin{fact}\label{TopoSoundness}
%The system {\bf LCD} is sound on topo-dependence models.
%\end{fact}
even LFD principles that remain valid as they stand now express something subtly different in a topological setting. In particular, the Transfer Axiom $(D_XY \land \dual_Y\varphi) \rightarrow \dual_X\varphi$  turns out to capture the \emph{continuity} of dependence.\footnote{In the topological reading, the Transfer Axiom literally tells us that, if there is a continuous function $F$ mapping $X$-values into the corresponding $Y$-values, then $s(Y)\in Int\{t(Y):t\models \varphi\}$ implies $F^{-1}(s(Y))=s(X)\in Int\{t(X):t\models \varphi\}= Int (F^{-1}\{t(Y):t\models\varphi\})$ for all $\varphi$. This is the syntactic counterpart of the topological definition of continuity: inverse images of open sets by continuous maps are open.}

\begin{rem} [Point-free alternatives] \, \emph{Dependence in the logic LCD strengthens the notion of dependence in LFD: the functions made explicit in Section \ref{Functions} are now to be \emph{continuous}. But the intuitions behind the topological view seem independent from the existence of point-to-point  functions. They rather talk about \emph{correlating evidence}, i.e., open sets, whether or not there is some underlying set of sharp limit points and functions between these. The better framework, then, might be a \emph{point-free} topology, with the notion of dependence suitably adapted to direct correlations between open sets that induce continuous functions under some appropriate mathematical construction of points.}
\end{rem}

\subsection{\textbf{Dynamical systems}}

\vspace{-2mm}

Many real-life dependencies have a temporal aspect. Even the simple propositional example of Remark 6.8 suggests a network dynamics  where propositions can become true or false, and dependencies involve a time delay. \footnote{The same is true for situation-theoretic scenarios of information flow, \cite{BarSel}, \cite{BentMart}, and for ubiquitous strategies in iterated strategic games, such as Tit-for-Tat or Copy-Cat: what you do now is what I will do next, \cite{O&R94}.} This suggests a temporal universe of assignments occurring over time, with dependencies such as
$$s_{t + 1} (y) = s_t(x).$$
%Here,
%dependence models are static structures, representing just all possible system states and possible transitions between them, but without giving the temporal evolution.
Now one might reduce this to a static setting by adding temporal variables, using function terms as in Section 6.1. But it seems more natural to turn dependence models into \emph{dynamical systems} where assignments are global states that can occur and repeat over the permissible evolutions of the system. A logic for this should combine LFD with a temporal language.
%\footnote{With a dynamical system map $F$ from points in time to assignments, object-denoting terms can involve a `Next' operator $Nx$ evaluated at $t$ as $F(t+1)(x)$. This suffices for defining the mentioned game-theoretic strategies.}

 Consider a dependence model $\bM=(M,A)$ with an  assignment-changing next-state map $g:A\to A$. The \emph{dynamical system defined by $(\bM,g)$} is the family of functions $\{g^n\}_{n\in N}$, with $g^n$  the $n$-fold composition of  $g$ with itself.
 %The above temporal notation $s_t$ can be now defined, for all $t\in N$ and $s\in A$, by putting $s_t:= g^n(s)$, so   $s_0=s$, $s_1=g(s)$, $s_2=g(g(s))$ etc.
 A simple language of dynamic dependence adds three items to the syntax of LFD: a \emph{next} operator $\bigcirc\varphi$, an \emph{$n$-th step dependence} operator $D_X^{(n)} y$ for each $n\in N$, and a \emph{henceforth} operator $*\varphi$. Their semantics has these clauses:

\vspace{-2mm}

$$s\models \bigcirc\varphi \, \, \mbox{ iff }\,\, g(s)\models \varphi$$

\vspace{-4mm}

$$s\models D_X^{(n)} y  \, \, \mbox{ iff }\,\, s=_X t \mbox{ implies } g^n(s)=_y g^n(t) \mbox{ for all $t\in A$}$$

\vspace{-4mm}

$$s\models *\varphi \, \, \mbox{ iff }\,\,  s\models \bigcirc^n\hspace{-0.05cm}\varphi \, \mbox{ for all $n\in N$} \quad (\mbox{with}  \bigcirc^{n}\hspace{-0.1cm}\varphi \, \,\mbox{the \emph{n}-th iteration of} \,\, \bigcirc)$$

In particular, $D_X^{(1)} y$ says that the current values of the variables in $X$ uniquely determine the next-step value of $y$. This is just what is needed to formalize Tit-for-Tat or Copy-Cat.\footnote{The syntax chosen here for purposes of illustration is a bit cumbersome. Direct functional notations, such as $Ox$ for the value of $x$ at the next state, will be more perspicuous in practice.}

As for valid reasoning, dynamic analogues of Reflexivity, Monotonicity and Transitivity are easy to formulate.
%Here is the valid principle for Transitivity:
%$$D_X^{(0)} x, \,\, \mbox{ provided that $x\in X$,}$$
%$$D_X^{(n)} y \to D_Z^{(n)} y \,\, \mbox{ provided that $X\subseteq Z$,}$$
%$$\left( D_X^{(n)} Y \wedge \bigcirc^n D_Y^{(m)} Z \right) \to D_X^{(n+m)} Z.$$
There is also a valid dynamic analogue of the Transfer Axiom:

\vspace{-2mm}

$$\left( D_X^{(n)} Y \wedge \bigcirc^n \dual_Y \varphi \right)\to \dual_X\hspace{-0.1cm}\bigcirc^n \hspace{-0.1cm}\varphi$$

\smallskip

\noindent{\bf Open problem} \quad  \emph{Axiomatize  dynamic dependence logic completely. Is this logic decidable?}\footnote{In response to an earlier version of this paper, completeness and decidability results for the temporal dependence logic of dynamical systems have been claimed in \cite{Dazhu}.}

\begin{rem}[Topology once more] \emph{
Dynamical systems usually have a  state space endowed with a topology. This richer setting gives rise to  `dynamical topo-dependence models' $(M, A, \tau)$ with a next-step map $g:A\to A$ that is  continuous.
%with respect to $\tau_V$: the coarsest topology that makes all $f_x$ continuous.
Then,  for instance, extending the system LCD with temporal operators, for a finite total set of variables $V$, the commutation axiom $\bigcirc \dual_V \varphi \to \dual_V\hspace{-0.1cm}\bigcirc\hspace{-0.1cm}\varphi$
expresses the continuity of the next-step function $g$.} %cf. \cite{KrMints}.}

%A topological perspective has independent motivations, and it will be discussed later on. Here is a  natural topological clause for dynamic dependence:
%$$s\models D_X^{(n)} y  \, \, \mbox{ iff }\,\,
%\forall V\in \tau (g^n(s)(y)) \exists U\in \tau (s(\mathbf{X})) \forall t\in A \left( t(\mathbf{X})\in U\to g^n(t)(y)\in V\right).$$
%It is easy to see that $\rotatebox[origin=c]{180}{$\forall$} D_X^{(n)} y$ holds globally on a model based on a $T_0$ topology iff there exists a continous function $F: O^{|X|}\to O$ such that
%$f_y\circ g^n= F\circ f_X$ (with $f_y$, $f_X$ defined as on topo-models).
%All the above validities on dynamic dependence models are still valid on dynamic topo-models. In addition, if the set of variables $V$ is finite, the following commutation principle is valid:
%$$\bigcirc \dual_V \varphi \to \dual_V\bigcirc \varphi.$$
%This  expresses the continuity of the next-step function $g$, cf. \cite{KrMints}.}
%\noindent{\bf Open problem} \quad Find a complete axiomatization for dynamic topo-dependence logic. Is this logic decidable?
\end{rem}

\vspace{-4mm}

\subsection{\textbf{Games}}

\vspace{-2mm}

Dependence also occurs in game theory, \cite{O&R94}, though with an additional flavor. While LFD speaks about dependence of \emph{values}, game theory talks about dependence of \emph{actions}. The notions are related, but games pose some interesting new features for logical dependence analysis, \cite{LiG}, \cite{BenKlein}.

\begin{exam}\label{game}\emph{(Choice and dependence).}\, \emph{Consider an extensive game of perfect information with two players $A$, $E$ that have two moves `left' and `right' at each turn. $A$ moves  first, then it is $E$'s turn. The four histories in the game tree can be viewed as assignments to two variables $x, y$, with $x$ the action chosen by  $A$, and $y$  by $E$. The available actions for each player are independent from those of the other: both $I_xy$ and $I_yx$ hold in the sense of Section 6.3.}

\emph{Now let $E$ choose a strategy, i.e., in this simple game: a move at each of her two possible decision nodes. This restricted play introduces a functional dependence: the LFD statement $D_xy$, which was false before, will now come to hold. Thus, committing to a choice, or a strategy in general, changes the current dependence model for the game to one where appropriate dependence statements come to  hold.}\footnote{It is often said that committing to a strategy makes one's actions independent from those by the other player, since the strategy was chosen beforehand. However, this seems a confusion between pre-game deliberation and in-game play: as the game is played, a strategy does follow the particular moves chosen by the other players.}
\end{exam}

Extensive game trees can be associated with dependence models whose variables stand for successive actions by the players.\footnote{To make this work, some issues have to be solved, since strategies produce forests rather than sub-trees, \cite{LiG}. Also, in extensive games, variables should depend on `earlier' variables, not on those for later stages.}
%These more detailed issues can be ignored here, since they do not concern the main points to be made.
Moreover, the action perspective introduces the dependence dynamics of Section \ref{DepDynamics}. Making a choice \emph{makes a dependence statement $D_Xy$ true} by removing some assignments from a given model $\bM$, to obtain a submodel {\bf N} satisfying (a) the statement $D_X y$, but also  (b) the following `$X$-richness' constraint, for the given set $X\subseteq V$ of variables:
$$\bM\hspace{-0.1cm}\restriction\hspace{-0.1cm}X  \, = \, {\bf N}\hspace{-0.1cm}\restriction\hspace{-0.1cm}X$$
\noindent (where $\bM\hspace{-0.1cm}\restriction\hspace{-0.1cm}X$ is the restriction of all assignments in $\bM$ to the domain $X$). The `$X$-rich submodels' {\bf N} satisfying this constraint can be viewed in both first-order and modal terms. An interesting question is which syntactic types of statement are preserved when moving from $\bM$ to {\bf N}.

\medskip

\noindent {\bf Open problem} \, Develop the dynamic dependence logic of strategic choice.
\medskip

Next consider extensive games with \emph{imperfect information}. Here is a simple illustration.

\begin{exam}[Imperfect information games] \emph{In Example \ref{game},  now assume that $E$ cannot observe $A$'s move. Then $E$'s epistemic uncertainty relation holds between the two mid-points of the game tree. This game has been discussed widely for its combination of action and knowledge, a typical feature of games with imperfect information and their links with modal logics, \cite{LiG}.}
\end{exam}

In game theory, a strategy must be \emph{uniform}, assigning the same move at points that $E$ cannot distinguish epistemically. In the above example, this leaves only strategies `always left' and `always right'.\footnote{In contrast, in IF Logic these strategies are said to act \emph{independently} from what player $A$ does, \cite{HintSand}.}
The intuition behind uniform strategies is that they can only appeal to things that players \emph{know}.
%, like the `knowledge programs' of \cite{FHMV}.
This knowledge is encoded by the current equivalence class of their epistemic equivalence relation. This is  LFD dependence combined with the epistemic representation in Section 3.4. The choice of move by a strategy depends on the variable for the agent's knowledge state. In general, this perspective  will work with distinct variables for agents and moves, corresponding to the knowledge and the action modalities whose interplay is crucial to reasoning about games with imperfect information.

But, if games are very regular, say, just choosing values for  stage variables $x, y, ...$ from some fixed set,
 epistemic uncertainty relations match up directly with dependence relations for sets of variables. Assume, as is common in epistemic-temporal logic, %\cite{FHMV},
\cite{PaRam}, that players can observe some events that have taken place, but not others. Then their equivalence relation on histories will be equality for the values of their \emph{observable variables} only. In games, strategies for a player now have to choose actions that depend in the LFD sense on the values of the observed variables for that player.  In general, in this sort of two-player game with imperfect observation, players $A$ and $E$ partition all the variables.\footnote{There are also several richer logical languages at the interface of imperfect information games and combined epistemic-dependence models: see \cite{LiG}, Ch. 21, and \cite{BenKlein}.}

%In the very simple game of Example 8.4, player $E$ has no observed variables. This at once explains the earlier tension between saying that $E$'s choice is dependent, or independent. In this particular case, the epistemic relation for $E$ can also be described as the
%relation giving the freedom of action for $A$.
%However, this connection is very special, and it is easy to give simple variations on the given game where $E$'s uniform strategies are correctly described through dependence on what $E$ knows, and incorrectly as being independent from what $A$ has chosen to do.
%\footnote{in the earlier technical terms, the changed dependence model for after the strategy choice makes dependence statements true, but there is no reason at all that it will make non-trivial independence statements true.}
\smallskip

This discussion by no means exhausts the topic of   games with either perfect or imperfect information from a  dependence-logical perspective, and in general, as stated before,  we will need  combinations of epistemic logic for players's knowledge and LFD for their actions.

\vspace{-2mm}

\subsection{\textbf{Causality}}

\vspace{-2mm}

 A final important arena for dependence is causality.
 %as studied in philosophy and computer science. %\cite{Pearl},
\emph{Causal graphs}, \cite{Pearl,HalpCau}, impose  correlations between variables, restricting the simultaneous assignments of values that represent possible states of world. This is reminiscent of the dependence graphs in Section 2, and indeed, one common notion  of `causal influence' of a variable $x$ on an endogenous variable $y$ found in \cite{Pearl} can be simply represented in LFD as $D_{V-\{y\}} y \land \neg D_{V-\{y, \, x\}} y$, with $V$  the set of all variables. But the match is not one-to-one. Not all relational facts in dependence graphs represent causal connections: singling out the truly causal ones requires a separate decision. Vice versa, causal graphs do not have a unique associated dependence model: they are schemata for  many models. Even so, the representation in Section 2 may extend to causality,  now also analyzing  various types of dependencies that can occur in dependence models.

Conversely, several themes in the  theory of causal graphs resonate in the present framework. For instance, LFD with function terms may be considered a modal companion to the logic for causality in \cite{HalpCau}, that manipulates explicit equations between variables in causal graphs. Also, the crucial notion of  `interventions' in causal graphs has an obvious counterpart in updates of dependence models that fix values, as in Section 6.5.
Even so, there may  be an essential surplus to the notion of causal dependence that transcends the resources of the LFD framework.
In this sense, see \cite{BarSan2019,Barbero} for a formalism that combines features of Dependence Logic with an interventionist approach to causality, and see \cite{Xie} for a similar combination of epistemic logic and causal models.
\bigskip

Many further concrete notions of (in-)dependence  occur in the literature. There is essential dependence and independence in  natural language, \cite{HintSand}, metaphysics, \cite{Fine}, \cite{Humber}, proof theory, \cite{T&S}, ceteris paribus reasoning, \cite{BGR}, social choice theory, logics of agency, and many other fields. A complete list is beyond the scope of this paper, but a confrontation with LFD seems worthwhile in many of these cases.
%Fine arbitrary objects, in proof theory, Meyer Viol Instantial Logic,  in  ceteris paribus reasoning, REF, and in theories of agency such as Social Choice Theory, REF, Coalition logic, REF, and STIT logic, REF. Taking LFD to such areas remains to be done.

\section{\large{\textbf{Related work}}}
In this section, some of many other approaches to dependence are listed in historical order, with comments on connections to the LFD framework.

\medskip

{\bf Armstrong axioms.} The basic structural properties of functional dependence used in this paper were identified by Armstrong \cite{Armstrong}, in the form of the postulates of Inclusion (cf. Definition \ref{struct prop}, Example \ref{LFD theorems}a), Transitivity (cf. Definition \ref{struct prop}) and Additivity (cf. Example \ref{LFD theorems}b). By Fact \ref{projection}, the first two together are equivalent with the conjunction of our Projection and Transitivity properties (as well as with the conjunction of Reflexivity, Monotonicity and Transitivity), while Armstrong's Additivity is absorbed into our definition of $D_XY$ as an abbreviation for $\bigwedge_{y\in Y} D_Xy$.  Armstrong gave a representation theorem showing that these axioms are complete for (global) database dependence. Section 2 of this paper presents a different proof, yielding a stronger representation theorem (Proposition \ref{Armstrong completeness}), for both global and local dependence. Similar  abstract structural axioms for independence,  given in \cite{PearleInd},  underlie the modal independence logic in Section 6.3.

\medskip

{\bf CRS logic.} As explained in our Introduction, LFD is a direct continuation of generalized assignment semantics CRS for first-order logic, for which we have given several references. The origins of CRS lie in relational and cylindric algebra, \cite{Nem85}. The decidability of CRS can be shown by first-order translation into the `Guarded Fragment' GF, \cite{ABN}, while the first-order translation for LFD in Section 3.2
does not map into GF. As we have noted, it is an open problem whether one can prove decidability for LFD via a known decidable fragment of FOL.

\medskip

{\bf Independence-friendly logic.} Dependence pervades game-theoretic semantics for logical systems. Strategies in evaluation games for FOL correspond with Skolem functions that express dependence in the sense of Section 2.3. A further innovation was `Independence-Friendly Logic' (IF-logic, for short), \cite{HintSand}, where the player for the existential quantifier may have imperfect information about the objects chosen by the player for the universal quantifier, cf. Section 7.5.
A compositional semantics for IF-logic uses evaluation on sets of assignments, \cite{Hodges}, allowing for choices of values  independently from the values for specified other variables. \footnote{As it happens, sets of assignments were used even earlier in dynamic semantics of natural language, in order to model  the meaning and anaphoric behavior of plural expressions, \cite{vdBerg}.} %For the basic logic of this scheme, see the Appendix of \cite{ELD}.
These sets are like LFD dependence models, but without designated single assignments and local dependence.  Moreover, in contrast with LFD, IF-logic is second-order and non-axiomatizable. For a complete mathematical development of IF-logic, see \cite{MaSaSe}.

A comparison between LFD and IF-logic  poses a challenge, already noted for CRS vs. IF-logic in \cite{ELD}. IF-logic sees first-order logic as tied to linear \emph{dependencies} between quantifiers, and incorporates `branching quantifiers', thereby moving up to second-order complexity. In contrast, CRS sees FOL as too much tied to \emph{independence}, and weakens it to a decidable logic that allows for both dependence and independence of variables. One obvious difference is that IF-logic takes standard FOL as is, and adds syntax for independence. We made some remarks on the connection of LFD with FOL in Section \ref{quantification},  and we have more precise results --  but a deeper treatment is a topic for a separate paper.  But perhaps more importantly here, in the terminology of Section 7.5, while LFD analyzes what might be called \emph{value dependencies} between variables, IF-logic describes what might be called \emph{choice dependencies} between quantifiers. It is easy to see formally that LFD cannot express choice dependencies,  and our brief discussion of games  showed that we would need additional modalities over dependence universes. Even so, LFD and IF-logic also share some traits, and cross-overs between the two are worth exploring.

For instance, one can \emph{enrich} LFD with natural forms of branching quantification on dependence models. For instance, the natural reading of the simplest Henkin formula
\begin{align*}
\left[\begin{array}{cc}
  \forall x_1 & \exists y_1 \\
  \forall x_2 & \exists y_2 \\
\end{array}\right]
P x_1 x_2 y_1 y_2
\end{align*}
in a dependence model $\bM=(O,I,A)$ is the assertion that
$$\exists F:O^{(x_1)}\to O\, \exists G:O^{(x_2)}\to O\, \forall (o_1, o_2)\in O^{(x_1, x_2)}\,\, (o_1, o_2, F(o_1), G(o_2))\in I(P)\cap O^{(x_1, x_2, y_1, y_2)}.$$
In other words: the witnessing functions have domains restricted to the corresponding admissible values and return a tuple combination that not only satisfies $P$, but  is actually realized by some admissible assignment of values to $(x_1,x_2, y_1,y_2)$. This seems to be the natural generalization to branching quantifiers of the semantic reading of the LFD formula $\forall_x \exists_y Pxy$ explained in Section \ref{quantification}. Similarly,  one can define LFD versions of the slash quantifiers from IF-logic. As happened in the case of CRS versus FOL as well as with other logical systems, \cite{ABBN}, this strategy of generalizing to models with admissible tuples of values might well lower the complexity of IF-logic. But again, we leave details to a further publication.

\medskip

{\bf Independence and randomness.} An innovative abstract first-order logic for probabilistic independence is presented in \cite{Lambal}, emerging from the study of randomness. The calculus contains several axioms at the abstraction level of LFD, but also more specialized principles such as the Steinitz Axiom for linear dependence discussed in Section 7.2. For another broad approach to elementary qualitative principles for dependence and independence, see the measurement-theoretical analysis of probabilistic reasoning in \cite{Narens}.

\medskip

{\bf Dependence logic.} V{\"a}{\"a}n{\"a}nen's dependence logic DL \cite{Vaana} was the first to introduce explicit dependence atoms (for global dependence), a crucial device that we have adopted in LFD. The language of DL is an extension of the language of first-order logic, but interpreted over \emph{sets} of assignments (called `teams'), instead of single assignments (as in FOL), or combinations of a current assignment and a surrounding
team (as in LFD). This `set lifting' semantics was first suggested in this context in \cite{Hodges}.\footnote{Set lifting as a general device has a long history in logical semantics, resulting in the theory of `complex algebras' which distinguish, amongst  other things, different `inner' and `outer` variants of Boolean operations, \cite{Brink93}. E.g., the disjunction of DL is the inner version of Boolean union.}
%\footnote{Set lifting has a long history, from interval semantics for temporal expressions \cite{LOT} and `complex algebra' in algebraic logics, \cite{Brink} to possiblity models, \cite{Holliday} and inquisitive logic, \cite{}.}
Interpreting on sets of assignments \emph{lifts} the meanings of standard propositional connectives, resulting in a richer vocabulary with a non-classical logic. Moreover, the DL interpretation of the first-order quantifiers ranges over sets of assignments, yielding a form of second-order quantification. By now, there is an extensive body of theory on variations, extensions, and fragments of the DL framework, which we cannot survey here. The reader is referred to the original source \cite{Vaana} and to the extensive survey article \cite{Galliani} in the  Stanford Electronic Encyclopedia of Philosophy.

Comparing LFD with DL, one striking difference is between the `team semantics' for DL on sets of assignments with global functional dependencies, and the local semantics of LFD with  single assignments inside teams, giving the central place to local dependence. This difference may seem slight, but as observed earlier, the set lifting brings with it some pressure towards non-classical logics.\footnote{For a system preserving classical semantics in a set-lifted setting, cf. the possibility semantics of \cite{Holliday}.} Another major difference is the view of quantifiers, as briefly discussed above in connection with IF-logic, and from a more classical angle, in Section \ref{quantification}. LFD quantifiers range over values available inside one dependence model, thus respecting all current dependence constraints. In contrast, DL quantifiers can evaluate in \emph{new} teams (i.e., \emph{other} dependence models), thus  `freeing' the quantified variables from the constraints of the old team. From the minimal standpoint of LFD, such `freeing quantifiers' are naturally viewed as composites
of \emph{two} different notions of logical interest: a dynamic modality for relevant kinds of model change in a
dependence universe, followed by a model-internal LFD quantifier.\footnote{Valid laws for   `freeing quantifiers' depend on the model change relation chosen plus the choice of dependence universes. E.g., quantifiers become second-order when the dependence universe contains all possible variants of the current dependence model, i.e., all sets of assignments. But as remarked in Section \ref{DepDynamics}, one might also allow gaps in dependence universes,  creating higher-level dependencies, and lowering complexity of the logic.}

%though subject to some analogies concerning the value ranges for the free variables of the formula
Both views have their attractions. The LFD quantifiers fit well with an
\emph{epistemic} interpretation (in which the current team comprises all the possibilities compatible with one's background knowledge), as well as with applications to \emph{complete databases} (where the current team stands for a complete state space, listing all the states that can be generated by some dynamic process). Other dependence models then only come in via informational update, or via process change. But in many other settings,
e.g., applications to more general \emph{partial} databases, dynamical updating, open systems, etc.,
the stronger DL quantifiers will be just what is needed, capturing interesting properties that
go beyond the resources of LFD.\footnote{Many further themes in this paper  have counterparts in the literature on more expressive DL formalisms. For instance, dependence plus independence in a sense close to that of Section 6.3 is studied in \cite{GraVa} and its follow-up literature, cf. \cite{Galliani}. In addition,  studies of dependence with a classical logic base are found in \cite{G&K}, \cite{Superve}.}

Finally, in terms of technical comparisons, there are various questions that can be asked. One is the point, already mentioned in the Introduction, that the lower complexity of LFD (inevitably) comes at a price of lower expressive power. Thus, it would be of interest to match LFD (minus local dependence atoms) and its extensions with some low-complexity fragments of DL, and in particular fragments with restricted forms of quantification. Given the modal nature of our formalism, it might seem at first sight that the appropriate comparison is with the system of `Modal Dependence Logic' in \cite{VaanaModal}, or other propositional logics of dependence \cite{FanVaana}. However, both of these are purely propositional languages, with no variables over objects and no quantification over them, however restricted. In fact, since CRS is known to have tight connections with the Guarded Fragment of FOL, it would be more natural to expect its extension LFD with dependence atoms to have interesting connections with some corresponding fragment of DL. As a particular instance, how is LFD related to the recent
Guarded Fragment versions of DL introduced in
 \cite{Aachen2}? Another interesting line to pursue might be the earlier-mentioned `deconstruction' of DL quantifiers into dynamic modalities plus LFD modalities, which could lead to a richer intermediate theory with various modalities over dependence universes. Indeed, it seems worthwhile to look for formalisms in between LFD and DL. The dependence universes in Section \ref{DepDynamics}  are an instance, since they represent a Henkin move of not considering all sets of assignments, but just certain subfamilies,  creating what might be called higher-order dependencies when moving between dependence models. We suspect that this will make the logic first-order, since one can describe this setting in a \emph{three-sorted first-order language} with variables over objects, assignments, and dependence models.\footnote{This is just one option. See \cite{Konti} for an alternative way of  reducing the complexity of DL.}

\smallskip

{\bf Logics of questions.}
We have noted at various places that LFD has informational interpretations in terms of implications between questions, as discussed in a general dynamic-epistemic setting in \cite{Baltag2016}. The logic of questions has a long history with classical sources such as \cite{B&S} and \cite{HintQu}. The handbook article \cite{Harrah} surveys many themes on the logic side, and \cite{G&S97} surveys  themes focusing on natural language semantics. For a  modern perspective on dependence in terms of inquisitive logic of questions, cf. \cite{DepQ}. For a comprehensive  treatment of inquisitive logic, we refer to the modern source \cite{CGRoe}. %As perhaps the most direct counterpart to our approach in the inquisitive literature, one may consider the system InqBQT in \cite{Ciardelli2016}, also studied in \cite{Yang}.

\smallskip

{\bf Extended epistemic logic.} Connections between LFD and epistemic logics were explained in Section 4. We cannot survey all points of contact, but general background can be found in \cite{LKB}. As  a special case, \cite{Lomuscio} is an early study of epistemic models with our equality-based accessibility relations. Many of our dependence themes are  reflected in analogies between LFD and recent work on `extended epistemic predicate logics' where agents can know not just propositions, but also objects, \cite{Plaza,Yanjing,Baltag2016}. These logics add `knowing wh'-constructions to propositional `knowing that', and can be seen as well-chosen often decidable fragments of epistemic predicate logic. Specific analogies include our theme of functional definability in Section \ref{Functions} and `knowledge of functions', studied in \cite{Yifeng}, the conditional knowledge of objects and facts by epistemic agents in Section \ref{DepDynamics} and \cite{Yanjing,Baltag2016}, and on logics for public `inspection' of values, \cite{vEGWang}.

%CONCLUSION NEEDED.

%{\bf And beyond.} The above list of research lines is not an exhaustive survey. There are many other formal approaches to either reasoning with dependence or general management of variables,
%\footnote{One suggestion for a congenial approach has been the syntactic framework of `nominal logic', Gabbay.}
%for which clarifying connections  to LFD may be a rewarding endeavor.

\vspace{-2mm}

\section{\large{\textbf{Conclusion}}}

Dependence has a ubiquitous semantic sense of determination of values for some variables by those of others. We have presented a decidable classical logic LFD for reasoning about functional dependence, together with complete axiomatizations. The proofs come in both first-order and modal style. Conceptually, these two complementary perspectives connect to the two manifestations of dependence highlighted throughout this paper: `ontic' in the world or in some dynamical system, and `informational' connecting to knowledge and questions. Further language extensions, as well as richer semantical settings, have been discussed in some detail.

\medskip

Many open problems  have been identified in this extension process, reflecting mainly its  semantic and model-theoretic spirit. But we have also shown that there is  room for a purely proof-theoretic analysis of LFD and its extensions, and perhaps as a compromise between the model theory and proof theory: an analysis in \emph{universal algebra} would be illuminating.

\medskip

Going beyond these standard logical perspectives, one can think of dependence \emph{information-theoretically}, in terms of values of dependent variables adding no Kolmogorov complexity to the given ones.
%But given the ubiquity and practical importance of dependence, the most important task may not be to produce more logic. In a few case studies, we have taken LFD to concrete areas such as linear algebra, dynamical systems and game theory, showing how the dependence semantics fits theses cases, and highlights what is general and what is special to the  area. These initial examples may at least have shown the interests of this second stage of the program proposed here.
But perhaps the greatest challenge left unaddressed here is tying the qualitative logical LFD analysis to \emph{probabilistic} notions of correlation and dependence.\footnote{There might even seem to be an essential mismatch, as \emph{independence} is complexity-increasing in LFD, while it is complexity-decreasing in probabilistic computation. Compare the undecidability of modal logics that have commutation axioms, with the beneficial use of commutation results like Fubini's Theorem in probabilistic reasoning. This mismatch dissolves, however, by making  a  distinction. Regular mathematical structures simplify computation, but their logical \emph{theory} is more complex than that of arbitrary structures.}

\medskip

A point of entry may be the analogy of dependence with consequence relations noted in Section 2. LFD-style dependence goes by universal quantification over all assignments. But as we observed, one can soften this, as in non-monotonic default logics, by going to models where the semantic dependence holds only in the most plausible cases, or only with high probability in some qualitative sense, \cite{DHI}. In that case, the agenda for LFD becomes wide open again.

\bigskip

\par\noindent
{\bf Acknowledgments} \,
\par\noindent
We thank various audiences in Amsterdam, Bayreuth, Beijing, Leiden, Moscow, Nice, Rennes, Pittsburgh, Stanford and Tbilisi for their feedback on this work, and in particular, Fausto Barbero, Adam Bjorndahl, Denis Bonnay, Jan van Eijck, Malvin Gattinger, David Gonzalez, Helle Hansen, Andreas Herzig, Kevin Kelly, Raoul Koudijs, Dazhu Li, Graham Priest, Phil P\"{u}tzst\"{u}ck, Valentin Shehtman, Gabriel Sandu, Chenwei Shi, Sonja Smets, Yde Venema, Jouko V\"{a}\"{a}n\"{a}nen, Yanjing Wang, and Dag Westerst\aa hl.

\newpage

%\small{

\section*{\large{\textbf{Appendix A: Modal proofs of LFD decidability and completeness}}}
To study LFD as a modal logic, we need to generalize the 'standard' relational models introduced in Section \ref{StandardRelationalSemantics} to a wider class of relational models. Viewing LFD as a modal language in the usual sense, with modalities $\dual_X\varphi$ and atomic formulas $P\mb{x}$ and $D_X y$, our general relational models will be just ordinary Kripke models for this language. This move allows us to apply to them well-known notions and methods in modal logic, such as $p$-morphisms, unraveling, and filtration. In the following we will assume familiarity with these standard modal techniques. See \cite{BRV} for definitions and explanations.

\medskip

So there are two main differences between general relational models and the standard models introduced earlier: (a) each relation $=_X$ for sets $X\subseteq V$ is taken as primitive, without being reduced to an intersection of basic relations $=_x$, and (b) $D_X y$ is treated as just another atom, whose semantics is given by a valuation (although one subject to restrictions).

\renewcommand{\thethm}{A.\arabic{thm}}

\subsection*{\textbf{A1. Relational semantics}}

%\setcounter{section}{1}
%Here is a standard modal relational semantics for LFD:

\begin{defi}
A \emph{relational model} is a structure $\bM=(A, =_X, D_X y, P\mb{x})$, where: $A$ is a set of possible worlds (``abstract assignments"); $=_X\subseteq A\times A$ are binary relations on worlds, one for each set $X\subseteq V$ of variables; $D^s_X y\subseteq {\mathcal P}(V)\times V$ are relations between sets of variables $X$ and variables $y$, one for each world $s\in A$; and $P^s$ are $n$-ary relations on variables, one for each $n$-ary predicate $P$ and each world $s$. These ingredients are required to satisfy four conditions:
\begin{description}
\item[(1)]\quad all relations $=_X$ are equivalence relations on $A$;
\item[(2)]\quad all relations $D^s$ satisfy Projection and Transitivity;
\item[(3)]\quad if $s=_X t$ and $D^s_X y$, then $s=_y t$ and $D^t_X y$;
\item[(4)]\quad if $s=_X t$ and $P^s \mb{y}$ for some $\{y_1, \ldots, y_m\}\subseteq X$, then $P^t\mb{y}$;
\item[(5)]\quad $=_\emptyset$ is the global relation on $A$ (relating every two worlds)
\end{description}
%A relational model is said to be \emph{standard} if it satisfies the following additional two conditions:
%\begin{description}
%\item[(5)] if $s=_X t$ and $s=_Y t$, then $s=_{X\cup Y} t$. (In other words, $=_{X\cup Y}$ is the intersection of $=_X$ and $=_Y$.)
%\item[(6)] if $s=_X t \mbox{ implies } s=_y t$ holds for all $t\in A$, then $D^s_X y$.
%\end{description}
The semantics of LFD on relational models is just as on dependence models, except that the abstract relations $s=_X t$, $D^s_X y$ and $P^s\mb{x}$ are used instead of their concrete counterparts.
\end{defi}

\medskip

\begin{fact}  \emph{Standard} relational models in the sense of Section \ref{StandardRelationalSemantics} are exactly those relational models satisfying the following two \emph{additional conditions}:
\begin{description}
\item[(5)]\quad if $s=_X t$ and $s=_Y t$, then $s=_{X\cup Y} t$.
%(In other words, $=_{X\cup Y}$ is the intersection of the relations $=_X$ and $=_Y$.)
\item[(6)]\quad if $s=_X t \mbox{ implies } s=_y t$ holds for all $t\in A$, then $D^s_X y$.
\end{description}
\end{fact}

\subsection*{\textbf{A2. Equivalence between relational models and dependence models}}

We now show that \emph{the logic of relational models is the same as the logic of dependence models}.

\medskip

To go from dependence models to relational models: we can just use the equivalence between dependence models and standard relational models (cf. Fact \ref{Standard1} and Fact \ref{Standard2}).

\medskip

But to go the other way, from relational models to dependence models, we need a representation of relational models in terms of standard ones:

\medskip

\begin{prop}\label{Representation}
Every relational model is a p-morphic image of some \emph{standard} relational model (in the sense of Section \ref{StandardRelationalSemantics}).
\end{prop}

\begin{proof}\quad \,\,
The proof is essentially a variation of modal \emph{unravelling}, making infinitely many copies of each world.\footnote{Note the similarity of the unraveled model in this proof and the tree construction in the proof of Theorem \ref{Rep}. Indeed, the two decidability proofs are based on similar ideas, but there are also notable differences. The proof in Section \ref{Decidability} is based on a syntactic construction (``type models") and is very elaborate; the proof of the key Truth Lemma is  a very syntactic complex induction on formulas. In contrast, the proof in this section is purely semantic, and it offers a shortcut, by relying on known results and techniques in Modal Logic.}

\medskip

Let $\bM=(A, =_X, D, P)$ be a relational model, and let $s_0\in A$ be any designated world. To construct a standard relational model $\bM^{st}$, take as  \emph{worlds} the set $A^{st}$ of all `histories', i.e. all finite sequences $h=(s_0, X^1, s_1, \ldots, X^n, s_n)$, with $n\geq 0$ and $s_o, \ldots, s_n\in A$ satisfying $s_{k-1} =_{X^k} s_k$ for all $k=1,n$.  We denote by $last(h):=s_n$ the last state in history $h$, and by $\to_{X}$ the natural \emph{one-step relation} on histories, given by $h\to_X h'$ iff $h'=(h, X, s')$ (with $last(h) =_X s'=last(h')$). The one-step relations structure $A^{st}$ can be viewed as a \emph{tree} with root $(s_0)$ (where $s_0$ is the designated world), in which any two nodes $h, h'$ are connected by a unique non-redundant path.

\medskip

To structure this as a relational model, we define a \emph{new one-step relation} $\stackrel{=}{\to}_{X}$, incorporating all the one-step relations labelled by sets that locally determine $X$:
$$h \stackrel{=}{\to}_{X}  h' \,\,\, \mbox{ iff } \,\,\, h\to_{Y} h' \mbox{ for some $Y$ with }
last(h)\models D_Y X.$$

Then the \emph{required equivalence relations} $=_X$ on worlds/histories in $A^{st}$ can be taken to be \emph{the reflexive-transitive-symmetric closure of the relations} $\stackrel{=}{\to}_{X}$. To check the claims below, it may be useful to note that $h=_X h'$ holds iff the unique non-redundant path from $h$ to $h'$ consists only of steps of the form $h_n \stackrel{=}{\to}_{Y^n} h_{n+1}$, or $h_n \stackrel{=}{\ot}_{Y^n} h_{n+1}$, with $last(h_n)\models D_Y X$.

\medskip

Finally, the \emph{valuation} on atoms is given by truth at the last world in the history (in the original model):
$$D^{h}_X y \,\,\, \mbox{ iff } \,\,\, last(h)\models D_X y, \quad\,
P^h \mb{x} \,\,\, \mbox{ iff } \,\,\, last(h)\models P\mb{x}.$$

The fact that this definition yields a standard relational model $\bM^{st}$ is an easy verification.

\medskip

To finish the proof, we define a map $f: A^{st}\to A$, by putting $f(h):= last(h)$ for all $h\in A^{st}$. It is easy to check that $f$ is a surjective p-morphism $f$ from $\bM^{st}$ to $\bM$. (Surjectivity follows from the fact that every world $s\in A$ satisfies $s_0 =_\emptyset s$, by condition 5 on relational models, hence $h=(s_0, \emptyset, s)$ is a history with $f(h)=last(h)=s$.)
\end{proof}

\medskip

Combining Fact \ref{Standard1}, Propositions \ref{Standard2} and \ref{Representation}, plus  the preservation of modal formulas under surjective p-morphisms (and so under surjective homomorphisms), yields the following:

\begin{cor}\label{Modal Equivalence} (\emph{Modal equivalence of relational and dependence models})
The same LFD formulas are valid on dependence models, relational models and standard relational models.
\end{cor}

\subsection*{\textbf{A3. Decidability via relational models}}

The preceding detour into abstract relational models and the above Corollary \ref{Modal Equivalence} on modal equivalence can be used to give a second, more general proof of decidability using the Modal Logic concept of \emph{filtration} \cite{BRV}.
%\footnote{There is also a third route, via  the limited form of cut elimination in Section 5.2, but this will be ignored here.}

\begin{prop}\label{RelFMP}
The language LFD has the Strong Finite Relational Model Property: if $\varphi$ is satisfied in some relational model $\bM$, then it is satisfied in a \emph{finite} relational model, whose size is bounded by a computable function of $\varphi$. As a consequence, the logic LFD is \emph{decidable}.
\end{prop}

\begin{proof}\quad \,\,  Start with the singleton  $F=\{\varphi\}$, and construct the finite set of formulas $\Phi=\Phi_F$ as in Section \ref{TypeModels} (whose size was bounded by a computable function of $\varphi$).

\medskip

The \emph{filtrated model} $\bM^f$ has as \emph{worlds} the equivalence classes $[s]$ of original worlds $s\in A$ modulo $\Phi$-equivalence $\equiv_{\Phi}$ (with respect to all formulas in $\Phi$). Note that there are only finitely many such classes (their number is bounded by a computable function $F(\varphi)$).

\medskip

To define the \emph{relations} $=_X$ in the filtrated model, we take the following `dependent filtration':
$$[s] =_X [t] \,\,\, \mbox{ iff } \,\, \, (s\models \theta \mbox{ iff } s\models \theta) \mbox{ for all
$\theta\in \Phi$ with } Free(\theta)\subseteq \{y\in V: s\models D_X y\}.$$
This is well defined (independent from the choice of representatives), and the definition implies that $\{y\in V: s\models D_X y\} =\{y\in V: t\models D_X y\}$ whenever $[s]=_X [t]$.

\medskip

As for \emph{valuation}: the truth values at $[s]$ for atoms $D_X y, P\mb{x}\in \Phi$ are inherited from the original truth values at $s$ in $\bM$. The resulting finite relational model $\bM^f$ is a filtration of $\bM$ in the usual sense. By the standard Filtration Lemma, $[s]$ will satisfy $\varphi$ in $\bM^f$.

\medskip

As usual, the Strong Finite Relational Model Property provides an obvious algorithm for deciding satisfiability on relational models (and thus by Corollary \ref{Modal Equivalence} also on dependence models). Given formula $\varphi$ generate all the relational models (up to isomorphism) of size $\leq F(\varphi)$; check whether $\varphi$ is satisfied in any of these models. If so, $\varphi$ is satisfiable; else, it is not.
\end{proof}

\bigskip

It should be noted that in general the filtrated model is typically a \emph{non-standard} relational model, not a dependence model.

\subsection*{\textbf{A4. Completeness via relational models}}

Completeness of LFD with respect to dependence models follows from Corollary \ref{Modal Equivalence} together with the following result:

\begin{lem}\label{Soundness+RelCompleteness} The calculus {\bf LFD} is sound and strongly complete wrt general relational models.
\end{lem}

\begin{proof}\quad \,\,  Soundness is immediate: the conditions on relational models were chosen to validate the matching axioms. For completeness, take the usual Henkin-style `canonical model' for LFD, considered as a basic modal logic. This canonical model is a relational model, and the calculus is strongly complete for this model.\end{proof}

\section*{\large{\textbf{Appendix B: Restricted cut elimination and subformula property}}}

\renewcommand{\thethm}{B.\arabic{thm}}

As announced, it is convenient to absorb Weakening into the logical rules (cf. \cite{T&S} for this technique), while simultaneously restricting Projection and Transitivity to variables that actually occur in the conclusion, and also restricting Cut to dependence atoms between actually occurring variables. This can be done by first modifying the axioms to

\medskip

 (a) $\, \, \Gamma, \varphi \, \vdash \, \varphi, \Delta \quad \quad (b) \,\, \Gamma\, \vdash D_X x, \Delta \,\, \, \mbox{ where} \, \,x\in X\subseteq Var(\Gamma\cup\Delta)$,

\medskip

\noindent while introduction rules are made `cumulative', by repeating principal formulas in the premises.

For instance, the left-introduction rule $(\dual_L)$ becomes
\begin{prooftree}
\def\fCenter{\ \vdash\ }
\Axiom$\dual_X\varphi,\varphi,\Gamma \fCenter \Delta$
\UnaryInf$\dual_X\varphi, \Gamma \fCenter \Delta$
\end{prooftree}

Transitivity needs special treatment: in addition to being made cumulative, it has to be restricted to relevant formulas, becoming the rule of `Restricted Transitivity':
\begin{prooftree}
\def\fCenter{\mbox{\ $\vdash$\ }}
\AxiomC{$\Gamma \vdash \Delta, D_X Y, D_X Z$}
\AxiomC{$\Gamma \vdash \Delta, D_Y Z, D_X Z$}
%\LeftLabel{(Restricted Transitivity)}
\RightLabel{\, where $Y\subseteq Var(\Gamma\cup\Delta)\cup X\cup Z$}
\BinaryInf$\Gamma \fCenter \Delta, D_X Z$
\end{prooftree}

Likewise, the right-introduction rule $(\dual_R)$ needs to be modified to:
\begin{prooftree}
\def\fCenter{\ \vdash\ }
\Axiom$\Gamma \fCenter \Delta,\varphi, \dual_X\varphi$
\RightLabel{\, where $Free(\Gamma\cup\Delta)\subseteq Y$}
\UnaryInf$D_X Y,\Gamma, \Gamma' \fCenter \Delta, \Delta', \dual_X \varphi $
\end{prooftree}

Finally, we replace Cut by a restricted version (in which we also absorbed Weakening):
\begin{prooftree}
\def\fCenter{\mbox{\ $\vdash$\ }}
\AxiomC{$\Gamma \vdash \Delta, D_X y$}
\AxiomC{$D_X y,\Gamma \vdash \Delta$}
\LeftLabel{(DA Cut)}
\RightLabel{where $X\cup \{y\}\subseteq Var(\Gamma\cup\Delta)$}
\BinaryInf$\Gamma \fCenter \Delta$
\end{prooftree}

\medskip

A \emph{restricted-cut proof} is a proof that uses only these modified rules. The following observation shows how LFD allows for a tighter management of variables than FOL:

\begin{lem}\label{Var} (\emph{Elimination of irrelevant variables})
\begin{itemize}
\item\quad If $\Gamma\vdash D_XY,\Delta$ has a restricted-cut proof, and $Z= X\cap (Var(\Gamma)\cup
    Y\cup Var(\Delta))$, then $\Gamma\vdash D_ZY,\Delta$ has a restricted-cut proof.
\item\quad If $\Gamma\vdash \dual_X \varphi,\Delta$ has a restricted-cut proof, and $Z= X\cap (Var(\Gamma)\cup Var(\varphi)\cup Var(\Delta))$, then $\Gamma\vdash  \dual_Z \varphi,\Delta$ has a restricted-cut proof.
\end{itemize}
\end{lem}

\medskip

Using this lemma and a cursory inspection of the above modified rules, we obtain:

\begin{lem}\label{SubfProp} (\emph{Subformula/Subterm Property})
Let ${\mathcal P}$ be a restricted-cut proof of the sequent $\Gamma \vdash\Delta$. Each formula $\theta$ in ${\mathcal P}$ is either of the form $D_X Y$ with  $X\cup Y\subseteq  Var(\Gamma\cup\Delta)$, or it is a subformula of some formula in $\Gamma\cup\Delta$. In particular, only variables $x\in Var(\Gamma\cup \Delta)$ occur in ${\mathcal P}$.
\end{lem}

\bigskip

Finally, we can prove our Restricted Cut Elimination theorem:

\medskip

\noindent \emph{Every provable sequent has a restricted-cut proof (which thus involves only subformulas of the sequent formulas, or dependence atoms for variables in the sequent)}

\begin{proof}\quad \,\,  To show this, first gradually eliminate Transitivity and Projection in favor of their modified versions, using the above lemma when necessary. Similarly replace all other rules except Cut by their cumulative versions. Finally,  eliminate unrestricted cuts in the usual way, by successively removing topmost maximal-rank cuts from a given proof of a sequent $\Gamma\cup\Delta$. Here, since DA Cut is permitted, one need not worry about cut-formulas of the form $D_X Y$, with all variables occurring in the original sequent $\Gamma\cup\Delta$. As a result, the additional axioms and rules for dependency are innocuous: the cut-formula to be removed was never introduced by such rules. The only case that presents any novelty is that of a dual-quantifier cut-formula $\dual_X\varphi$ that is principal in both antecedent and succedent: having been freshly introduced on both sides.\end{proof}
%}


\vspace{-2mm}

\begin{thebibliography}{}
\normalsize{

\bibitem{AbitHV} S. Abiteboul, R. Hull \& V. Vianu, 1994, \emph{Foundations of Databases}, Pearson, London.

\bibitem{Abr} S. Abramsky \&  J. V{\"a}{\"a}n{\"a}nen, 2009, From IF to BI: A Tale of Dependence and Separation, \emph{Synthese} 167:2, 207--230.

\bibitem{ABN} H. Andr{\'e}ka, J. van Benthem \& I. N{\'e}meti, 1998, Modal Languages and Bounded Fragments of Predicate Logic, \emph{Journal of Philosophical Logic}, 27:3, 217--274.

\bibitem{ABBN} H. Andr{\'e}ka, J. van Benthem, N. Bezhanishvili \& I. N{\'e}meti, 2014, Changing a Semantics: Opportunism or Courage?, in M. Manzano, I. Sain \& E. Alonso, eds., \emph{The Life and Work of Leon Henkin}, Birkhaueser Verlag, 307--337.


\bibitem{Armstrong} W. Armstrong, 1974, Dependency Structures of Database Relationships, \emph{Proceedings IFIP Conference}, 580--583.

%%%\bibitem{Baltag2016} Blind Review.

\bibitem{Baltag2016} A. Baltag, 2016, To Know is to Know the Value of a Variable, \emph{Adv. in Modal Logic 2016}, 135--155.

%%%\bibitem{BMS} Blind review.

\bibitem{BMS} A. Baltag, L. Moss \& S. Solecki, 1998, The Logic of Public Announcements, Common Knowledge, and Private Suspicions, \emph{Proceedings TARK 98}, 43--56.

%%%\bibitem{BvB2020} Blind review.

\bibitem{BvB2020} A. Baltag, A \& J. van Benthem, 2020. The Logic of Continuous Dependence and Knowability. Manuscript, ILLC, University of Amsterdam.



\bibitem{BarSan2019}  F. Barbero \& G. Sandu. Interventionist Counterfactuals on Causal Teams. Proceedings 3rd Workshop on
Formal Reasoning about Causation, Responsibility, and Explanations in Science and Technology, \emph{Electronic Proceedings in Theoretical Computer Science} 286: 16--30, 2019.


\bibitem{Barbero} F. Barbero \& G. Sandu. Team semantics for interventionist counterfactuals:
observations vs. interventions. To appear in the \emph{Journal of Philosophical Logic}, 2020.


\bibitem{BarSel} J. Barwise \& J. Seligman, 1995, \emph{Information Flow. The Logic of Distributed Systems}, Cambridge University Press, Cambridge UK.

\bibitem{B&S} N. Belnap \& Steele, 1976, \emph{The Logic of Questions and Answers}, Yale University Press, New Haven.

%\bibitem{Ben70s} J. van Benthem, 1974, Hintikka on Analyticity, \emph{Journal of Philosophical Logic}, 3:4, 419--431.

%\bibitem{Bent79} J. van Benthem, 1979, Universal Algebra and Model Theory: Two Excursions on the Border, Report ZW79-08, Mathematical Institute, University of Groningen.

%%%\bibitem{ELD} Blind review.

\bibitem{ELD} J. van Benthem, 1996, \emph{Exploring Logical Dynamics}, CSLI Publications, Stanford University.

%\bibitem{MFPL} J. van Benthem, 1997, Modal Foundations of Predicate Logic, \emph{Logic Journal of the IGPL}, 5:2, 259--286.

\bibitem{Bent2005} J. van Benthem, 2005, Guards, Bounds, and Generalized Semantics, \emph{Journal of Logic, Language and Information}, 14:3, 263--279.

%%%%\bibitem{LDII} Blind review.

\bibitem{LDII} J. van Benthem, 2011, \emph{Logical Dynamics of Information and Interaction}, Cambridge University Press, Cambridge UK.

%%%\bibitem{LiG} Blind review.

\bibitem{LiG} J. van Benthem, 2014, \emph{Logic in Games}, The MIT Press, Cambridge MA.

\bibitem{BGR} J. van Benthem, P. Girard \& O. Roy, 2009, Everything Else Being Equal: A Modal Logic for Ceteris Paribus Preferences, \emph{Journal of Philosophical Logic} 38:1, 83--125.

\bibitem{BenKlein} J. van Benthem \& D. Klein, 2019, Logics for Analyzing Games, \emph{Stanford On-Line Encyclopedia of Philosophy}.

\bibitem{BentMart} J. van Benthem \& M-C Martinez, 2008, The Stories of Logic and Information, \emph{Handbook of the Philosophy of Information}, Elsevier, Amsterdam, 2017--2080,

%\bibitem{BentIE} J. van Benthem, 2019, Implicit and Explicit Stances in Logic, \emph{Journal of Philosophical Logic}, 48:3,  571--601.

%\bibitem{BentBez} J. van Benthem \& N. Bezhanishvili, 2019, Some Recent Perspectives on Filtration, to appear in \emph{Kit Fine on Modal Logic}, Springer, Dordrecht.

%\bibitem{BentBon} J. van Benthem \& D. Bonnay, 2008, Modal Logic and Invariance, \emph{Journal of Applied Non-Classical Logics}, 18:2/3, 153--173.

%\bibitem{vBvEK} J. van Benthem, J. van Eijck \& B. Kooi, 2006, Logics of Communication and Change, \emph{Information and Communication}, 204:11, 1620--1662.


%\bibitem{BentMart} J. van Benthem \& M-C Martinez, 2008, The Stories of Logic and Information, in \emph{Handbook of the Philosophy of Information}, Elsevier, Amsterdam, 217--280.

%\bibitem{BentMin} Blind review.

\bibitem{BentMin} J. van Benthem \& S. Minica, 2012, Toward a Dynamic Logic of Questions, \emph{Journal of Philosophical Logic}, 41:4, 633--669.

\bibitem{vdBerg} M. van den Berg, 1996, \emph{The Internal Structure of Discourse}, Dissertation 96-03, ILLC, University of Amsterdam.

\bibitem{BRV} P. Blackburn, M. de Rijke \& Y. Venema, 2000, \emph{Modal Logic}, Cambridge University Press, Cambridge.

\bibitem{Brewka} G. Brewka, 1991, \emph{Nonmonotonic Reasoning: Logical Foundations of Commonsense}, Cambridge University Press, Cambridge.

\bibitem{Brink93} Ch. Brink, 1993, Power Structures, \emph{Algebra Universalis} 30, 177--216.

\bibitem{DepQ} I. Ciardelli, 2016, Dependency as Question Entailment, in \emph{Dependence Logic: Theory and Applications}, Springer,  129--182.

%\bibitem{Ciardelli2016} I. Ciardelli, 2016, \emph{Questions in Logic}, Dissertation, ILLC, University of Amsterdam.



\bibitem{CGRoe} I. Ciardelli, J. Groenendijk \& F. Roelofsen, 2019, \emph{Inquisitive Semantics}, Oxford University Press, Oxford.

%\bibitem{tCKol} B. ten Cate \& Ph. Kolaitis, 2014, Schema Mappings, in \emph{Johan van Benthem on Logic and Information Dynamics}, Springer, Dordrecht, 67--100.

\bibitem{LKB} H. van Ditmarsch, J. Halpern, W. van der Hoek \& B. Kooi, 2015, \emph{Handbook of Epistemic Logic}, College Publications, London.


\bibitem{DHK} H. van Ditmarsch, W. van der Hoek \& B. Kooi, 2007, \emph{Dynamic Epistemic Logic}, Springer Science Publishers, Dordrecht.


\bibitem{DitmarschEtAl} H. van Ditmarsch, W. van der Hoek \& B. Kooi, 2009, Knowing More - from Global to Local Correspondence,    \emph{Proc. of IJCAI-09}, 955--960.


\bibitem{Yifeng} Y. Ding, 2016, Epistemic Logic with Functional Dependency Operator, \emph{Studies in Logic}, 9:4, 55--84.

\bibitem{DHI} Y. Ding, W. Holliday \& Th. Icard, 2020, Logics of Imprecise Comparative Probability, to appear in \emph{Journal of Approximate Reasoning}.

%\bibitem{DHK} H. van Ditmarsch, W. van der Hoek \& B. Kooi, 2007, \emph{Dynamic Epistemic Logic}, Springer, Dordrecht.

\bibitem{vEGWang} J. van Eijck, M. Gattinger \& Y. Wang, 2017, Knowing Values and Public Inspection, \emph{Proceedings 7th Indian  Conference on Logic and its Applications}, Kanpur, 77--90.

\bibitem{FHMV} R. Fagin, J. Halpern, Y. Moses \& M. Vardi, 1995, \emph{Reasoning About Knowledge}, The MIT Press, Cambridge MA.

\bibitem{Fine} K. Fine \& N. Tennant, 1983, A Defense of Arbitrary Objects, \emph{Aristotelean Society Supplementary Volume} 57:1, 55--90.

\bibitem{Galliani} P. Galliani, 2018, Dependence Logic, \emph{Stanford Encyclopedia of Philosophy}, Stanford Univ.

\bibitem{PearleInd} D. Geiger, A. Paz \& J. Pearl, 1991, Axioms and Algorithms for Inferences Involving Probabilistic Independence, \emph{Information and Computation}, 91:1, 128--141.

\bibitem{Gonza} D. Gonzalez, 2019, An Exploration of Matroids and Modal Logic, Dept of Philosophy, Stanford Univ.

\bibitem{G&K} V. Goranko \& A. Kuusisto, 2018, Logics for Propositional Determinacy and Independence, \emph{The Review of Symbolic Logic}, 11:3, 470--506.

\bibitem{Graedel99} E. Graedel, 1999, On the Restraining Power of Guards, \emph{Journal of Symbolic Logic}, 64, 1719--1742.

\bibitem{Aachen2} E. Graedel \& M. Otto, 2020, Guarded Teams: The Horizontally Guarded Case, \emph{Proceedings 28th Annual Conference on Computer Science Logic}, Leibniz Center for Informatics, Schloss Dagstuhl, 22:1--22:17.

\bibitem{GraVa} E. Graedel \& J.  V{\"a}{\"a}n{\"a}nen, 2013, Dependence and Independence, \emph{Studia Logica}, 101:2, 399-410.

\bibitem{G&S97} J. Groenendijk \& M. Stokhof, 1997, Questions, \emph{Handbook of Logic and Language}, Elsevier, Amsterdam, 1009--1053.

\bibitem{HalpCau} J. Halpern, 2016, \emph{Actual Causality}, The MIT Press, Cambridge MA.

\bibitem{Harrah} D. Harrah, 2002, The Logic of Questions, \emph{Handbook of Philosophical Logic}, Springer, Dordrecht, 61--145.

\bibitem{HMT} L. Henkin, D. Monk \& A. Tarski, 1971, \emph{Cylindric Algebra}, Part I,  North-Holland, Amsterdam.

\bibitem{HintQu} J. Hintikka,
1976, \emph{The Semantics of Questions and the Questions of Semantics}, North-Holland, Amsterdam.

\bibitem{HintSand} J. Hintikka \& G. Sandu, 1997, Game-Theoretical Semantics, \emph{Handbook of Logic and Language}, Elsevier, Amsterdam, 361--410.

\bibitem{Hodges} W. Hodges, 1997, Compositional Semantics for a Language of Imperfect Information, \emph{Logic Journal of the IGPL}, 5:4, 539--563.

\bibitem{Holliday} W. Holliday, 2020, Possibility Semantics, in \emph{New Directions in Logic}, College Publications, London.

%\bibitem{Hoogland} E. Hoogland, 2001, Definability and Interpolation, Dissertation DS 2001-05, ILLC, University of Amsterdam.

\bibitem{Humber} L. Humberstone, 2019, Explicating Logical Independence, \emph{Journal of Philosophical Logic},
49 (2020), 135–-218.


\bibitem{Superve} L. Humberstone, 2019, Supervenience, Dependence, Disjunction, \emph{Logic and Logical Philosophy}, 28:1, 3--135.

%\bibitem{KrMints} P. Kremer \& G. Mints, 2007, Dynamic Temporal Logic, in \emph{Handbook of Spatial Logics}, Springer, Dordrecht, 565-606.


\bibitem{deJChagrova} D. de Jongh \& L. Chagrova, 1995, The Decidability of Dependency in Intuitionistic Propositional Logic, \emph{Journal of Symbolic Logic}, 60:2, 498--504.

\bibitem{Konti} J. Kontinen \& F. Yang, 2019, Logics for First-order Team Properties, \emph{Proceedings 26th WOLLIC}, LNCS 11541, 392--413.

\bibitem{Koudijs} R. Koudijs, 2020. Characterization Theorems for LFD, working paper, Institute for Logic, Language and Computation, University of Amsterdam.


\bibitem{Lambal} M. van Lambalgen, 1992, Independence, Randomness, and the Axiom of Choice, \emph{Journal of Symbolic Logic}, 57:4, 1274--1304.

\bibitem{Dazhu} D. Li, 2020, LFD for Dynamical Systems, working paper, Institute for Logic, Language and Computation, University of Amsterdam.

\bibitem{Lomuscio} A. Lomuscio \& M. Ryan, 1998, Ideal Agents Sharing (Some!) Knowledge, \emph{Proceedings ECAI 2008}, John Wiley, Hoboken NJ, 557--561.

\bibitem{MaSaSe} A. Mann, G. Sandu \& M. Sevenster, 2011, \emph{Independence-Friendly Logic}, Cambridge University Press, Cambridge UK.

\bibitem{Marx} M. Marx, 2001. Tolerance Logic, \emph{Journal of Logic, Language and Information}, 10:3, 353--373.

\bibitem{Marx2006} M. Marx, 2006, Complexity of Modal Logic,   \emph{Handbook of Modal Logic}, Elsevier Science, Amsterdam, 139--179,

\bibitem{MarxVenema} M. Marx \& Y. Venema, 1997. \emph{Multi-Dimensional Modal Logic}, Springer, Dordrecht.



\bibitem{Narens} L. Narens, 2007, \emph{Theories of Probability. An Examination of Logical and Qualitative Foundations}, World Scientific, Singapore.


\bibitem{Nem85} I. N{\'e}meti, 1985, The Equational Theory of Cylindric Relativized Set Algebras is Decidable, Preprint 63/85, Mathematical Institute, Hungarian Academy of Sciences, Budapest.

\bibitem{O&R94} M. Osborne \& A. Rubinstein, 1994, \emph{A Course in Game Theory}, MIT Press, Cambridge MA.

%\bibitem{Oxley} J. Oxley, 1992, \emph{Matroid Theory}, Oxford University Press, Oxford.

\bibitem{PaRam} R. Parikh \& R. Ramanujam, 2003, A Knowledge-Based Semantics of Messages,  \emph{Journal of Logic, Language and Information}, 12, 453--467.

\bibitem{Pearl} J. Pearl, 2009, \emph{Causality: Models, Reasoning and Inference}, Cambridge University Press.

\bibitem{Plaza} J. Plaza, 2007, Logics of public communications, \emph{Synthese}, 158:2, 165–179.

\bibitem{Aachen1} P. Pützstück, 2020, \emph{Exploring LFD}, Bachelor Thesis, Department  of Informatics, RWTH Aachen.

%\bibitem{Pearl} J. Pearl, 2000, \emph{Causality}, The MIT Press, Cambridge MA.

\bibitem{Scott71} D. Scott, 1971, On Engendering an Illusion of Understanding, \emph{Journal of Philosophy}, 68:21, 787--807.

\bibitem{T&S} A. Troelstra \& H. Schwichtenberg, 2000,  \emph{Basic Proof Theory}, Cambridge University Press, Cambridge.

\bibitem{Vaana} J. V{\"a}{\"a}n{\"a}nen, 2007, \emph{Dependence Logic: A New Approach to Independence Friendly Logic}, Cambridge University Press, Cambridge.

\bibitem{VaanaModal}, J. V{\"a}{\"a}n{\"a}nen, 2008, Modal Dependence Logic, \emph{New Perspectives on Games and Interaction (Texts in Logic and Games)}, Amsterdam University Press, 237-–254.

\bibitem{FanVaana} F. Yang \& J. V{\"a}{\"a}n{\"a}nen, 2016, Propositional Logics of Dependence, \emph{Annals of Pure and Applied Logic}, 167(7): 557–-589.

\bibitem{Venema} Y. Venema, 1995, Cylindric Modal Logic, \emph{Journal  of Symbolic Logic} 60:2, 591--623.

%\bibitem{Venema} Y. Venema, 1995, Cylindric Modal Logic, \emph{Journal of Symbolic Logic}, 60:2, 591--623.

\bibitem{Yanjing} Y. Wang, 2018, Beyond Knowing That: A New Generation of Epistemic Logics,
     \emph{Jaakko Hintikka on Knowledge and Game Theoretical Semantics}, Springer, Dordrecht, 499-533.

%\bibitem{Yang} F. Yang, 2014, \emph{On Extensions and Variants of Dependence Logic}, Dissertation, University of Helsinki.

\bibitem{Xie} K. Xie, \emph{Where Causality, Conditionals and Epistemology Meet}, PhD dissertation, ILLC, University of Amsterdam, 2020.

}
\end{thebibliography}
\end{document}